\documentclass[fleqn,usenatbib]{mnras}

\usepackage{newtxtext,newtxmath}

\usepackage[T1]{fontenc}

\usepackage{booktabs, multirow}
\usepackage{tabularx}
\usepackage{longtable}

\DeclareRobustCommand{\VAN}[3]{#2}
\let\VANthebibliography\thebibliography
\def\thebibliography{\DeclareRobustCommand{\VAN}[3]{##3}\VANthebibliography}

\usepackage{graphicx}	
\usepackage{amsmath}	
\usepackage[dvipsnames]{xcolor} 
\usepackage{arydshln} 
\usepackage{gensymb} 

\newcommand{\Msol}{M_{\odot}}
\newcommand{\ang}{\textup{\AA}}

\title[NLTE KNa Spectra]{NLTE Spectra of Kilonovae}

\author[Q. Pognan et al.]{
Quentin Pognan$^{1}$\thanks{E-mail: quentin.pognan@astro.su.se},
Jon Grumer$^{2}$,
Anders Jerkstrand$^{1}$,
Shinya Wanajo$^{3}$
\\
$^{1}$The Oskar Klein Centre, Department of Astronomy, Stockholm University, AlbaNova, SE-10691 Stockholm, Sweden\\
$^{2}$Theoretical Astrophysics, Department of Physics and Astronomy, Uppsala University, Box 516, SE-751 20 Uppsala, Sweden \\
$^{3}$Max-Planck-Institut f\"ur Gravitationsphysik (Albert-Einstein-Institut), Am M\"uhlenberg 1, D-14476 Potsdam-Golm, Germany\
}

\date{Accepted XXX. Received YYY; in original form ZZZ}

\pubyear{2023}

\begin{document}
\label{firstpage}
\pagerange{\pageref{firstpage}--\pageref{lastpage}}
\maketitle

\begin{abstract}
The electromagnetic transient following a binary neutron star merger is known as a kilonova (KN). Owing to rapid expansion velocities and small ejecta masses, KNe rapidly transition into the Non-Local Thermodynamic Equilibrium (NLTE) regime. In this study, we present synthetic NLTE spectra of KNe from 5 to 20 days after merger using the \texttt{SUMO} spectral synthesis code. We study three homogeneous composition, 1D multi-zone models with characteristic electron fractions of $Y_e \sim 0.35, 0.25$ and $0.15$. We find that emission features in the spectra tend to emerge in windows of reduced line blocking, as the ejecta are still only partially transparent even at 20 days. For the $Y_e \sim 0.35$ (lanthanide-free) ejecta, we find that the neutral and singly ionised species of Rb, Sr, Y and Zr dominate the spectra, all with good potential for identification. We directly test and confirm an impact of Sr on the $10000~\ang$ spectral region in lanthanide-free ejecta, but also see that its signatures may be complex. We suggest the Rb\,\textsc{i} $\rm{5p^{1}}$- $\rm{5s^{1}}$ 7900 \AA~transition as a candidate for the $\lambda_0 \sim$ 7500--7900 $\ang$ P-Cygni feature in AT2017gfo. For the $Y_e \sim 0.25$ and $0.15$ compositions, lanthanides are dominant in the spectral formation, in particular Nd, Sm, and Dy. We identify key processes in KN spectral formation, notably that scattering and fluorescence play important roles even up to 20 days after merger, implying that the KN ejecta are not yet optically thin at this time.
\end{abstract}




\section{Introduction}
\label{sec:introduction}

Binary neutron star (BNS) mergers are accepted to produce transients known as kilonovae (KNe), powered by the radioactive decay of heavy elements synthesised by rapid neutron capture (r-process) \citep[][]{Symbalisty.Schramm:82,Eichler.etal:89,Li.Paczynski:98,Freiburghaus.etal:99,Rosswog.etal:99,Metzger.etal:10}. The properties of BNS merger ejecta, such as mass, velocity, and composition, have been studied in detail with hydrodynamical simulations and nuclear network calculations \citep[e.g.][]{Rosswog.etal:99,Martinez.etal:07,Metzger.etal:10,Wanajo.etal:14,Rosswog.etal:18}. Of particular interest is the question as to whether KNe alone can reproduce the measured r-process solar distribution \citep[see e.g.][]{Lodders.etal:09,Rosswog.etal:18,Prantzos.etal:20,Nedora.etal:21,Lodders:21}. The first, and thus far only complete KN observation, of AT2017gfo \citep[see e.g.][for a review]{Abbott.etal:17,Margutti.Chornock:21}, has allowed some initial answers to be given to the question of BNS mergers as the dominant source of r-process elements \citep[][]{Rosswog.etal:18,Metzger:19,Cote.etal:19,Arcones.Thielemann:2023}.

In order to conclusively answer this question, positive identification of individual elements and determination their abundance in KN ejecta will have to occur. The first strong candidate identification was that of strontium (Sr), as identified from a photospheric phase P-Cygni line \citep[][]{Watson.etal:2019,Domoto.etal:21}, though a possible contribution from helium (He) has also been suggested for this feature \citep{Perego.etal:22,Tarumi.etal:23}. Subsequently, plausible signatures have been identified for lanthanum (La) and cerium (Ce) \citep{Domoto.etal:22}, yttrium (Y) \citep{Sneppen.Watson:23Y} and tellurium (Te) \citep{Gillanders.etal:23,Hotokezaka2023}. A comprehensive analysis and review of possible candidate species for various spectral features has been carried out by \citet{Gillanders.etal:21,Gillanders.etal:22,Gillanders.etal:23}. Robust element identification is a difficult challenge due to the complex composition and morphology, extensive but in parts not well known line lists, and high expansion velocities ($\gtrsim 0.1~$c) leading to complex radiative transfer and line blending. Identifying the presence or absence of groups of elements based on properties such as expansion opacity is an also an avenue currently being explored \citep[see e.g.][]{Tanaka.Hotokezaka:13,Tanvir.etal:17,tanaka:opacities:2020,Banerjee.etal:20,Banerjee2022,Domoto.etal:22,Fontes2023,Deprince2023,CarjavalGallego2023}. 

Spectral identification of species requires knowledge of the atomic properties of that species, notably its level structure and associated radiative transitions. Recent studies in theoretical atomic data for heavy r-process elements have made great advances in the completeness and accuracy of energy levels and transition probabilities (Einstein A-values) \citep[see e.g.][]{Kasen.etal:13,Gaigalas.etal:19,Gaigalas2020,Banerjee.etal:20,Fontes.etal:20,Radziute.etal:20,Bromley2020,Carvajal2021,Rynkun2022,Domoto.etal:22,McCann2022,Flors.etal:23}. These advances, and in particular recent efforts to calibrate theoretical data to experimental values, are extremely useful for the spectral modelling of KNe across all epochs.  

Almost all light curve and spectral modelling so far has been conducted for the early epochs during the photospheric phase \citep[see e.g.][]{Tanaka.etal:18,Wollaeger2018,Barnes2021,Domoto.etal:22,Kawaguchi.etal:21,Gillanders.etal:22,Just2022,Bulla:2023,Viera2023}. For the most part, the modelling method used in these studies is almost invariably time-dependent Monte Carlo radiative transfer. Although the 3D aspects of the problem have been addressed and modelled from the outset, these first generation of models rely on an LTE assumption for the gas state, and radiative transfer using expansion opacities with thermal resampling. The recent paper by \citet{Shingles2023} is the first one extending the transfer treatment to include also fluorescence which is found to be important in KNe. One should also be aware that fundamental equations used can vary between LTE models. For example, some codes calculate temperature by balancing heating with cooling \citep{Kasen.etal:13,Wollaeger2018,Wollaeger.etal:21}, while others compute a characteristic radiation field temperature and then equate this to the electron temperature \citep{Tanaka.Hotokezaka:13,Kawaguchi.etal:21,Bulla:2023}. The results can vary quite dramatically; the reader is referred to the supernova code comparisons in \citet{Blondin2022} for some illustration.

The KN ejecta transition to the NLTE regime around $\sim 5$ days after the merger \citep{Hotokezaka.etal:21,Hotokezaka.etal:22,Pognan.etal:22b}. In NLTE, the ionisation and excitation structure of the ejecta are found by solving rate equations, with many different processes, including non-thermal ionization, thermal collisional excitation, photo-excitation, and recombination, all playing a role. As such, modelling of KNe in this regime requires significantly more atomic data than in the LTE phase, in which only energy levels and A-values are needed. For many of these processes, only sparse r-process data are available so far \citep[e.g.][for electron collision strengths]{McCann2022}, so modelling such processes relies to a large extent on generic formulae. Detailed radiative transfer is needed to properly account for the effects of photoionisation (PI) and photo-excitation (PE), with the latter process leading to scattering and fluorescence. In the low density, fast moving dynamical ejecta, time-dependent effects on recombination and cooling processes may also play a role \citep[][]{Pognan.etal:22a}, in addition to the time-dependent thermalisation of radioactive decay products \citep{Barnes.etal:16,Waxman.etal:18,Kasen.Barnes:19,Hotokezaka.etal:20}. Accurate spectral modelling in this regime represents both a physical and computational challenge, with only a few studies making use of NLTE physics to varying degrees existing so far \citep[][]{Hotokezaka.etal:21,Hotokezaka.etal:22,Pognan.etal:22a,Pognan.etal:22b,Tarumi.etal:23,Gillanders.etal:23,Hotokezaka2023}.

In this study, we present 1D, NLTE KN spectra calculated by the spectral synthesis code \textsc{sumo} \citep[][]{Jerkstrand:11,Jerkstrand.etal:12}, adapted to KN simulations as described in \citet{Pognan.etal:22a}. We compute spectra at 5--20 days post-merger; the first epoch reflecting the results found in \citet{Pognan.etal:22b}, which showed that the bulk of the KN ejecta transition to NLTE conditions around 5 days after merger, and the last one reflecting a plausible final optical/NIR detectability of future events. We construct three ejecta models with varying compositions using the data of \citet{Wanajo.etal:14}, all with the same ejecta mass and density profile. The goal of this study is not to identify which species give rise to which specific features in the observed spectrum of AT2017gfo (between 5--10 days), as both the simplified ejecta structure, and insufficient wavelength accuracies of our line lists prevent doing this robustly. Instead, our goal is to identify key processes in post-diffusion phase spectral formation (including information on when an 'optically thin' limit is reached), and key elements that play important overall roles in shaping the emergent spectrum. By this, we provide guidance to the hydrodynamic,  nuclear, atomic, and radiative transfer communities for the most important diagnostic aspects of KNe at 5--20 days, and in particular which atoms and ions appear to be the most crucial to obtain better atomic and nuclear data for.

The paper is structured as follows. In Section \ref{sec:models}, we present the models employed in this study, including the composition and associated energy deposition. We present the spectral synthesis methodology in Section \ref{sec:sumo_sim}. We investigate the resulting thermodynamic state of the ejecta in Section \ref{sec:thermo_results} and emergent spectra in Section \ref{sec:spectral_results}. In section \ref{sec:AT2017gfo} we discuss the models in the context of AT2017gfo. We discuss the implications of our findings, and summarise future directions, in Section \ref{sec:discussion}. 

\section{Ejecta Models}
\label{sec:models}

\begin{figure}
    \includegraphics[trim={0.cm 0.1cm 0.4cm 0.3cm},clip,width = 0.49\textwidth]{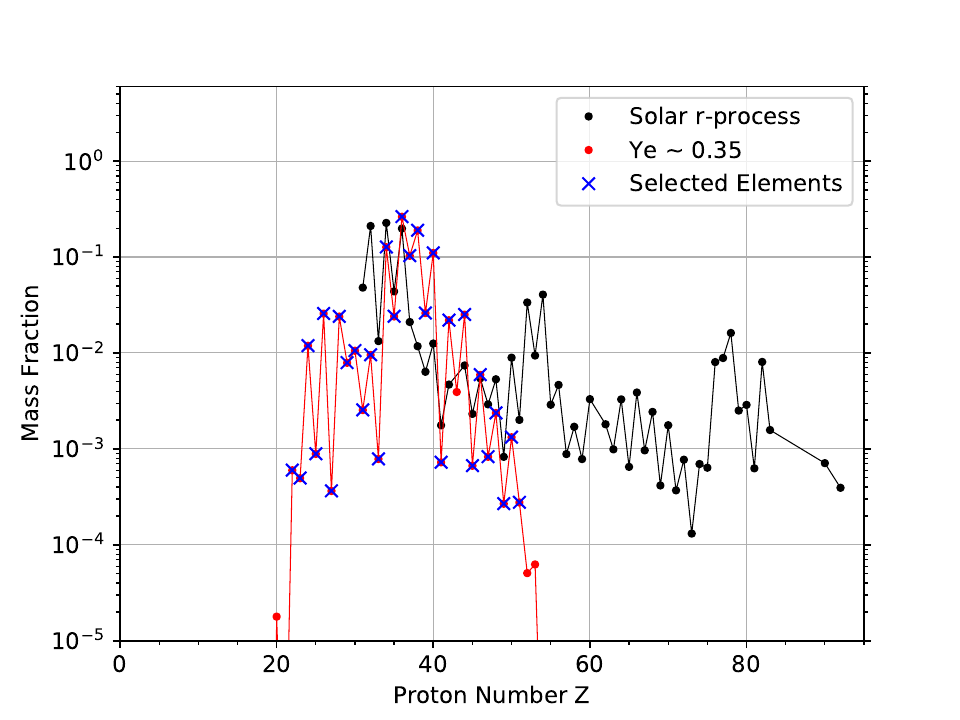} 
    \includegraphics[trim={0.cm 0.1cm 0.4cm 0.3cm},clip,width = 0.49\textwidth]{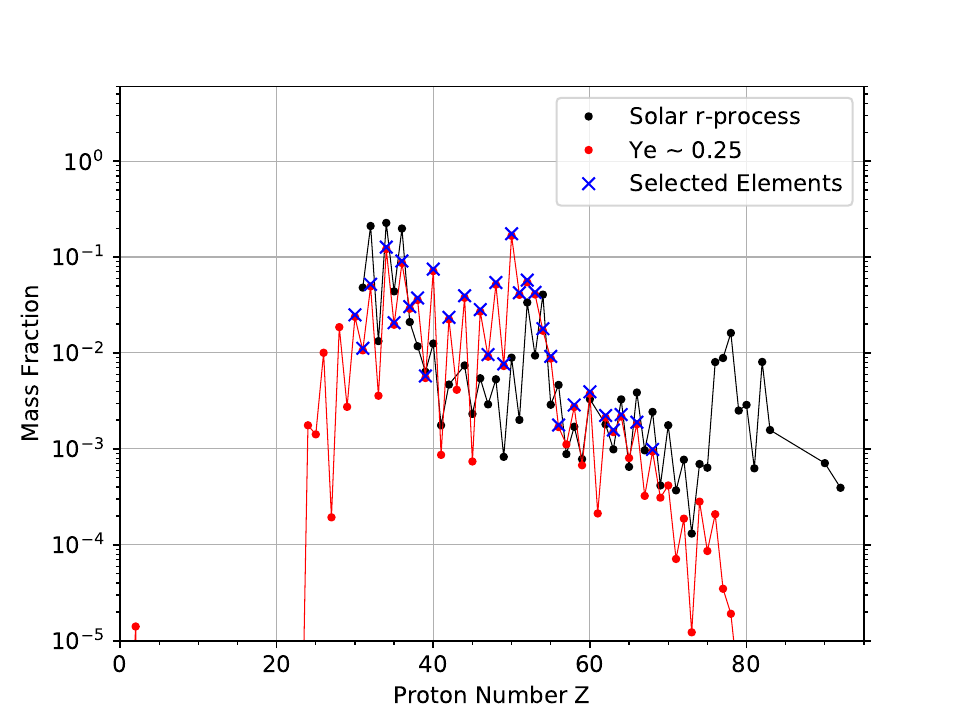}
    \includegraphics[trim={0.cm 0.1cm 0.4cm 0.3cm},clip,width = 0.49\textwidth]{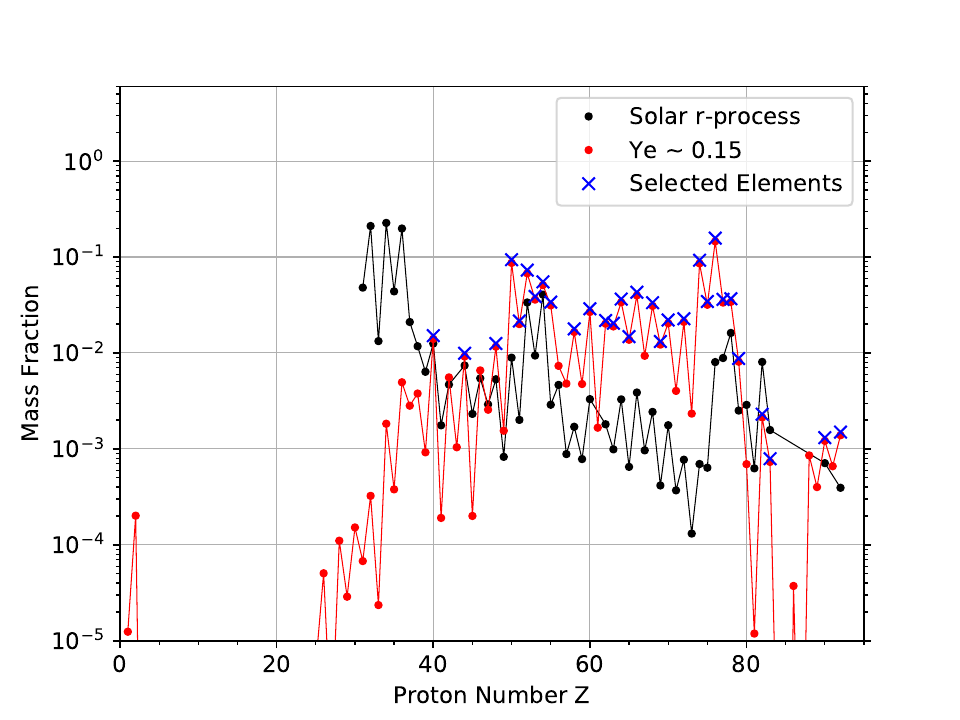}
    \caption{Abundance patterns for the three different $Y_e$ models studied here. The black points represent the normalised solar r-process residual pattern taken from \citet{Prantzos.etal:20}. The red points are the outputs resulting from the nuclear network calculations of \citet{Wanajo.etal:14}, taken at 10 days after merger, while the blue crosses are the 30 chosen elements by model. Our $Y_e \sim 0.15$ and $Y_e \sim 0.25$ models contain only trans-iron elements, whereas the $Y_e \sim 0.35$ model contains also some lighter elements down to Ti.}
    \label{fig:compositions}
\end{figure}

We study three uniform composition, spherically symmetric multi-zone models, characterised by light, intermediate, and heavy r-process elements. The compositions of the models can be found in Fig. \ref{fig:compositions} and are tabulated in Table \ref{tab:compositions}. Each model has a total ejecta mass of $M_{\rm{ej}} = 0.05~\rm{\Msol}$, and a density profile of $\rho \propto v^{-4}$, consistent with bulk ejecta from early time dynamical ejecta, and later viscosity driven disc wind ejecta \citep[see e.g.][]{Kawaguchi.etal:21}. Homologous expansion $v = r/t$ is assumed. The models consist of 5 radial zones, with the inner boundary at $v_{\rm min} = 0.05~$c and the outer boundary at $v_{\rm{max}} = 0.3~$c. The zones are spaced linearly with a uniform velocity width $v_{\rm{step}} = 0.05~$c. With this setup, the mass distribution in the model zones is 60, 20, 10, 6 and 4 per cent, from the innermost to outermost zone. The models are placed at a distance of 40 Mpc from the observer in order to simplify comparisons to AT2017gfo, which was measured at this distance \citep[e.g.][]{Abbott.etal:17,Pian2017,Smartt.etal:17,Margutti.Chornock:21}.  

\citet{Wanajo.etal:14} present composition and radioactive decay power in various channels, versus time, for different $Y_e$ values. Using this data, we generate three compositions, and their associated radioactive decays, from Gaussian $Y_e$ distributions described by  $\mu_{Y_e} = 0.35, 0.25, 0.15$, and $\sigma_{Y_e} = 0.04$. By using a small range of $Y_e$ values we avoid dependency on specifics, while still staying focused on the characteristic $Y_e$ values \citep[see also e.g.][]{tanaka:opacities:2020}. These are still essentially 'single component' compositions, and so will not reproduce the solar r-process residual pattern, or the total composition in individual KNe which simulations indicate contain multiple components with different $Y_e$ values. We also consider a fourth model equivalent to the $Y_e \sim 0.35$ model, but with strontium (Sr) removed, in order to study its effect in the lanthanide-free ejecta 
case (Section \ref{subsec:SrII_focus}).

We limit the composition input to \textsc{sumo} for each model to 30 elements, as shown in Fig. \ref{fig:compositions}, each with four ionisation stages from neutral to triply ionised. The included elements are chosen mainly with respect to their abundance in the model, but also from consideration to cover different parts of the periodic table. For example, the composition with $Y_e \sim 0.15$ is the only one to have a significant actinide abundance at 5--20 days, and so we include both thorium (Th) and uranium (U) even though their abundances are quite low (0.13 and 0.15 per cent, respectively). We limit ourselves to stable elements due to the the lack of atomic data for heavy elements with only unstable isotopes. This omission is expected to only have any impact for our $Y_e \sim 0.15$ model, which synthesises such heavy elements. In order to avoid changing our model compositions for every epoch, we take the composition of each model to be that of the nuclear network abundance at 10 days, corresponding roughly to the middle of the range of epochs that we study (5--20 days). We note that this does imply a small inconsistency between the energy deposition and the abundances put into \textsc{sumo}, as the radioactive power is calculated from an evolving composition \citep{Wanajo.etal:14}. However, there is relatively little variation in composition for our models during this time range, so this simplification should not have any significant impact, and certainly much smaller than that arising from uncertainties in the theoretical atomic data, described below. We note that the $Y_e \sim 0.15$ model has $25$ per cent of lanthanides by mass fraction, a composition which is too lanthanide-rich to represent the entire KN ejecta. Rather, this composition is more relevant to dynamical ejecta which avoids any neutrino irradiation from disc or remnant winds.

\subsection{Atomic Data}
\label{subec:atomic_data}
In order to achieve a consistent and complete set of atomic energy levels and processes, fundamentally bound-bound radiative transitions, for all elements and relevant ions from Fe to U, it is necessary to restrict the complexity of the applied atomic structure method significantly. To this end, in this work, we employ a spectroscopic configuration-interaction (SCI) model where the configuration space is limited to include only those that represent the spectroscopic (physical) states. This can be thought of as a correlation-limited model that only includes the most fundamental many-electron effects beyond the single-configuration Dirac--Hartree--Fock approach, but practically allows for computations of all relevant energy levels and processes in a single calculation for each ion. The accuracy of the data computed with such small-scale ab-initio atomic structure models is not good enough for e.g. spectral identifications of individual lines from a given ion, but this approach allows for a statistically complete physical model for each relevant ion in the periodic table.

The SCI calculations were performed with the Flexible Atomic Code (\textsc{Fac}) \citep{Gu:08}, which is based on a Dirac--Fock--Slater scheme for the orbital optimizations, and employs a standard Dirac--Coulomb Hamiltonian for the structure calculations. In this work, we also include the Breit interaction in the low-frequency limit and leading quantum-electrodynamical effects (vacuum polarization and electron self-energy). The Dirac--Fock--Slater method works on an average configuration to optimize a common, local central potential. We refer to the code documentation\footnote{For further information and documentation of the \textsc{Fac} code, consult the open-source GitHub repository: \url{https://github.com/flexible-atomic-code/fac}.} for further details. 

The present SCI model includes spectroscopic configurations compiled from comparisons with the NIST Atomic Spectra Database (ASD) \citep{NIST_ASD}, the Vienna Atomic Line Database (VALD3\footnote{VALD is hosted on e.g. \url{http://vald.astro.uu.se/}), which contains further information and documentation.} database \citep{VALD3_2015} that, for rare earth elements in particular, includes the data compiled in the DREAM database \citep{DREAM_1999}, and with earlier similar works targeting consistent multi-element calculations (most notably the early \textsc{Autostructure} calculations by \cite{Kasen.etal:13} and the extensive \textsc{Hullac} calculations by \cite{Tanaka.etal:18}). For each ion, we start by optimizing the common central potential on the ground configuration and extend to additional low energy configurations when deemed necessary to balance screening effects between states of varying subshell structures. This procedure is made possible by comparison to the energy level tables of the NIST ASD. Following the energy level computations, we remove all states with ab-initio energies above the first ionization limit. The remaining set of states is then used in subsequent computations of all relevant radiative bound-bound rates, including allowed electric dipole (E1), and also forbidden magnetic dipole (M1) and electric quadrupole (E2) transitions which, in our investigations, have proved important for the thermodynamic properties of the ejecta. The fundamental parameters defining the model of each ion included in this work are summarized in Tab. \ref{tab:atomic_data} of Appendix \ref{app:atomic_data}.

Following the ab-initio calculations described above, the energy levels of Sr\,\textsc{ii} were rescaled to match those found in the NIST database \citep{NIST_ASD}. Since Sr\,\textsc{ii} is a species of key interest in KN modelling due to its claimed detection in previous works \citep{Watson.etal:2019,Domoto.etal:22}, accurate modelling of the spectral features arising from Sr\,\textsc{ii} is of particular importance, specifically the emission and absorption arising from the strong transitions of the $4000~\ang$ doublet, the $6800~ \ang$ doublet, and $10000~\ang$ triplet. In order to do so, 6 levels from the NIST ASD were added to the theoretical data, bringing the total number up to level 27 when energy ordered, with level 27 corresponding to the highest lying state with available A-value data. The levels were then rescaled to their experimental values, and the transition wavelengths correspondingly adjusted. The theoretical A-values were kept, apart from transitions where NIST A-values were available to a precision of 'B' or better, corresponding to an estimated accuracy of $\leq 10$ per cent. We note that the theoretical transition strengths for these were all within an order of magnitude to the values found in NIST, supporting the accuracy of our model atom for Sr\,\textsc{ii}. 

The first 8 levels of Y\,\textsc{i} were also rescaled to the values found in the NIST database \citep{Palmer:77} in order to improve the wavelength accuracy of low lying strong transitions found to be important in our models. The higher lying levels above this were not modified. The A-values for Y\,\textsc{i} were kept to be those of the ab-initio calculations as no transition probabilities were found for the relevant transitions from these low lying states. 

\section{Spectral Simulation Methods}
\label{sec:sumo_sim}

We use the NLTE spectral synthesis code \textsc{sumo} \citep[][]{Jerkstrand:11,Jerkstrand.etal:12} to generate spectra of KNe in 1D. The majority of the physics used in the code is described in the papers above, and adaptations to KNe in \citet{Pognan.etal:22a}. We summarise here the processes particularly important for this study.

\subsection{Energy Deposition}
\label{subsec:Energy_dep}

The raw radioactive power per unit mass, $\dot{Q}$, for each model is constructed from the relevant weighting of different single $Y_e$ components from the nucleosynthesis calculations of \citet{Wanajo.etal:14}, self-consistently with the evolving abundance. The total raw power is a sum of all decay products: neutrinos, $\alpha$-decay (He nuclei), $\beta$-decay (electrons and positrons), $\gamma$-decay (gamma rays), and spontaneous fission (SF) (heavy nuclei). Each of these decay products thermalise differently and depending on the the ejecta density as well as the ordering of the magnetic fields. For $\alpha$, $\beta$ and $\gamma$-decay, we follow the thermalisation prescription of \citet{Kasen.Barnes:19}, with additional considerations from \citet{Waxman.etal:19} for $\beta$-decay. Here, the calculation is conducted for mildly relativistic electrons as opposed to non-relativistic electrons, such that the thermalisation efficiency drops off in a steeper manner (see Appendix \ref{app:model_extra}), which is found to a better analytical fit to numerical calculations \citep[see e.g.][]{Hotokezaka.etal:20}. For spontaneous fission, the formalism of \citet{Barnes.etal:16} is adopted. Neutrinos are assumed to completely escape the ejecta at all times. The details of our thermalisation treatment can be found in Appendix \ref{app:model_extra}. The thermalisation efficiency of charged particles will depend on the geometry of magnetic fields inside the ejecta \citep[][]{Barnes.etal:16, Waxman.etal:19}, and we here assume an arbitrary configuration which gives local particle trapping. The total energy input in a given zone, per unit mass, is given by:

\begin{equation}
    \dot{q}_{\rm{tot}} (t) = \sum_i \dot{Q}_i f_i(t) \; \rm{erg \; s^{-1} \; g^{-1}}
    \label{eq:energy_dep}
\end{equation}

\noindent where the subscript $i$ indicates a decay product and $f_i$ is the corresponding thermalisation factor.

The deposited power $\dot{q}_{\rm tot}$ is channeled into heating and ionisation by solving the Spencer--Fano equation \citep[][]{Spencer.Fano:54,Kozma.Fransson:92} for the cascading of high-energy electrons. By this treatment, we make an assumption that $\alpha$-particles and fission fragments create a population of high-energy electrons in a similar manner as leptons and gamma-rays. Leptons and atomic nuclei have relatively similar ionization loss rates \citep{Longair2011}, and in addition, the ensuing cascade distribution is almost independent of the initial injection energy, motivating this. Since we currently lack high-energy collisional excitation cross sections for r-process elements, we do not include non-thermal excitations in the solution. The omission of this channel is expected to have limited impact on the spectral formation, as in the relatively ionized conditions of KN ejecta ($x_e \sim$ 1--2), the vast majority of the energy ($\gtrsim 99$ per cent) goes towards heating. The treatment of non-thermal collisional ionisation is addressed below in Section \ref{subsec:ionisation}. The energy deposition for each model versus time can be visualised in Fig. \ref{fig:energy_dep}.

For the $Y_e \sim 0.35$ and $Y_e \sim 0.25$ compositions, $\beta$-decay is the only important channel at 5--20 days, whereas for the $Y_e \sim 0.15$ model $\alpha$-decay and spontaneous fission also contribute, with fission becoming as important as $\beta$-decay at 20 days. The total deposition varies quite significantly between the models, e.g. at 10 days it is $\sim 4\times 10^{39}$ erg s$^{-1}$ in the $Y_e\sim 0.35$ model, $1 \times 10^{40}$ erg s$^{-1}$ in the $Y_e\sim 0.25$ model, and $2.4\times 10^{40}$ erg s$^{-1}$ in the $Y_e\sim 0.15$ model (see Fig. \ref{fig:energy_dep}).

\subsection{Temperature Calculation}
\label{subsec:temperature}

The temperature in each zone is found from solving the time-dependent first law of thermodynamics, with heating due to radioactivity, free-free absorption, and photoionisation, and cooling due to net thermal electron bound-bound deposition (giving line emission), free-free, free-bound, and bound-free (collisional) cooling, and (in time-dependent mode) adiabatic cooling. For $t\leq$ 10d we use the steady-state approximation instead of solving the full time-dependent equation. The continuum-involving cooling channels are found to be unimportant ($\lesssim 0.1$ per cent), while adiabatic cooling may start to play a role for low density zones at late times, following the results found in \citep[][]{Pognan.etal:22a}. 

The evolution of KN temperatures in the post-diffusion phase has previously been studied by \citet{Hotokezaka.etal:20,Pognan.etal:22a}, who showed that temperatures in the NLTE regime generally increase with time, at least as long as steady-state conditions hold. This is qualitatively different to LTE-model temperatures, in which both thermal balance \citep[e.g.][]{Wollaeger2018,Wollaeger.etal:21} and $T_e=T_r$ \citep[e.g.][]{Tanaka.Hotokezaka:13,Gillanders.etal:21,Bulla:2023,Collins2023} approaches typically give monotonic temperature decreases with time.

\subsection{Ionisation Structure}
\label{subsec:ionisation}
The NLTE ionisation structure of the ejecta is calculated by solving the rate equations for each ionisation state. The ionisation processes considered are non-thermal (NT) collisional ionisation, thermal collisional ionisation (TI), and photoionisation (PI), such that $\Gamma_{\rm{tot}} = \Gamma_{\rm{NT}} + \Gamma_{\rm{TI}} + \Gamma_{\rm{PI}}$, where thermal collisional ionisation has been added for this study. We treat this by the formalism of \citet{Shull.Steenberg:82}:

\begin{equation}
\begin{split}
\Gamma_{\rm{TI}} =& 1.3 \times 10^{-8} F \xi I_{ev}^{-2} T^{1/2} \left(1+ a \frac{kT}{I}\right)^{-1} \\
& \times \exp{\left(-I/kT\right)}~\mbox{cm}^3~\mbox{s}^{-1}
\end{split}
\end{equation}

\noindent where $I$ is the ionization potential, and we take $F=1$, $\xi=1$, and $a=0.1$ \citep[see][for a discussion of these parameters]{Shull.Steenberg:82}. PI and NT ionisation are treated as in \citet{Pognan.etal:22a}. Due to lack of atomic data for these processes, we use a hydrogenic cross-section for PI \citep{Rybicki.Lightman:79}, and the \citet{Lotz:67} formalism for NT cross-sections. We only apply the NT cross-section to the valence shell, as treatment of inner shell ionisation requires modelling of Auger processes and X-ray fluorescence not currently included for r-process elements. Though these contributions to ionisation should be smaller than that of the valence shell, they may add up to be non-negligible. As such, it is possible that our NT ionisation rates are somewhat underestimated.

We find that PI typically dominates ionisation rates for neutral and single ionised species, while NT ionisation dominates for doubly ionised species. This is largely due to the higher ionisation potentials for more highly ionized ions, making photoionisation by the moderate-energy radiation field less effective. Thermal collisional ionisation is never found to dominate, consistent with the analytic estimates of \citet[][]{Pognan.etal:22a}. 

For recombination, we use a constant recombination rate $\alpha = 10^{-11} \; \rm{cm^3 \; s^{-1}}$, as in \citet{Pognan.etal:22a,Pognan.etal:22b}. As there is a critical lack of data for total (radiative and dielectronic) rates for r-process elements, the usage of a constant value for recombination is an assumption that was made by comparing known rates of light elements (see Appendix C of \citet{Pognan.etal:22a}), alongside a limited calculation for Nd \citep{Hotokezaka.etal:21}. The total recombination rates for Nd for temperatures of $T \gtrsim 10^3~$K relevant to this study were found to be $\alpha \lesssim 5 \times 10^{-10} ~ \rm{cm^3 \, s^{-1}}$, with values decreasing as temperature increases. As such, for the temperatures we find ($T \sim 3000 - 35000~$K), our fiducial rate is typically within an order of magnitude of calculated rates.

Time-dependent effects (steady-state breakdown) may begin to affect the ionisation solution when the recombination time, $t_{\rm{rec}} = 1/(\alpha n_e)$, becomes a significant fraction of the evolutionary time \citep[][]{Pognan.etal:22a}. Analytic considerations indicate that the low density outer ejecta layers may become prone to time-dependent effects at $t \gtrsim 10$ days for our model. As such, we run \textsc{sumo} in time-dependent mode from 10 days onwards. 
 
\subsection{Excitation Structure and Radiation Field}
\label{subsec:excitation}

The NLTE excitation structure within the ejecta is calculated by solving the rate equations for ground and excited states. Certain (de)-excitation processes, such as collisional processes and spontaneous radiative de-excitation by allowed channels, will always be faster than the evolutionary time-scale. As such, the excitation structure of the ejecta is always calculated under the steady-state assumption. 

Since KNe are expected to mainly cool by spontaneous line emission, thermal collision strengths have a direct impact on the ejecta temperature. The effective collision strength from lower level \textit{l} to upper level \textit{u}, $\Upsilon_{l,u}$, is treated differently depending on whether the transition is allowed or forbidden, distinguished by the (dimensionless) oscillator strength $f_{osc}$ \citep[e.g. equation 2.68 of][]{Rutten:03}. Allowed transitions are taken to have $f_{osc} \geq 10^{-3}$, for which the effective collision strength is found using the formula derived by \citet{Regemorter:62}, while forbidden transitions with $f_{osc} < 10^{-3}$ are instead calculated using the formula from \citet{Axelrod:80}. Values calculated by this formula typically range between 0.01 -- 1 depending on the transition. Both of the above equations are commonly used in NLTE radiative transfer codes \citep[see e.g.][]{Botyanski.Kasen:17,Shingles.etal:20,Hotokezaka.etal:21}, though some KN studies have taken transition strengths for forbidden transitions instead to be fixed \citep[e.g.][]{Hotokezaka.etal:21}. 

The radiative transfer used by \textsc{sumo} is described in great detail in \citet{Jerkstrand.etal:11,Jerkstrand.etal:12}, and has not been significantly modified since. We describe here some details particularly relevant for this study. The radiation field is likewise not treated in a time-dependent fashion, but the assumption of $c = \infty$ is made (except for in Doppler shift terms). This steady-state assumption for the radiation field is sometimes called the stationarity approximation. This approximation is formally only valid when the photon transport time is short compared to the evolutionary time, which may not necessarily be the case for KNe in the range of epochs studied here. Quantification of the diffusion phase from theory is difficult as it is influenced by fluorescence which has so far not been included in KN modelling, except for in the recent model presented by \citet{Shingles2023}. Diffusion effects in general lead to the emergent bolometric luminosity not tracking the instantaneous energy deposition (after thermalisation). As the optical depth of the ejecta drops in post-peak times, the radiation transport time approaches the free-streaming limit $v_{ej}t/c$, extended by path enhancements due to scattering caused by remaining line opacity \citep[see][for a discussion]{Jerkstrand.etal:16}. The effect of a steady-state radiation field approximation has been previously conducted in the context of Type Ia supernovae, where it was found that while flux levels are naturally not reproduced over the diffusion phase, the spectrum remains accurate \citep{Kasen.etal:06,Shen.etal:21}. We will return to a discussion of possible diffusion effects when comparing our models to AT2017gfo in section \ref{subsec:LC_evolution}. 

Since our photon packets do not track time, we also do not capture the effects of different travel times from last interaction point to the observer. That time-scale is given by $t_{\rm travel} \sim 2 v_{\rm ej}t/c$, where $v_{\rm ej}$ is the characteristic ejecta expansion velocity. For our model, the mass-weighted mean velocity is 0.11c, so $t_{\rm travel} \sim 0.22t$. As long as radioactive power follows the canonical $t^{-1.3}$, power levels change on a time-scale $t_{\rm power}=t/1.3 \approx 0.8t$ (or somewhat shorter if the decreasing thermalisation efficiency is also considered). Density changes on the homology time-scale $0.33t$. Thus, $t_{\rm travel} < t_{\rm power},t_{\rm density}$, though it is not much smaller, so some mild/moderate effects can be expected. This however, mostly pertains to the specific shape of individual line profiles which is not the focus of study here.

Processes like fluorescence and scattering may play important roles in the spectral formation of KN in the 5 -- 20 day period \citep{Shingles2023}. We treat these processes in full detail using the standard SUMO line-by-line transfer method with full fluorescence. This is a hybrid Monte Carlo/ray tracing method, in which packets can be either fully or partially absorbed depending on the nature of the opacity. Lines are divided into the groups of coupled or uncoupled to the NLTE solutions, with an adjustable cut-off for each ion. Here, we compute all models with all lines fully coupled, such that all are available as (de)excitation channels, therefore allowing all fluorescence and resonance transitions. In these line interactions, photon packets are attenuated upon passing a line, with the corresponding power increasing the PE estimators. Radiative deexcitation following a PE is resonance scattering if it is by the same transition, and fluorescence if by other transitions. For continuum interactions, a random draw determines whether the packet Thomson scatters or not; if not the packet is attenuated by all continuum opacities and the corresponding PI estimators (and PI heating) are updated.

\section{Thermodynamic Evolution}
\label{sec:thermo_results}

\begin{figure*}
    \centering
    \includegraphics[trim={0.4cm 0.cm 0.4cm 0.3cm},width = 0.47\textwidth]{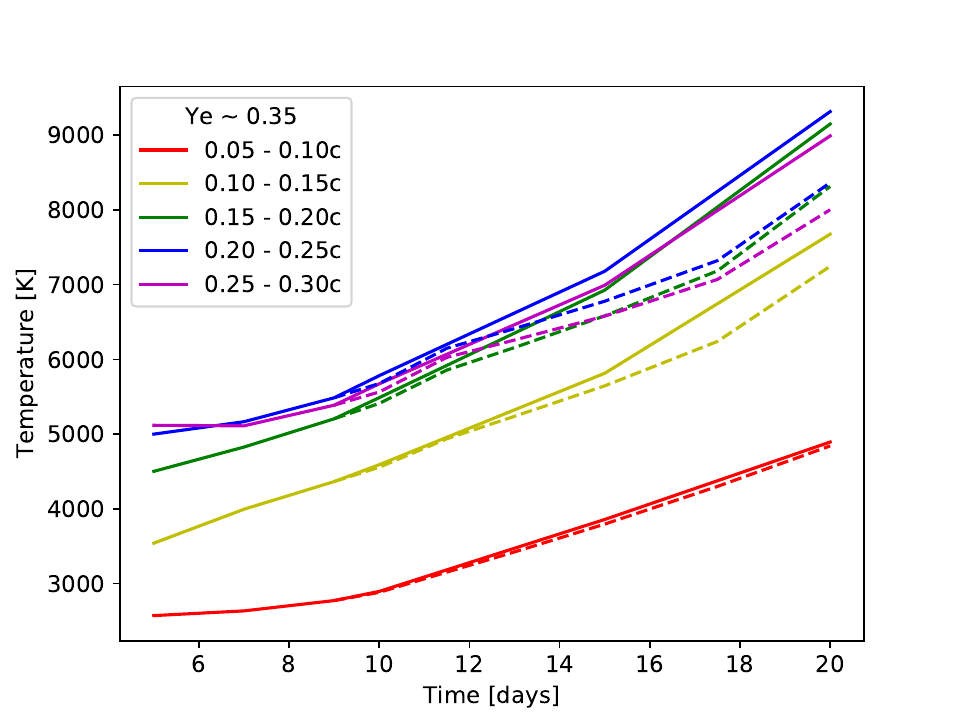} 
    \includegraphics[trim={0.4cm 0.cm 0.4cm 0.3cm},width = 0.47\textwidth]{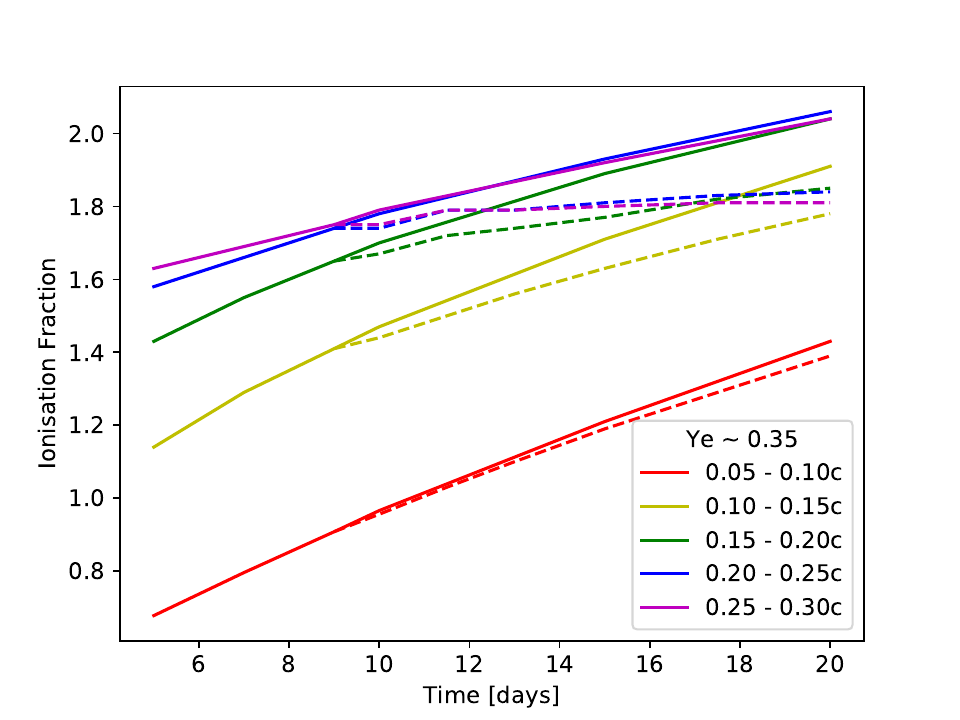}
    \includegraphics[trim={0.4cm 0.cm 0.4cm 0.3cm},width = 0.47\textwidth]{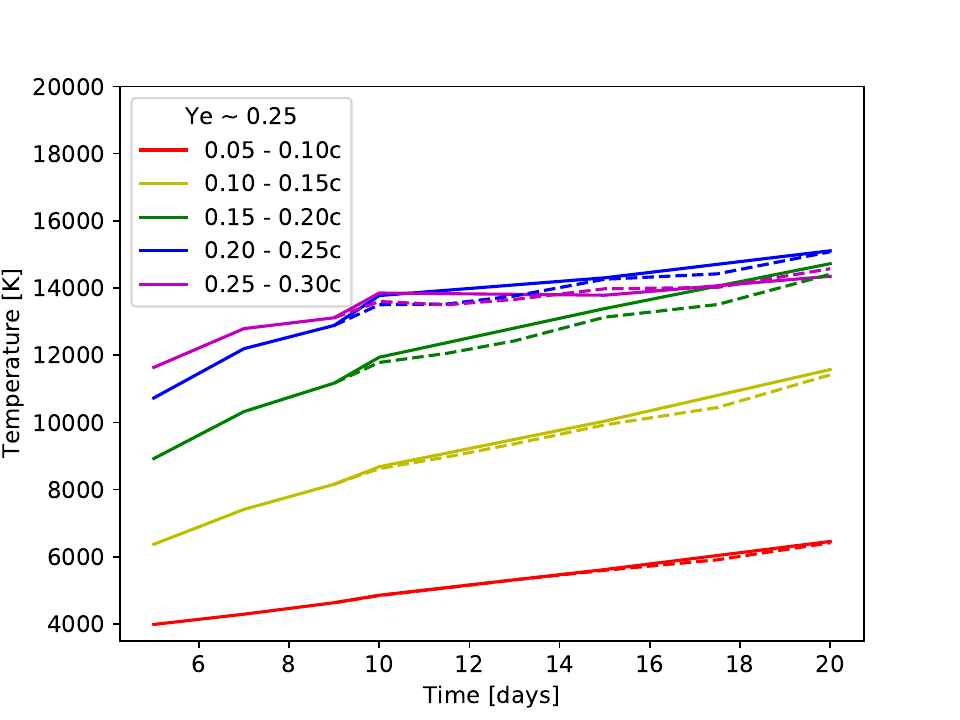}
    \includegraphics[trim={0.4cm 0.cm 0.4cm 0.3cm},width = 0.47\textwidth]{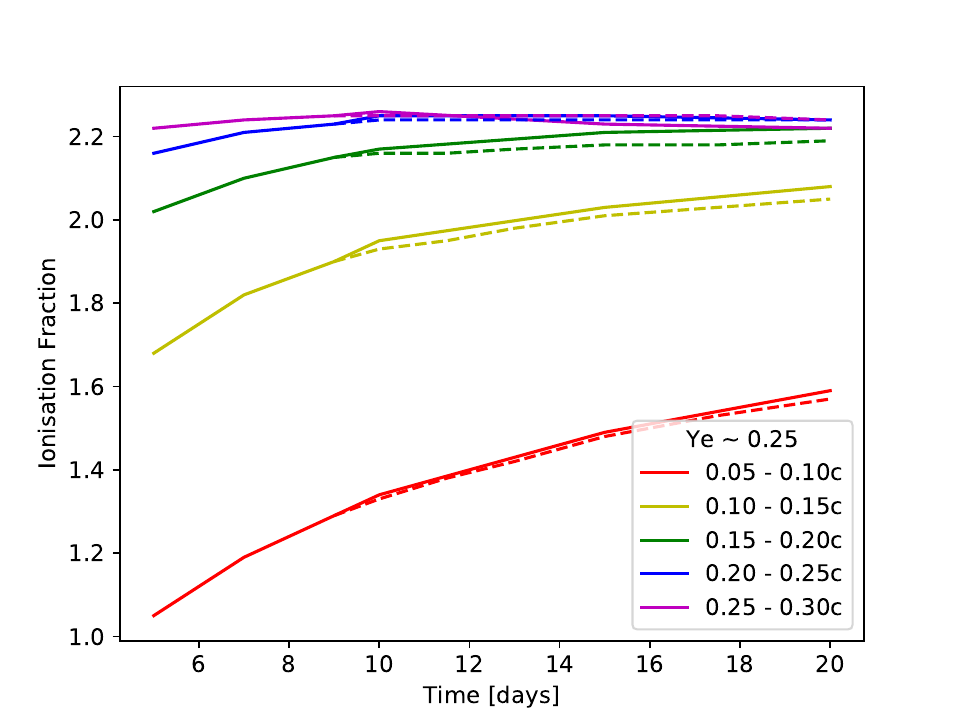}
    \includegraphics[trim={0.4cm 0.cm 0.4cm 0.3cm},width = 0.47\textwidth]{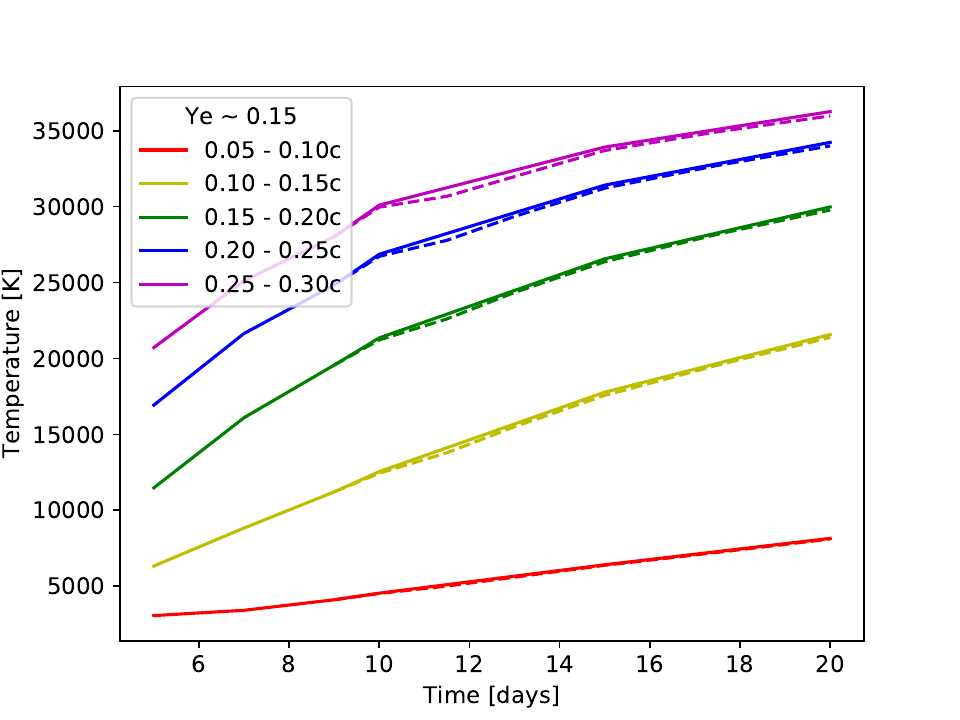}
    \includegraphics[trim={0.4cm 0.cm 0.4cm 0.3cm},width = 0.47\textwidth]{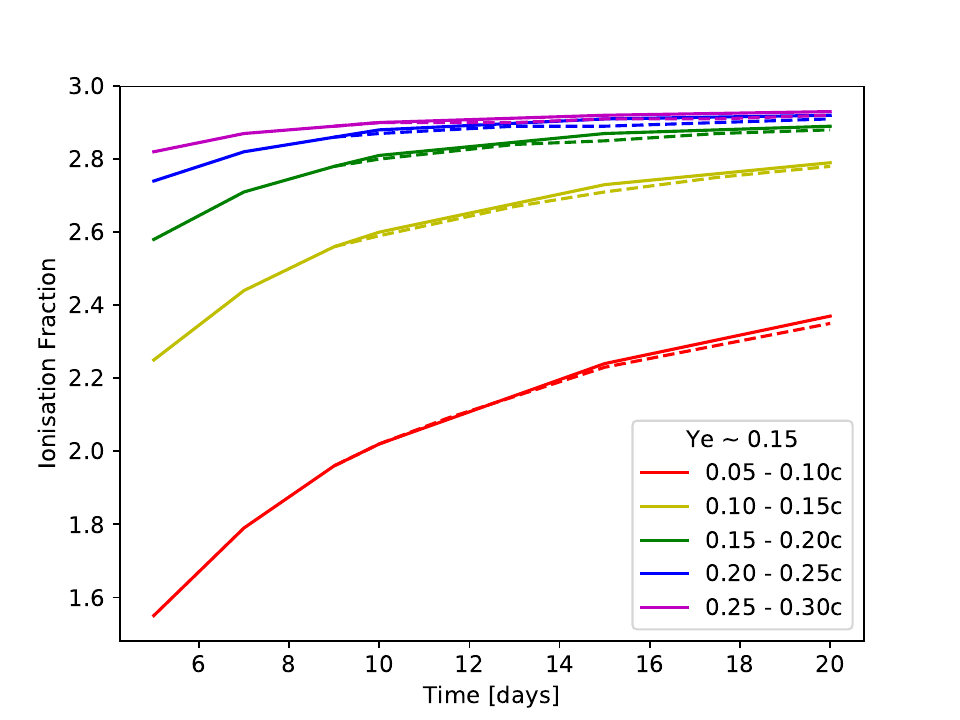}
    \caption{The evolution of temperature (left panels) and ionisation degree (electron fraction $x_e$, right panels) by zone of the models. The solutions with time-dependent ionisation and temperature physics are indicated by the dashed lines, steady-state solutions with solid. Note that time-dependent mode is only run from 10 days onwards.}
    \label{fig:thermo_evolution}
\end{figure*}

The three models, with $Y_e \sim 0.35, 0.25$, and $0.15$ respectively, are evolved from 5 to 20 days following the methodology described in Section \ref{sec:sumo_sim}. The models are initially computed with a steady-state approximation for all epochs, and then again from 10 days onwards using the full time-dependent equations for temperature and ionisation. We begin by considering the structure of the thermodynamic properties of the ejecta, and the evolution with time. The evolution of temperature and electron fraction ($x_e$ indicating degree of ionisation) are shown in Fig. \ref{fig:thermo_evolution}, where the solid lines represent the steady-state solutions, and the dashed lines represent the time-dependent solutions (computed from 10 days onwards).

Beginning with the evolution of temperature as shown in the left side panels of Fig. \ref{fig:thermo_evolution}, the zone temperatures increase in each model with time, consistent with calculations from previous studies \citep{Hotokezaka.etal:21,Pognan.etal:22a}. From 10 days onwards, we see slightly lower temperatures for the time-dependent solutions than for the steady-state ones, with the effect being more significant for the outer ejecta layers. The effect is maximised at our final epoch of 20 days, where for the outermost layer, we find a temperature drop from $\sim 9000$~K down to $\sim 8000$~K in the $Y_e  \sim 0.35$ model, and an adiabatic cooling contribution of $\sim 4$ per cent. This contribution is too small to solely account for a change in temperature of $\sim 11$ per cent, implying that effects arising from an altered ionisation solution are also occurring (see below).

Regarding the spatial gradient of temperatures (Fig. \ref{fig:thermo_profile}), at early times the outer layers of the ejecta are systematically hotter than the inner layers, consistent with reduced line cooling efficiency due to lower densities, while thermalisation is still largely effective and thus not density-dependent. This general trend is not ubiquitous, however, and we see a more complex temperature structure from $\sim$10 days onwards in the $Y_e \sim 0.35, 0.25$ models, where the hottest ejecta layer is not always the outermost layer. When thermalisation starts to become inefficient, decreasing density leads to a competing effect between reduced thermalisation efficiency, which will lower the heating, and reduced line cooling efficiency.

Looking at the ionisation structures in the right side panels of Figs. \ref{fig:thermo_evolution} and \ref{fig:thermo_profile}, we see a similar trend as for temperature. There is a stratification, with the inner layers of the ejecta being less ionised than the outer ones, consistent with more efficient recombination at higher densities. For the $Y_e \sim 0.35, 0.25$ models, the ionisation degree in the outer layers approach each other at late times, with the second outermost layer becoming the most ionised one at 20 days. As for temperature, density variation gives a competing effect between ionisation and recombination, as lower density will mitigate both processes. In general, the overall ionisation state increases with time, just as temperature does. Ultimately, this has its origin in the slower decline of the input power ($\sim t^{-(1.3-2.8)}$) compared to the density evolution ($\sim t^{-3}$).

As for temperature, the $Y_e \sim 0.35$ model experiences significant time-dependent effects in the degree of ionisation, the maximal deviation occurring in the outermost layer at our last epoch of 20 days, where the ionisation fraction is $x_e \sim 1.8$ compared to the steady-state value of $x_e \sim 2.0$. This change in ionisation structure also affects the temperature solution, as different ions have different line-cooling capacities \citep{Pognan.etal:22a}. The ionisation structure 'freezes out' at around 10 days in the outer ejecta layers. This freeze-out effect occurs when the time-scales for both ionisation (by all processes), and recombination become comparable to the evolutionary time \citep{Pognan.etal:22a}.

Comparing the models to each other, we see that the lower the electron fraction $Y_e$, the hotter and more ionised the ejecta typically are at all epochs. This is consistent with a larger energy deposition, notably in the case of the $Y_e \sim 0.15$ model which also has contributions from $\alpha$-decay and spontaneous fission, and has a factor 2--5 times more energy deposition than the $Y_e \sim 0.35$ model at every epoch (see Fig. \ref{fig:energy_dep}). Since the thermalisation by decay product is the same for each model (the zone densities are identical), this larger energy deposition arises from the inclusion of $\alpha$-decay and fission, the products of which thermalise more effectively than $\beta$-decay electrons in the 5--20 day range. The gradients of temperature and ionisation throughout the ejecta are likewise more significant for the models with heavier compositions and enhanced energy deposition. For example, at 20 days, the innermost layer of the $Y_e \sim 0.15$ model has a temperature of $\sim 6000$~K, while the outermost layer reaches $\sim 35 000$~K. Conversely, the values for the $Y_e \sim 0.35$ model only range from $\sim 5000$~K (inner boundary) to $\sim 8000$~K (outer boundary). The ionisation gradient follows the same trend as the temperature gradient, with the $Y_e \sim 0.15$ model having $x_e =$ 2.3--2.9 at 20 days and the $\sim{Y_e} \sim 0.35$ model a significantly lower $x_e =$ 1.4--1.8. 
 
\section{Spectral Features}
\label{sec:spectral_results}

In this section, we examine our emergent spectra and determine which species are the main drivers of spectral formation in our models, focusing on the 5, 10 and 20 day epochs, shown in Figs. \ref{fig:Ye035_species}, \ref{fig:Ye025_species} and \ref{fig:Ye015_species} for the $Y_e \sim 0.35,0.25.0.15$ models, respectively. The colours in the spectra tag the last element with which the photon packet interacted, including by scattering/fluorescence. The spectra have all been smoothed by a Gaussian with a full-width half-max velocity of $3500~ \rm{km~s^{-1}}$ in order to reduce Monte Carlo noise. To aid in the analysis, we show the optically thick lines in each model at the aforementioned epochs in Figs. \ref{fig:Ye035_lines}, \ref{fig:Ye025_lines} and \ref{fig:Ye015_lines}. In order to clarify whether emission features arise from cooling (i.e. spontaneous radiative de-excitation following a collisional excitation by a thermal electron), or from processes such as scattering and fluorescence, we show the contribution of key elements to the total cooling compared to their total flux contribution in the emergent spectra in Fig. \ref{fig:coolvflux}. We note here that the purpose of this section is not to conclusively identify potential features observed in AT2017gfo, but rather to understand which species play key roles in the spectral formation, and which processes are important. We add that every element's contribution to both emission and absorption has been checked, and the species presented in the following section are found to be the most significant in our models.

\subsection{The \texorpdfstring{$Y_e \sim 0.35$}{Ye ~ 0.35} Model}
\label{subsec:highYe_species}

\begin{figure}
    \includegraphics[trim={0.cm 0.1cm 0.4cm 0.3cm},clip,width = 0.47\textwidth]{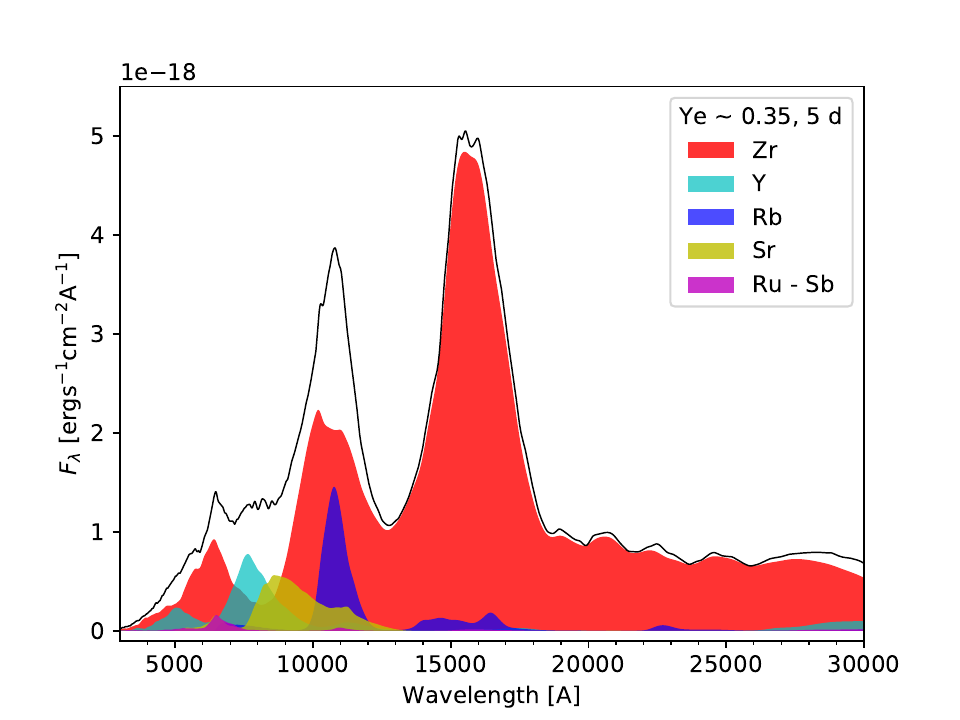} 
    \includegraphics[trim={0.cm 0.1cm 0.4cm 0.3cm},clip,width = 0.47\textwidth]{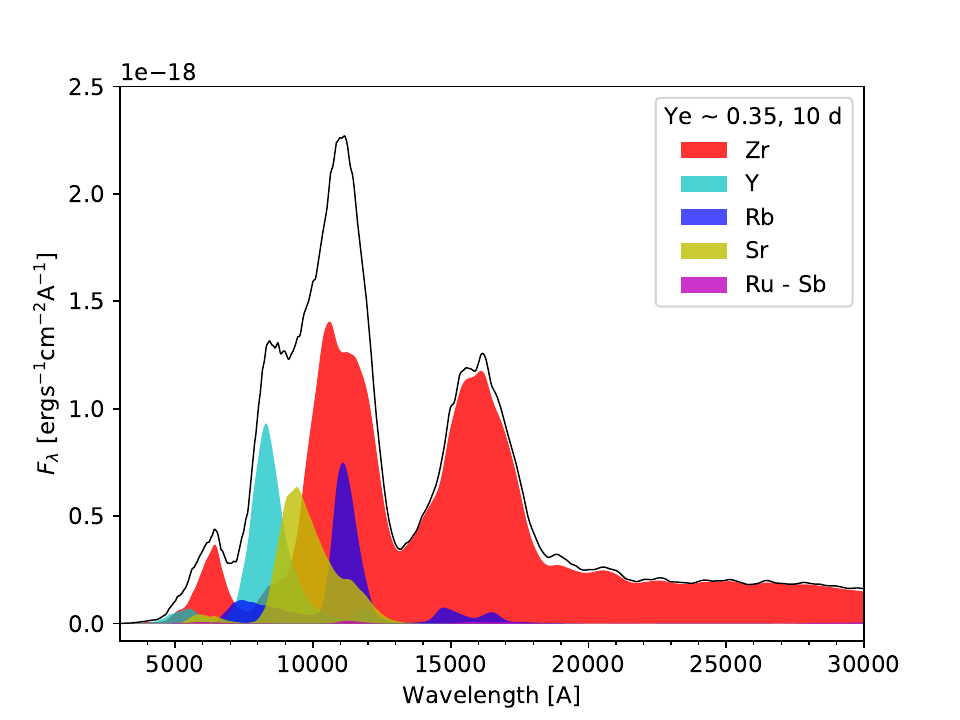}
    \includegraphics[trim={0.cm 0.1cm 0.4cm 0.3cm},clip,width = 0.47\textwidth]{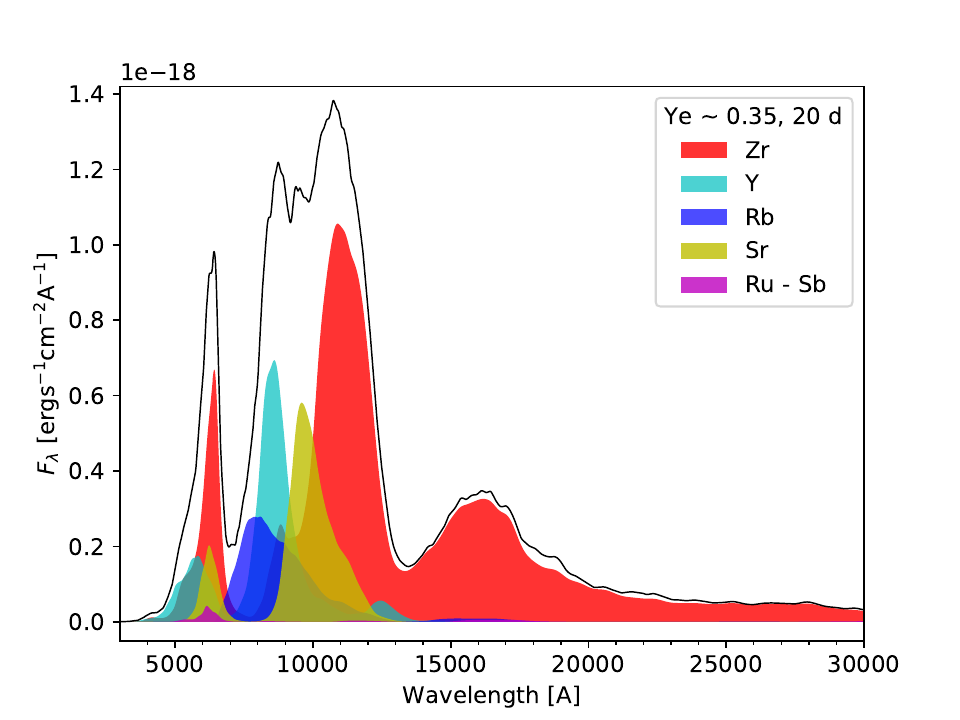}
    \caption{The spectrum of the $Y_e \sim 0.35$ model at 5, 10 and 20 days, with key emitting species marked. The magenta filled area represents total second r-process peak element contribution in this model (Ru--Sb), which is seen to be small.}
    \label{fig:Ye035_species}
\end{figure}

We first take a closer look the key species in the $Y_e \sim 0.35$ model, shown at 5, 10 and 20 days in Fig. \ref{fig:Ye035_species}. We find that the spectral features are dominated by only a few first r-process peak elements from groups I--IV of the periodic table: Rb, Sr, Y and Zr. These first r-process peak elements are quite abundant in the model composition (see Fig. \ref{fig:compositions}, and Table \ref{tab:compositions}), but their atomic structure also plays an important role in their domination. These elements all have relatively few valence electrons, providing them with strong, allowed transitions between thermally accessible low lying states, enabling both powerful absorption and emission channels, as was previously found for the neutral and singly ionised species by \citet{Domoto.etal:22}.

Of particular interest are the Zr\,\textsc{i} and Zr\,\textsc{ii} ions, which have closely packed low lying multiplets, all of the same parity \citep[see e.g.][]{Moore:71a,NIST_ASD,Lawler.etal:22}. Due to this structure, the transitions between these low lying states are (semi)-forbidden transitions often with relatively low energies, providing blanket features across the NIR at late times when densities are low enough. However, the first excited states of opposite parity, and therefore with strong allowed transitions, are still well within range of thermal excitation and/or optical/NIR scattering, and so while Zr\,\textsc{i} and Zr\,\textsc{ii} have many weak NIR transitions, they also possess some strong allowed transitions like Rb--Y. We find domination of Zr\,\textsc{i} past $1.8~\mu$m, both in terms of emission and absorption. 

\begin{figure}
    \includegraphics[trim={0.cm 0.1cm 0.4cm 0.3cm},clip,width = 0.47\textwidth]{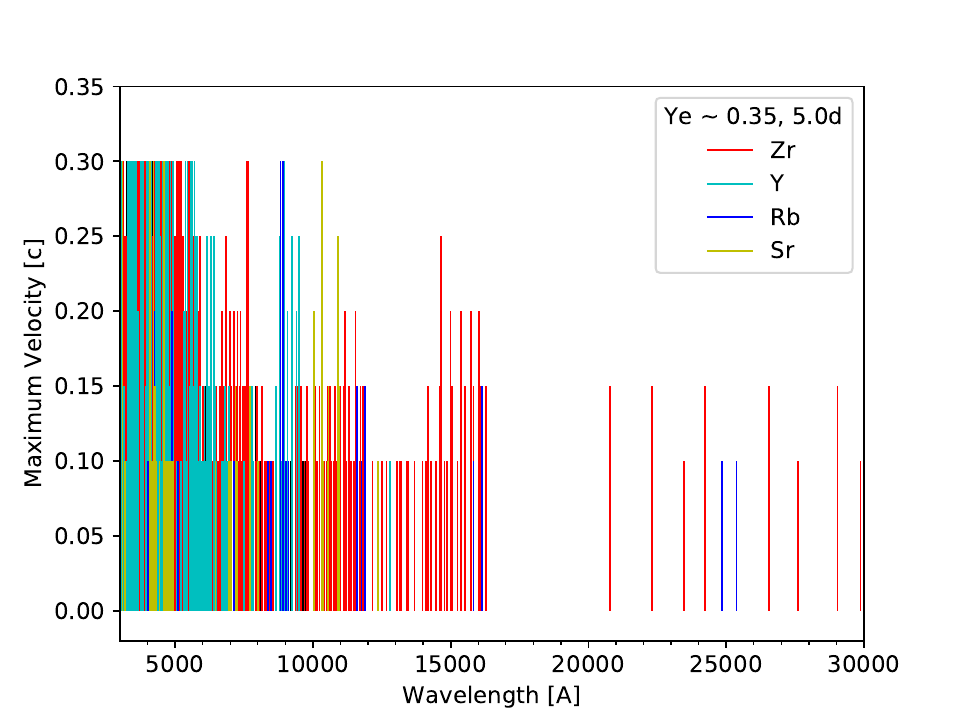} 
    \includegraphics[trim={0.cm 0.1cm 0.4cm 0.3cm},clip,width = 0.47\textwidth]{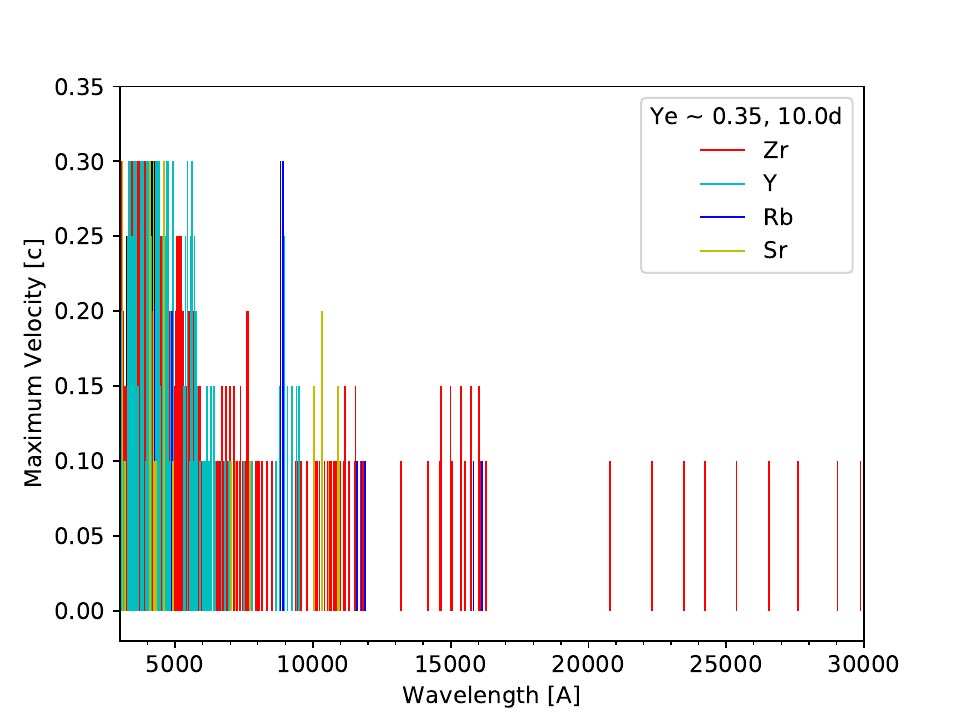}
    \includegraphics[trim={0.cm 0.1cm 0.4cm 0.3cm},clip,width = 0.47\textwidth]{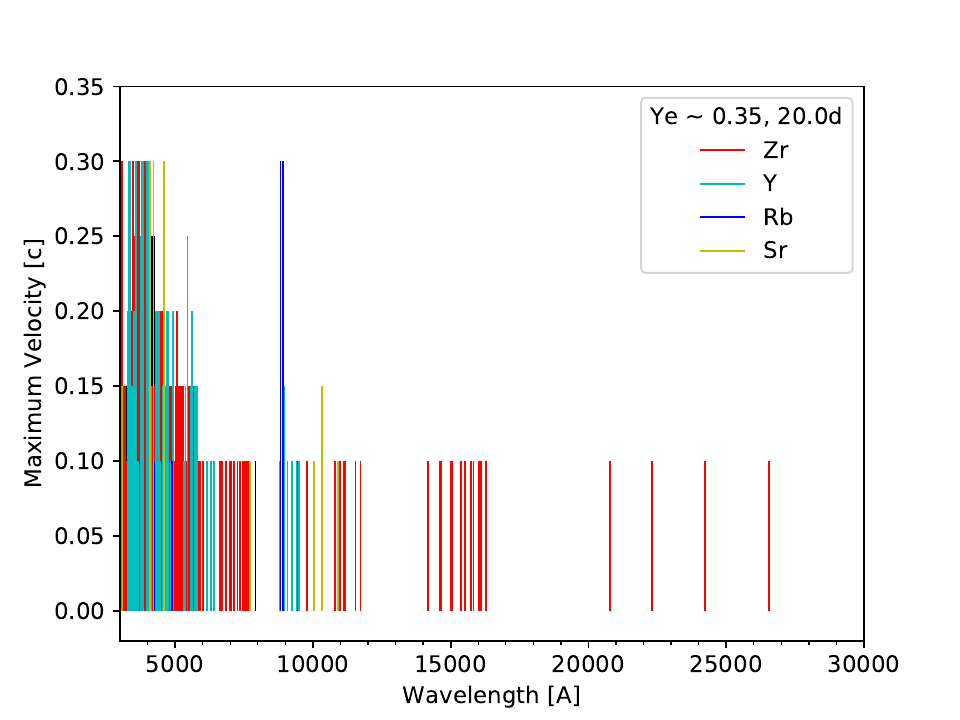}
    \caption{Optically thick lines in the $Y_e \sim 0.35$ model at 5, 10 and 20 days. The velocity up to which the line is optically thick is shown. Line wavelengths are plotted as rest wavelengths. Major contributing elements are marked in colours.  Contributions from other elements are also plotted, but are typically minimal compared to the marked elements.}
    \label{fig:Ye035_lines}
\end{figure}

We find that the Zr emission is mostly from neutral Zr\,\textsc{i} at early times, with some contribution by Zr\,\textsc{ii} to the $\sim 6000~\ang$ feature and at $\sim 1.7~\mu$m at 20 days. The three Zr emission peaks seen in the top panel of Fig. \ref{fig:Ye035_species} arise from the blending of many allowed transitions, occurring from the higher lying odd parity multiplets down to the many even parity low lying multiplets. Owing to the closely packed nature of these multiplets, this yields many similar strength lines of similar wavelengths, which were found to play an important role also in the context of photospheric absorption and opacity \citep{tanaka:opacities:2020,Domoto.etal:22}. As such, the widths of the Zr emission features are determined both from velocity broadening as well as line blending. At 5 days, the Zr emission past $\sim 1.8~\mu$m is due to allowed transitions, with P-Cygni formation arising from the optically thick lines seen in Fig. \ref{fig:Ye035_lines}. At 20 days, we no longer see these small absorption features, and the few optically thick lines are only in the innermost ejecta layer. The emission at this epoch is likely a combination of these now optically thin allowed transitions alongside some contribution from the (semi)-forbidden transitions between the lowest lying states. 

The emission from the other species (Rb, Y, Sr) arises from strong transitions. At all epochs, we see emission from the Sr\,\textsc{ii} triplet around $10000~\ang$ ($\lambda_0 = 10039,10330,10918~\ang$), and at 20 days, the appearance of emission also from the $\sim 6800~\ang$ doublet ($\lambda_0 = 6740,6870~\ang$). We discuss Sr line formation in more detail in section \ref{subsec:SrII_focus}. 

The Y emission is initially from Y\,\textsc{i}, with contributions from Y\,\textsc{ii} at later epochs. The persistent emission at $\sim 8500~\ang$ arises from several strong transitions to the ground doublet from low lying states in Y\,\textsc{i}, and is wavelength accurate. The bluer feature at $\sim 5500~\ang$ arises from a strong transition in Y\,\textsc{ii} between the first opposite parity state and the ground state, but is slightly too red in our model atom and should be located at $\sim 4400~\ang$. We note that this feature arises from different transitions than those identified as possibly responsible for a $\sim 7600~\ang$ P-Cygni feature in AT2017gfo \citep{Sneppen.Watson:23Y}. 

The Rb emission originates from various strong transitions in the low lying multiplets of Rb\,\textsc{i}. Rb\,\textsc{ii} is a closed-shell noble ion and therefore participates negligibly in the spectral formation. Notably, the Rb\,\textsc{i} emission at 20 days is found to be associated to the strong transition from the first excited doublet down to the ground state ($\rm{4p^{6}5p}$ to $\rm{4p^{6}5s}$), located at $\lambda_0 = 8827,8920~\ang$ in our model atom, slightly redder than the measured $\lambda_0 = 7802,7950~\ang$ \citep{Volz:96,Simsarian.etal:98}. This line remains optically thick throughout the whole ejecta at all epochs here, therefore contributing to producing an absorption trough of width $\sim 0.3~$c bluewards of $\sim$8900~\ang\  ($\sim$7900~\ang\ if wavelength corrected). This trough is most distinctly seen in the model spectrum at 20 days, the reason being that at this epoch there is no other optically thick line within $\sim 0.3$~c bluewards of the Rb\,\textsc{i} line that can link in further absorption troughs, as happens at earlier epochs. This line remains important even at this late epoch when the abundance of Rb\,\textsc{i}I is only $\sim$ 0.04 per cent of the model by mass fraction. 

The Rb\,\textsc{i} doublet has experimentally determined rest wavelengths of $\lambda_0 = 7802,7950~\ang$, quite close to the P-Cygni feature identified in AT2017gfo at rest wavelength of $\sim 7500--7900 ~\ang$, and proposed by \citet{Sneppen.Watson:23Y} to be due to Y\,\textsc{ii} 4d5p -- 4d$\rm{^{2}}$ transitions, the most prominent of which has rest wavelength $\lambda_0 = 7882~ \ang$. Our model atom for Y\,\textsc{ii} includes these lines with transition strengths of similar values, albeit at inaccurate wavelengths ($\sim 11000~\ang$). We see little activity in these lines, with the transitions being optically thin ($\tau_s \leq 0.1$) in the entire ejecta, at all epochs, implying that in our model these lines are unable to produce features by P-Cygni formation. The Rb\,\textsc{i} doublet could be an alternative candidate for the $\sim$ 7500--7900 $\ang$ P-Cygni feature found in the early AT2017gfo epochs, as we see strong scattering in this transition at all epochs.

Some more clues to the spectral formation processes can be found from \ref{fig:coolvflux} (top panel), showing the contribution to cooling and to the total emergent flux, for different elements. Rb is an important coolant, especially at 5 and 10 days, doing 30--55 per cent of the cooling, almost entirely by the collisional excitation of the $\rm{4d^2}$ to $\rm{4d5p}$ doublet transition mentioned above. However, we only see emission from this transition emerging as time progresses, implying that the $\sim 10$ per cent contribution of Rb to the emergent flux initially arises from other lines by scattering/fluorescence processes. At 20 days, we see clear emission from this doublet, and find that Rb lies along the diagonal in Fig. \ref{fig:coolvflux}, implying that the 20d emission feature is a combination of scattering (e.g. P-Cygni formation), and cooling emission following collisional excitation. Y is less important for cooling (5--10 per cent) but contributes somewhat more to emergent flux (10--20 per cent). Sr becomes the dominating source of cooling at late times (40 per cent at 20 days), which we find to be driven almost entirely by collisional excitation of the $\sim 4000~\ang$ ($\lambda_0 = 4078,4216~ \ang$) doublet. We find that while Zr dominates in terms of emitted flux (55--75 per cent of total), its contribution to cooling is much less (20--35 per cent), implying that most of its emission arises from scattering/fluorescence rather than cooling (assuming that recombination cannot reach such levels of emission\footnote{The recombination emission from element i, ion j, in zone k, is, roughly, $V_k n_e n_{k,i,j} \alpha \left(\chi + kT\right)$. If we look at the innermost zone (with the highest recombination rate), this becomes about $10^{37}$ erg/s at 10 days which is less than 1\% of the bolometric luminosity. }). The Zr\,\textsc{i,ii} structures of closely packed multiplets with many transitions of similar strength over a wide range of wavelengths allows for effective scattering/fluorescence, as many different channels of approximately equal probabilities are available. 

In terms of absorption, we see in Fig. \ref{fig:Ye035_lines} that there are many optically thick lines throughout the whole ejecta at $\lambda \lesssim 7000~\ang$ even up to 20d. Initially at 5d, we see that the majority of the ejecta are line blocked up to $\lambda \sim 1~\mu$m. With time, the ejecta layers gradually become more transparent for $\lambda \gtrsim$7000~\AA. However, the innermost core region at $\leq$ 0.1~c (which contains 60 per cent of the ejecta mass), remains largely blocked even at 20d, due to many optically lines from Y and Zr, and to a lesser extent Sr.

At 5 days, the figure shows that two escape windows in the line blocking exist at rest wavelengths $\lambda_0 \sim$1.2--1.4$~\mu$m and $\sim 1.6 ~\mu$m. The first one is partial, in the sense that the $\leq$0.1~c core is still absorbing, whereas the second window is fully transparent. It is in these windows that much of the radiation that has been scattering and fluorescing from line to line can finally escape, producing the two distinct peaks in the spectrum at 1.1 and 1.6 $\mu$m. We find that the emission at $1.6~\mu$m also coincides with the strongest cooling channels of Zr\,\textsc{i}, which contribute $\sim 5$ per cent to the total Zr cooling of $\sim 40$ per cent at 5d (top panel of Fig. \ref{fig:coolvflux}). However, since Zr emission in this model is overwhelmingly produced by scattering/fluorescence, it is likely that the $1.6\ \mu$m feature is partially a P-Cygni feature, even at 20d after merger. As the line blocking in the $\lambda_0 = $6000--9000$~\ang$ range is reduced with time, a larger fraction of the flux starts to escape at these wavelengths, with Y\,\textsc{i} and Sr\,\textsc{ii} being dominant emitters. Eventually, at 20d, a window becomes fully open also at $6000~\ang$, giving a third peak.

The Sr\,\textsc{ii} $10000~\ang$ triplet is optically thick throughout most of the ejecta at 5 days, and remains thick up to 0.2c at 10d and 0.15c at 20d. Fig. \ref{fig:Ye035_lines} does reveal several other absorption lines in that regime, however, which may complicate definitive association of an observed feature in that wavelength range to Sr\,\textsc{ii} alone. In the model here, we find that Sr is an important coolant at late times (see Fig. \ref{fig:coolvflux}), mainly by the $\sim 4000~\ang$ doublet transition. Photons in that transition are however strongly resonance trapped, and the $\sim10000~\ang$ branching triplet (same parent multiplet) provides an alternate de-excitation channel. Considering the radiative deexcitation flows ($A\times \beta_{Sob}\times h\nu)$, we estimate that at 5d, $\sim 95$ per cent of cooling excitations along the $4000~\ang$ doublet channel radiatively de-excite in the $\sim 10000~\ang$ triplet channel. However, the Sr emergent flux contribution at at 5d is only about 1/3 of this, so a significant fraction of the photons emitted by the $10000~\ang$ triplet ($\sim$2/3) are then reprocessed by other species via scattering/fluorescence. At 20 days, we find that $\sim 50$ per cent of cooling excitations of the $4000~\ang$ doublet are radiated away by the $10000~\ang$ triplet, and at this epoch a larger fraction of this directly escapes. The other $\sim 50$ per cent (emitted by the 4000 \AA\ lines) are likely reprocessed by the many thick lines of Y and Zr at $\sim 4000~\ang$, and re-emitted redwards. These findings suggest that the $\sim 10000~ \ang$ Sr\,\textsc{ii} feature in this model likely arises from a combination of scattering and cooling emission, and as such is not a pure P-Cygni feature. We discuss further the impact of Sr on lanthanide-free ejecta in more detail in Section \ref{subsec:SrII_focus}.

The most abundant and third-most abundant elements in the $Y_e \sim 0.35$ model are krypton (Kr) and selenium (Se),(see Fig. \ref{fig:compositions}, Table \ref{tab:compositions}). Despite their high abundances, neither of them give any significant contribution to the emergent spectra here. This is due to the atomic structure of their ions, where the first excited states of opposite parity to the ground multiplet lie at relatively high energies difficult to reach by thermal collisional excitation. As such, the transitions between the lowest lying states are typically (semi)-forbidden lines, and either extremely weak, or far into the mid infra-red (MIR) \citep[e.g. the Se\,\textsc{iii} $4.5\mu$m line studied in][]{Hotokezaka.etal:22}. We note that calcium (Ca) has been previously suggested as important due to its co-production alongside Sr \citep{Domoto.etal:21}, but it is not included here as its mass fraction is below our cut-off limit of $10^{-4}$ (see Fig. \ref{fig:compositions}).

The general view is that prominent emission features can arise from both cooling and from scattering/fluorescence, emerging within limited escape windows in the optical/NIR. We tend to see emission peaks associated with optically thick lines (e.g. the Rb\,\textsc{i} feature at rest wavelength $\sim 8900~\ang$, and partially the $\sim 10000~\ang$ Sr\,\textsc{ii} triplet), even up to 20 days, suggesting a continued important role of P-Cygni like line formation well after the diffusion phase has ended. Fluorescence is also important, e.g. for the Zr contributions. The importance of radiative transfer processes, also at 20 days, implies that the KN is not yet in a purely optically thin ('nebular') phase by this time for its optical/NIR emission, even for our lightest model which does not include high opacity elements like lanthanides or actinides.

\subsection{The \texorpdfstring{$Y_e \sim 0.25$}{Ye ~ 0.25} Model}
\label{subsec:midYe_species}

\begin{figure}
    \includegraphics[trim={0.cm 0.1cm 0.4cm 0.3cm},clip,width = 0.47\textwidth]{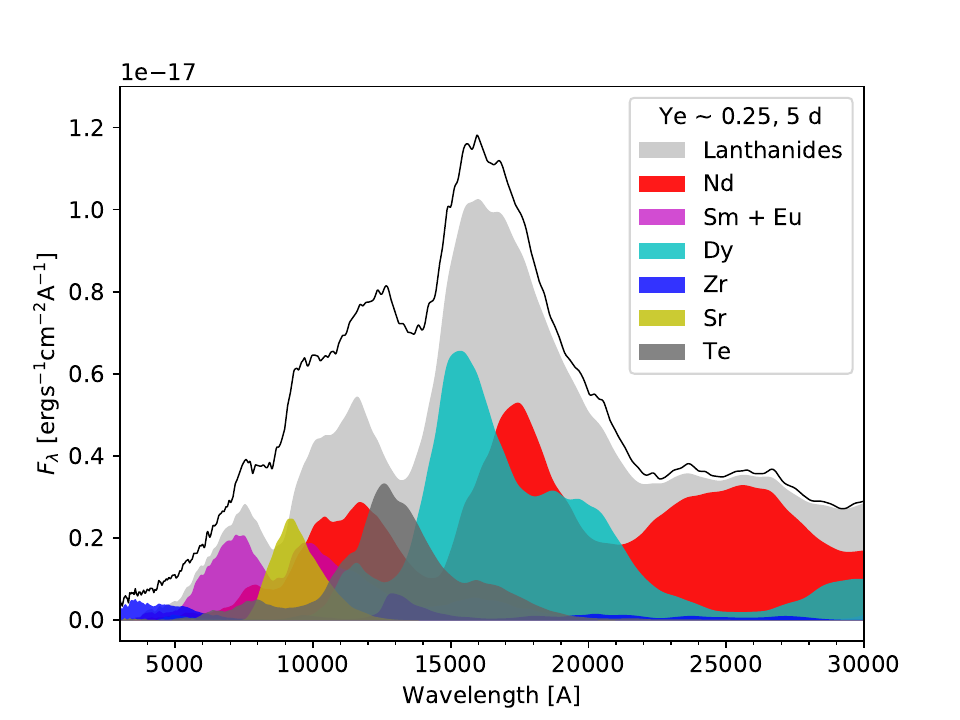} 
    \includegraphics[trim={0.cm 0.1cm 0.4cm 0.3cm},clip,width = 0.47\textwidth]{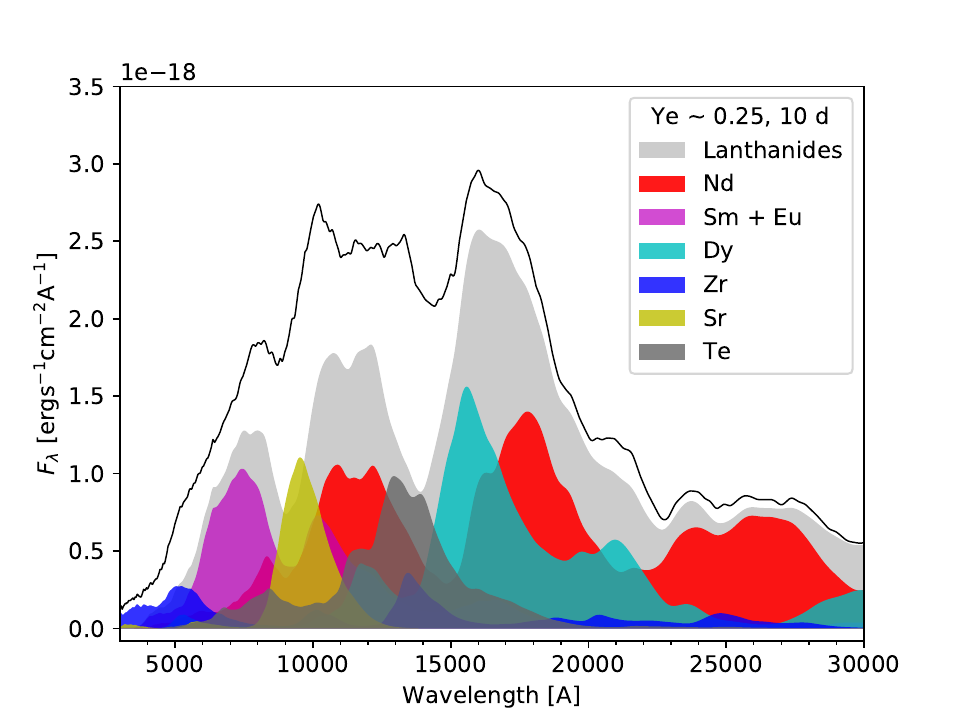}
    \includegraphics[trim={0.cm 0.1cm 0.4cm 0.3cm},clip,width = 0.47\textwidth]{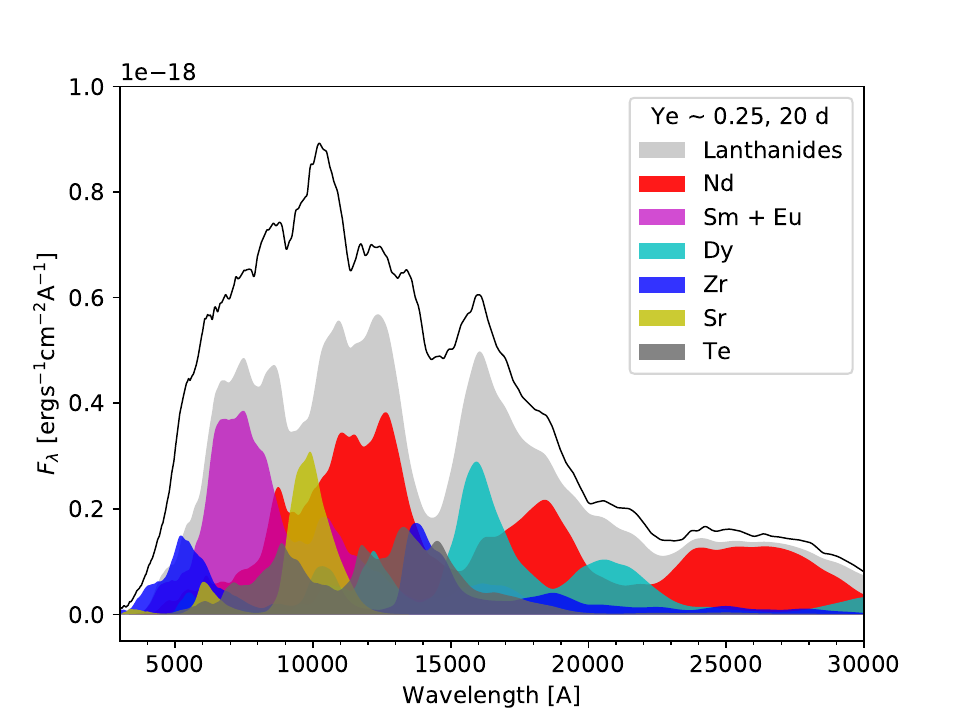}
    \caption{The spectrum of the $Y_e \sim 0.25$ model at 5, 10 and 20 days, with key emitting species marked. The light grey filled area represents the total lanthanide element contribution in this model (note that this also contains also the individually marked lanthanides Nd, Sm, Eu and Dy).}
    \label{fig:Ye025_species}
\end{figure}

The elemental contributions to the spectra of the $Y_e \sim 0.25$ model are shown in Fig. \ref{fig:Ye025_species} at 5, 10 and 20 days after merger. Consistently across every epoch, we see strong lanthanide contributions (light grey shading) to the spectral emission across all wavelengths. The lanthanides achieve this prominence despite having a mass fraction of only 0.015 in the model composition. Of the lanthanides, the most prominent specific elements are Nd (neodymium), Sm (samarium), Eu (europium), and Dy (dysprosium), marked out with individual colours. The $\sim$ 1.6--1.7~$\mu$m feature seen in this model is produced by a mix of Nd and Dy, with Zr (which produced a strong feature at this wavelength in the $ Y_e \sim 0.35$ model) now being negligible. Past $1.5~\mu$m, the IR emission is entirely dominated by Nd and Dy.

From the plot of optically thick lines at 5d, one may identify an escape window at $\lambda_0 \sim$1.8 $\mu$m (the innermost zone at $\leq$0.1~c remains opaque but cannot stop all radiation); the formation of a peak at this wavelength may thus be explained in a similar manner as the 1.1 and 1.6 $\mu$m peaks in the $Y_e \sim 0.35$ model. It is Dy which provides the last optically thick lines before the window, which explains its prominent emission at those wavelengths. Indeed, Dy lies well down below the diagonal in the cooling vs. flux contribution plot (Fig. \ref{fig:coolvflux}), which shows that this is scattering/fluorescence emission. The windows in this model are, however, less pronounced, which gives a smoother overall spectrum with less dramatic peaks and troughs.

\begin{figure}
    \includegraphics[trim={0.cm 0.1cm 0.4cm 0.3cm},clip,width = 0.47\textwidth]{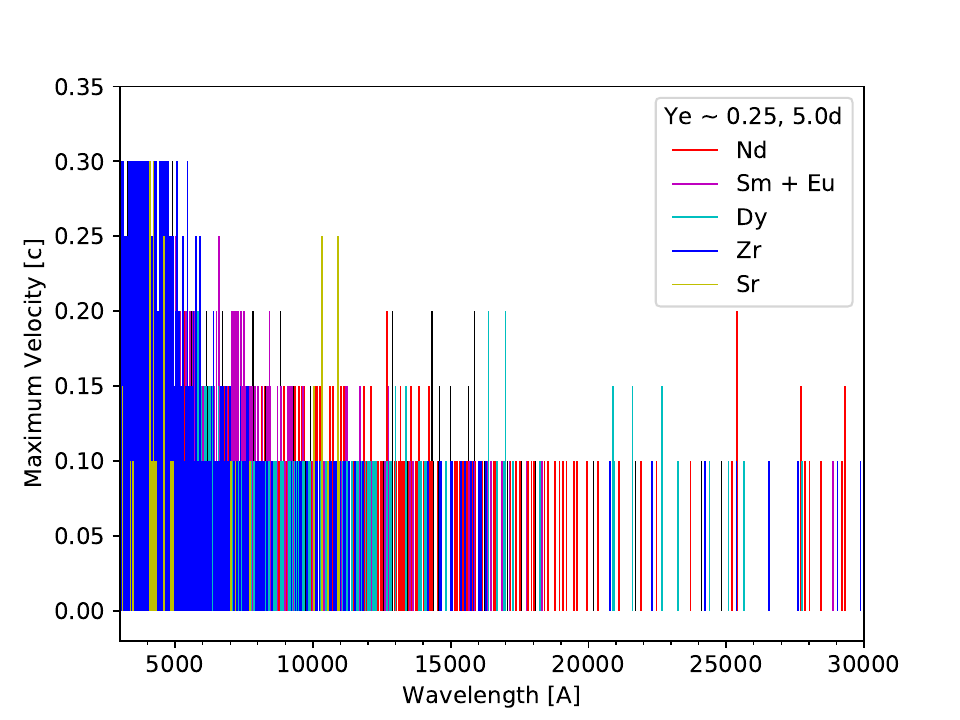} 
    \includegraphics[trim={0.cm 0.1cm 0.4cm 0.3cm},clip,width = 0.47\textwidth]{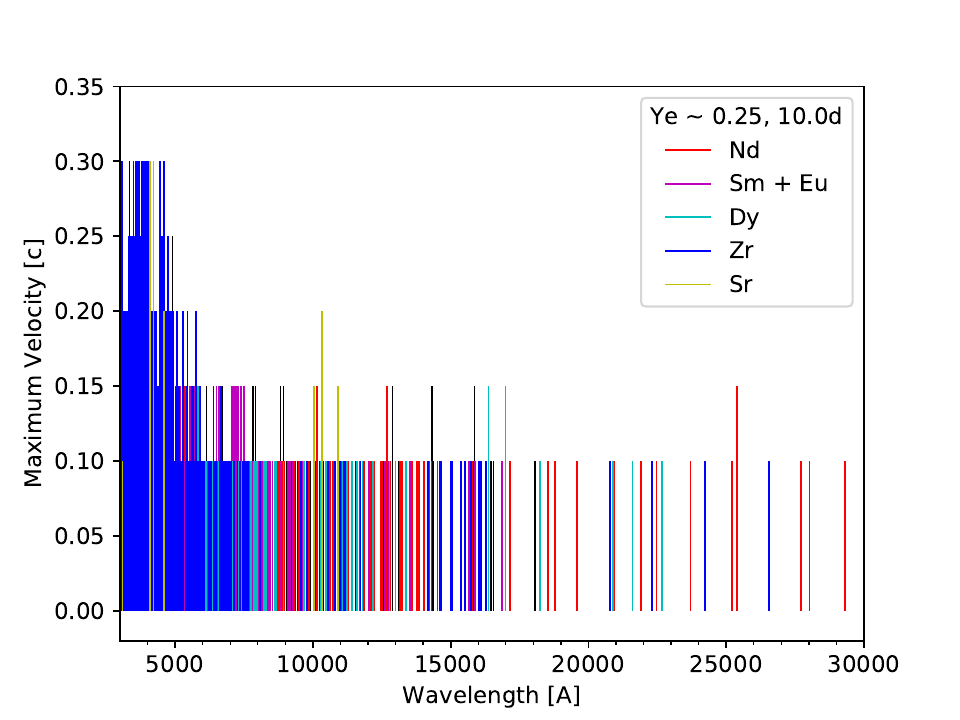}
    \includegraphics[trim={0.cm 0.1cm 0.4cm 0.3cm},clip,width = 0.47\textwidth]{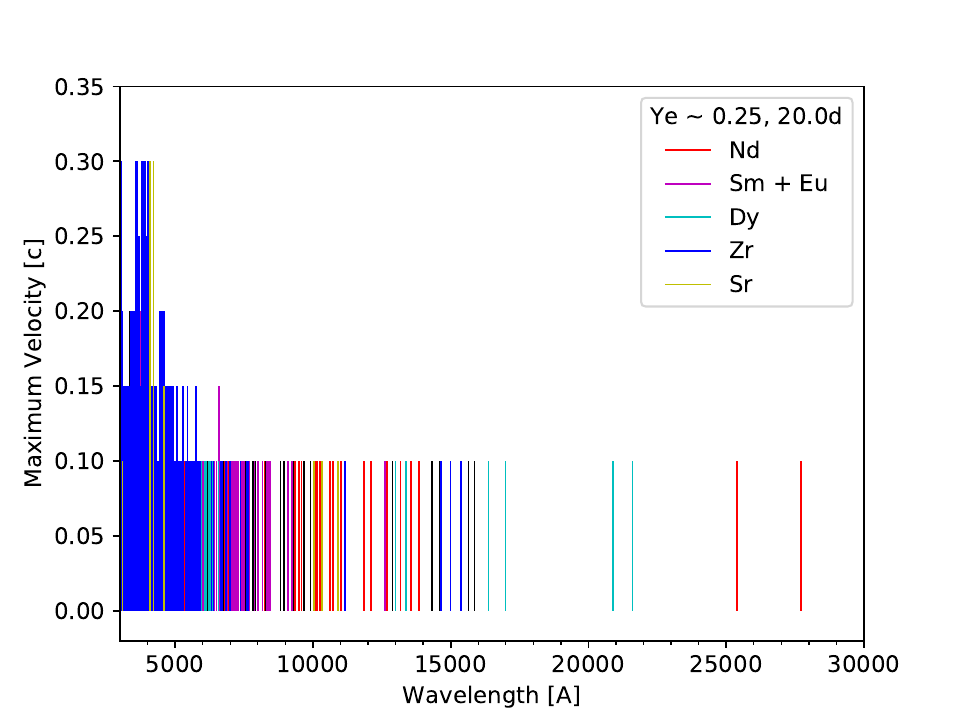}
    \caption{Optically thick lines in the $Y_e \sim 0.25$ model at 5, 10 and 20 days. The velocity up to which the line is optically thick is shown. Line wavelengths are plotted as rest wavelengths. Major contributing elements are marked in colours. Remaining contributions from other elements are also plotted, but are typically minimal compared to the marked elements.} 
    \label{fig:Ye025_lines}
\end{figure}

Neodymium is even more extreme in its dominance by scattering/fluorescence (Fig. \ref{fig:coolvflux}). It provides a rich set of optically thick lines throughout the optical/NIR range at all epochs (Fig. \ref{fig:Ye025_lines}), enabling it to absorb and reprocess a significant amount of radiation, playing a similar role here to the one of Zr in the $Y_e \sim 0.35$ model.

The landscape below $1.5 \mu$m is more complex, with several different species contributing to the emission and absorption. The dominating lanthanides in this range are Nd, Sm, and Eu. Around $10 000~\ang$ we recover the Sr\,\textsc{ii} triplet feature with the corresponding absorption at $\sim 8000~\ang$, with Fig \ref{fig:Ye025_lines} showing that in this model Sr\,\textsc{ii} is the only ion that provides optically thick lines in the outermost layers, at 5 and 10d. A partial escape window can be seen between $\lambda_0 \sim$ 8000--10000~$\ang$ at 5d and 10d, which allows for formation of a feature approaching that of a classical P-Cygni profile by Sr\,\textsc{ii}. Thus, for epochs up to 10 days or so, an absorption trough by the Sr\,\textsc{ii} triplet becomes distinct in the overall spectrum, even though the Sr mass-fraction abundance at this $Y_e$ is only 0.03, compared to 0.19 in the $Y_e \sim 0.35$ model. 

Bluewards of $\lambda \sim 7000~\ang$, emission is dominated by Zr. As such, while the optical red and NIR emission is dominated by the lanthanides, the bluer optical regime does retain important contributions from first r-process peak elements. The abundant elements Sn (0.17), Se (0.13), Kr (0.091), Cd (0.054), and Ge (0.052) produce no significant emission or absorption. The reasons discussed above for Kr and Se also apply to the other elements listed here, which have similar atomic structures, i.e. a ground multiplet followed by low lying states of the same parity, with opposite parity states at thermally inaccessible energies.

In Fig. \ref{fig:coolvflux}, we see that Sr and Zr play important cooling roles, as for the $Y_e \sim 0.35$ model. However, Zr is now outdone by the lanthanides for scattering and fluorescence, having the last interaction for a minority of the photons, and moving left in the diagram. As discussed above, the dominating lanthanides Nd and Dy are firmly within the scattering/fluorescence regime. This suggests that species with particularly strong transitions in optical wavelengths, corresponding to the more energetic thermal electrons, are key for cooling. Species with many closely packed multiplets providing weaker, but still allowed transitions at redder optical and NIR wavelengths represent ideal scattering/fluorescence conditions, as photons are able to be absorbed and re-emitted over a broad range of wavelengths. In both Nd and Dy, we see their contribution to the overall emitted flux decreasing with time, which is consistent with expectations for the evolution of scattering/fluorescence processes. It is possible that the KN will transition to a fully optically thin thermal nebular phase, which for these models would occur later than 20d after merger. However, increasing temperatures may bolster emission of photons in the UV where line blocking may still be effective, thus leading to a 'fluorescence' nebular phase similar to SNe after many years \citep{Jerkstrand:17}.

\subsection{The \texorpdfstring{$Y_e \sim 0.15$}{Ye ~ 0.15} Model}
\label{subsec:lowYe_species}

\begin{figure}
    \includegraphics[trim={0.0cm 0.1cm 0.4cm 0.3cm},clip,width = 0.47\textwidth]{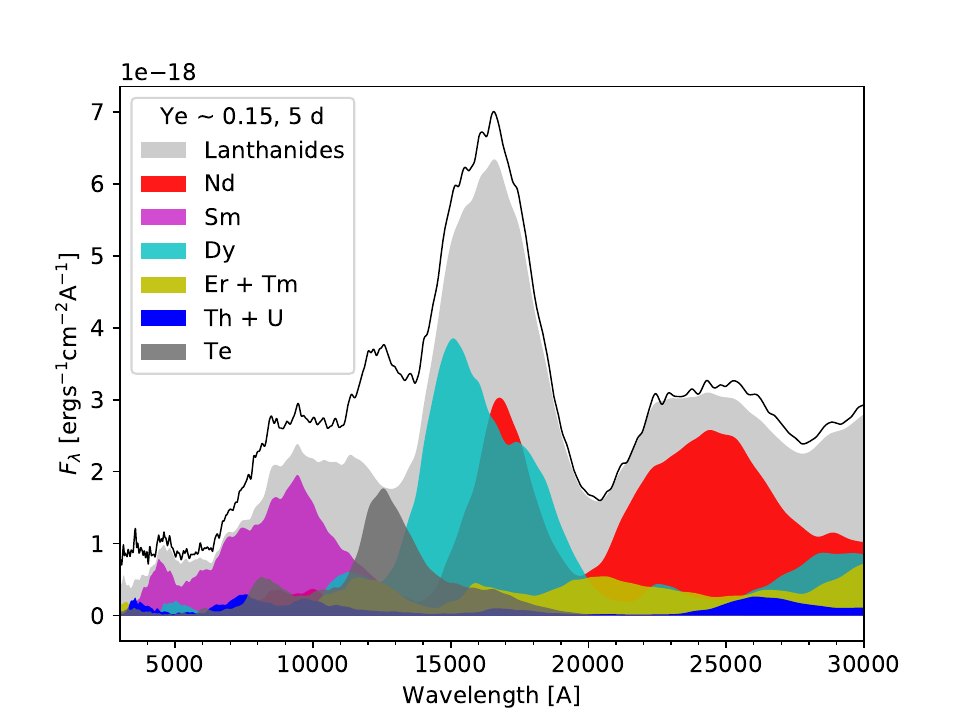} 
    \includegraphics[trim={0.0cm 0.1cm 0.4cm 0.3cm},clip,width = 0.47\textwidth]{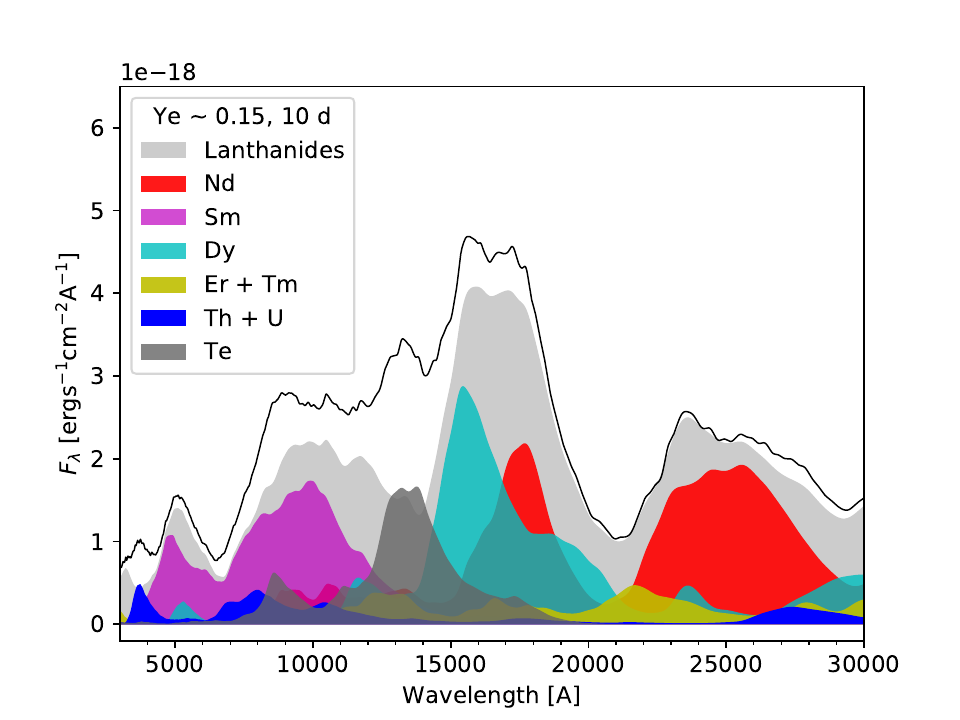}
    \includegraphics[trim={0.0cm 0.1cm 0.4cm 0.3cm},clip,width = 0.47\textwidth]{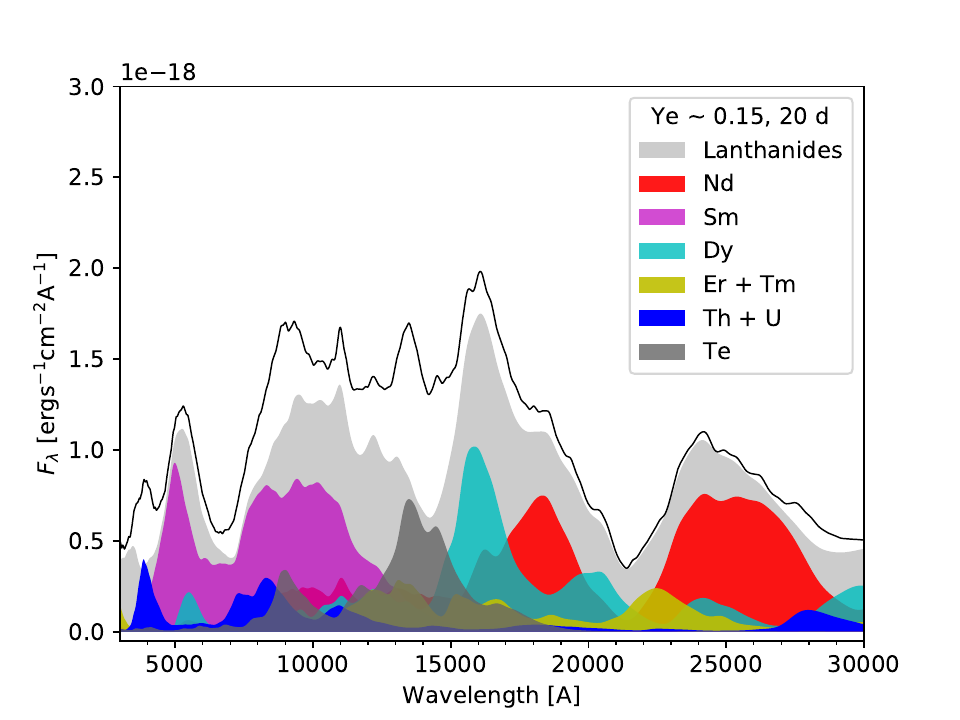}
    \caption{The spectrum of the $Y_e \sim 0.15$ model at 5, 10 and 20 days, with key emitting species marked. The light grey shaded area represents the total lanthanide emission (note that this contains also marked individual lanthanide contributions of Nd, Sm, Dy, Er, Tm), whilst key individual elements are marked out in colours.}
    \label{fig:Ye015_species}
\end{figure}

Looking now at the $Y_e \sim 0.15$ model shown in Fig. \ref{fig:Ye015_species}, we see that the emergent spectra are almost entirely dominated by lanthanide emission across the entire wavelength range. The shape of the emergent spectra of this model remains relatively similar across all epochs, with the amplitude of individual features varying in different ways. In particular, we recover also here the distinct Nd and Dy feature at $1.7~\mu$m, which is the dominant emission feature at all epochs. As for the $Y_e \sim 0.25$ model, the emission past $1.5~\mu$m is dominated by Nd and Dy, with some additional contributions from Er and Tm. The actinides Th and U, with a small abundance of $\sim$0.001 each, contribute a little bit of emission between 2.5 and 3~$\mu$m.

\begin{figure}
    \includegraphics[trim={0.0cm 0.1cm 0.4cm 0.3cm},clip,width = 0.47\textwidth]{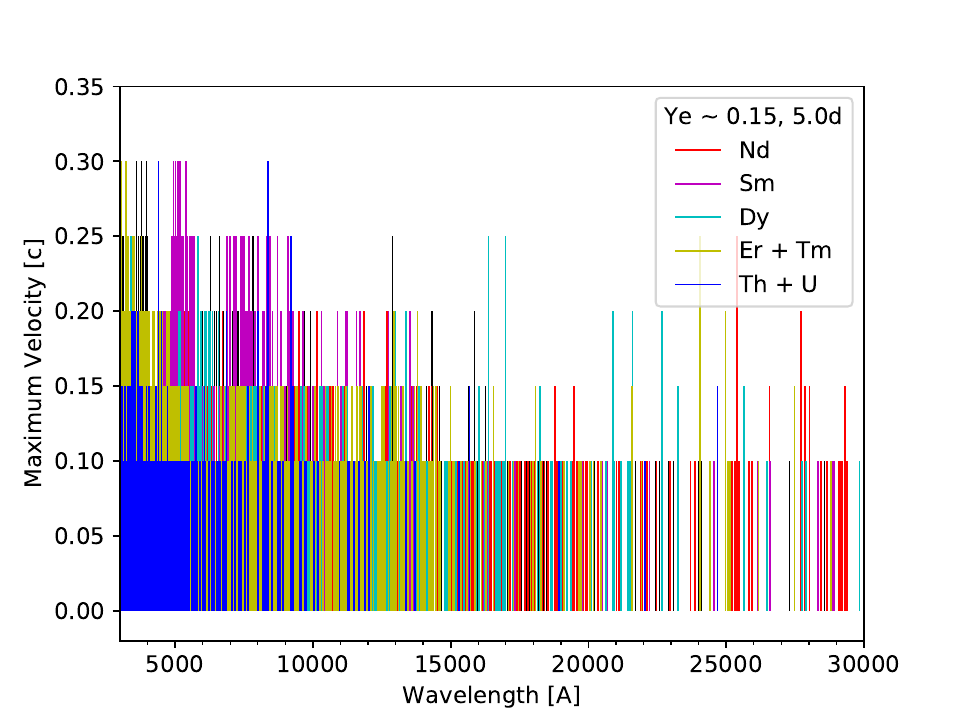} 
    \includegraphics[trim={0.0cm 0.1cm 0.4cm 0.3cm},clip,width = 0.47\textwidth]{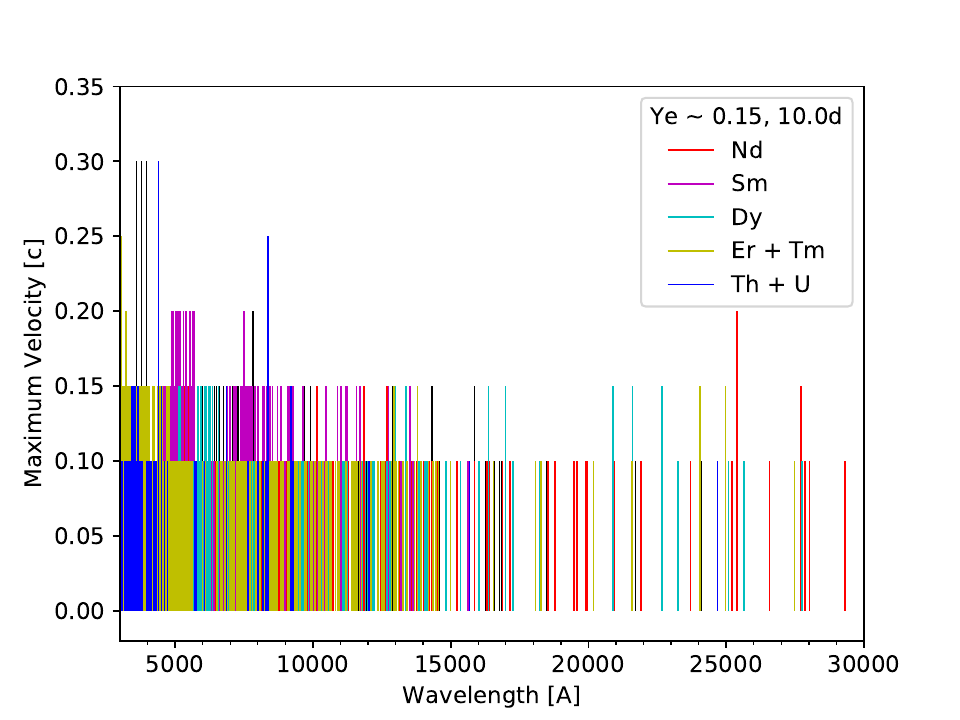}
    \includegraphics[trim={0.0cm 0.1cm 0.4cm 0.3cm},clip,width = 0.47\textwidth]{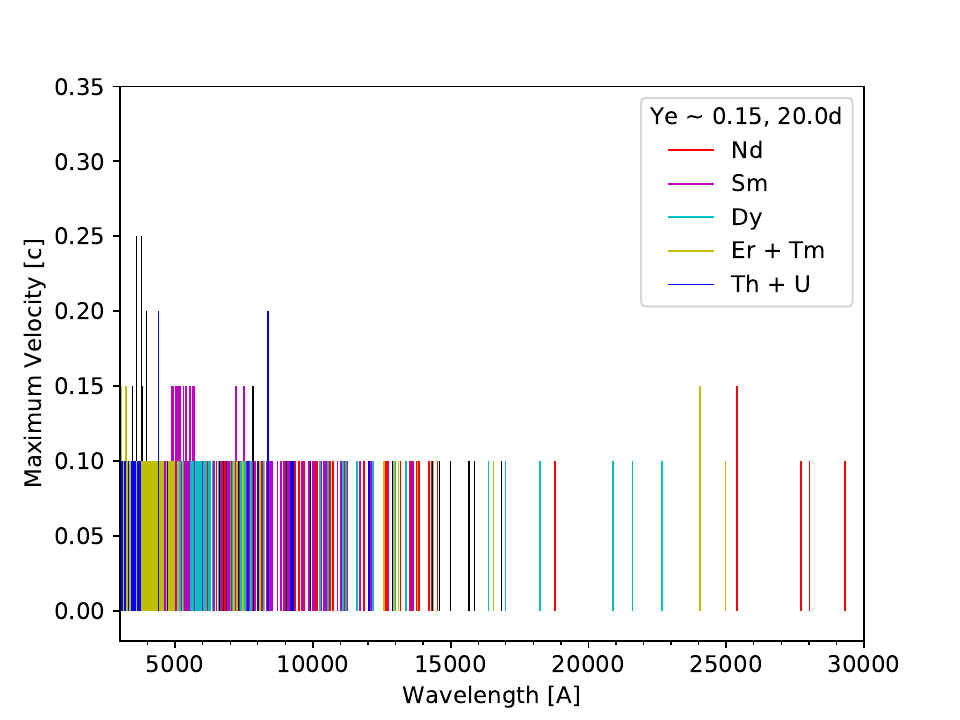}
    \caption{Optically thick lines in the $Y_e \sim 0.15$ model at 5, 10 and 20 days. The velocity up to which the line is optically thick is shown. Line wavelengths are plotted as rest wavelengths. Major contributing elements are marked in colours. Remaining contributions from other elements are also plotted, but are typically minimal compared to the marked elements.}
    \label{fig:Ye015_lines}
\end{figure}

\begin{figure}
    \includegraphics[trim={0.cm 0.1cm 0.4cm 0.3cm},clip,width = 0.47\textwidth]{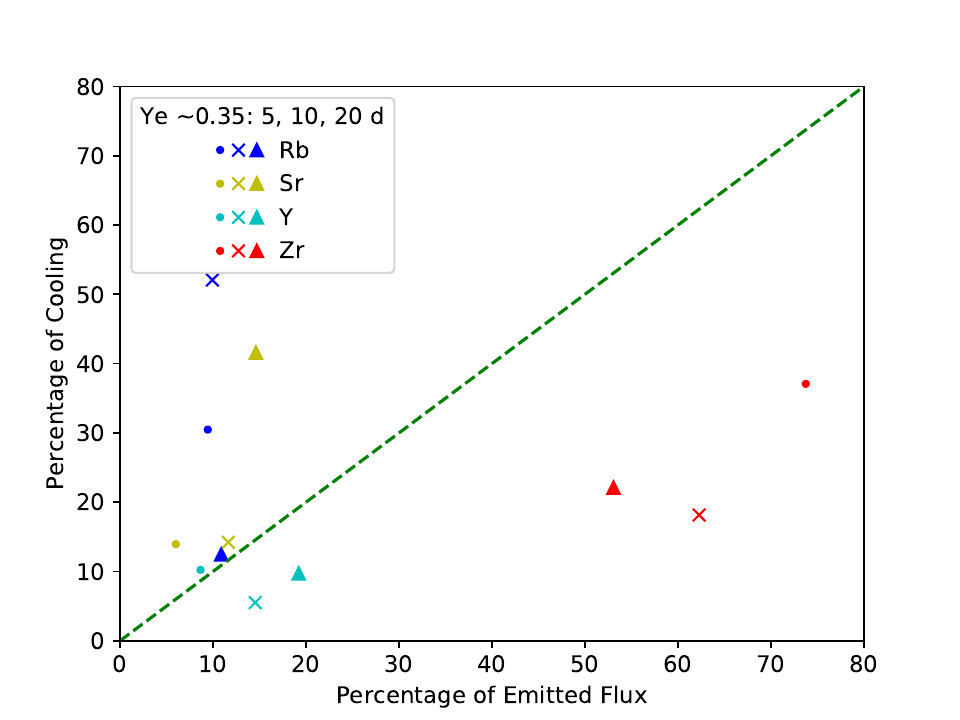} 
    \includegraphics[trim={0.cm 0.1cm 0.4cm 0.3cm},clip,width = 0.47\textwidth]{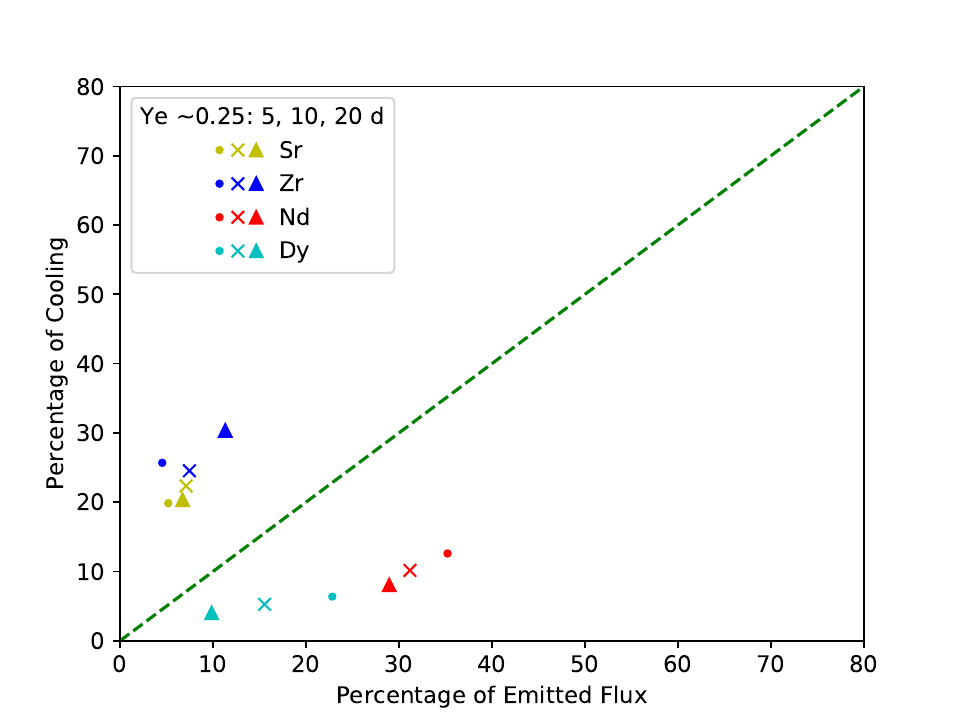}
    \includegraphics[trim={0.cm 0.1cm 0.4cm 0.3cm},clip,width = 0.47\textwidth]{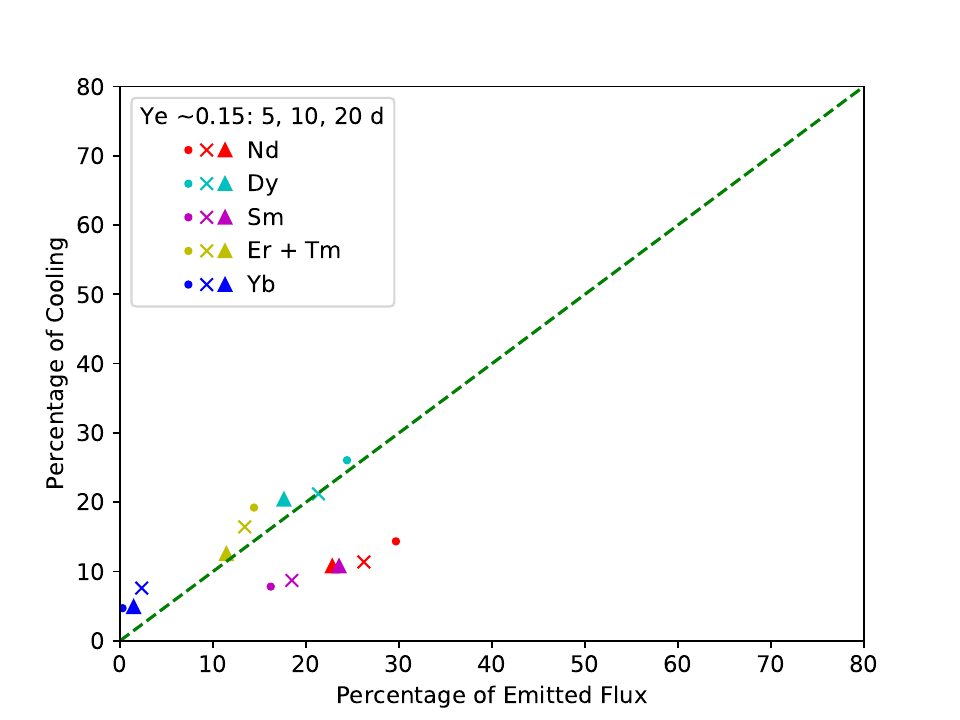}
    \caption{Comparison of the cooling (y-axis) vs emergent flux contribution (x-axis) for key elements at 5d (points), 10d (crosses), and 20d (triangles). The diagonal dashed line represents equal contribution to both. Points below the diagonal indicate the emission by that element is dominated by  scattering/fluorescence.}
    \label{fig:coolvflux}
\end{figure}

The strong Nd feature at $2.5~\mu$m is persistent across all epochs, as well as the apparent trough on its blue side. Combining the information that Nd is strongly scattering/fluorescence dominated (Fig. \ref{fig:coolvflux}) and that there is a thick Nd line at $\lambda_0 \sim $2.55 $\mu$m in many ejecta layers at all epochs (Fig. \ref{fig:Ye015_lines}), we conclude that this line is probably formed primarily by the P-Cygni mechanism.

We also see the emergence of a prominent Te feature at $\sim 1.3~\mu$m, which is distinct in the spectra across all epochs. Looking closer at the levels and transitions responsible for this emission we cannot, however, clearly link these to experimentally validated levels, and so choose not to analyse this feature in more detail due to its unreliability.

In the optical, the lanthanide samarium (Sm) is the element contributing most of the flux. The reason for this is its dominant line blanketing in the optical (Fig. \ref{fig:Ye015_lines}), and indeed this yields scattering/fluorescence emission (Fig. \ref{fig:coolvflux}). Dy, Th and U also provide some contributions to the optical flux. Conversely, Zr, which was significant in the $Y_e \sim 0.25$ model, is only at 0.015 abundance here, and does not contribute much to the spectral formation. From Fig. \ref{fig:Ye015_lines}, we see a large amount of line blocking at 5d, with a partial escape window around $\lambda_0 \sim 1.7$--$2.1~\mu$m, which allows the emergence of the large $1.7~\mu$m peak. At 10 days, we also identify a partial escape window at $\lambda_0 \sim 7000~\ang$, which leads to the Sm peak at $\sim 5000~\ang$. We also see a strong U\,\textsc{iv} absorption line at rest wavelength $\lambda_0 = 8400~\ang$, which drives the trough formation at $\lambda \sim 7000~\ang$, most noticeable at 10 and 20 days. Given the U emission just redwards of this feature, and finding that this transition does negligible cooling, we suggest this to be a P-Cygni like feature. However, due to the lack of atomic data for U\,\textsc{iv}, we cannot verify the accuracy of this particular transition. As in previous models, we see how spectral peaks arise within escape windows in the line blocking, rather than corresponding to any particular transition.

\citet{Hotokezaka2023} identify [Te\,\textsc{iii}] 2.1 $\micron$ as a candidate for the observed emission-like line in AT2017gfo at epochs 7.5--10 days. Our $Y_e \sim 0.15$ model here has a similar composition as their $A>88$ model, both with little to no material from the first r-process peak, and a large  Te mass fraction (0.07 in our model). Our model atom has the transition at theoretical wavelength $\lambda = 2.21~\mu$m and with transition rate A = 1.6 s$^{-1}$, both close to the values calculated by \citet{Madonna2018}. However, our effective collision strength for this transition follows the prescription for forbidden transitions from \citet{Axelrod:80}: $\Upsilon = 0.004g_ig_j = 0.012$, for upper level \textit{i} and lower level \textit{j} respectively. This value is much lower than the value calculated by \citet{Madonna2018} of $\Upsilon \gtrsim 5$ (for $T \lesssim 5000~$K). The significance of this depends on whether the line is close to LTE or not. In our model, we find that the population of the upper level of this transition has a departure coefficient of $\sim 10^{-2}$ \citep[see also fig. A2 in][]{Pognan.etal:22b}, implying that this line is far from LTE, and as such our smaller collision strength has a direct impact. On the other hand, for our model at 10 days, we have higher temperature solutions ($T =$ 4000--24000~K) than that assumed by \citet{Hotokezaka2023} of 2000~K, which gives a higher emissivity for a given departure coefficient. These two differences in physical conditions, combined,  give a total (angle-integrated) emissivity from this line of $2.2 \times 10^{38} ~ \rm{erg~s^{-1}}$, about an order of magnitude smaller than the observed value of $\sim 2\times10^{39}~ \rm{erg~s^{-1}}$ \citep{Hotokezaka2023}. As such, it is possible that a too low intrinsic emissivity of the [Te]\,\textsc{iii} $2.1~ \mu$m line is at least partially responsible for its lack of emergence in our model.

Another aspect to consider is to what extent an emission line at $2.1 ~ \mu$m is free to directly escape at 5--10 days. In our model, the inner 2--3 zones (corresponding to 90 per cent of the ejecta mass) still have some line blocking at these epochs, such that any [Te\,\textsc{iii}] emission at that wavelength will be at least partially absorbed and re-emitted at longer wavelengths. However, in contrast to the 'forest' of blocking lines below $\sim$ 1.5 $\mu$m, the opacity here is provided by relatively few lines. Given the current limitations to the accuracy of wavelengths and A-values, one cannot yet make any real robust statements about the degree of optical thinness around 2.1 $\mu$m in the 5--10 day epochs.

Looking at the bottom panel of Fig. \ref{fig:coolvflux}, we see that the dominating species, all of which are lanthanides, typically lie close to the dividing line between cooling driven and scattering/fluorescence driven emission, with Nd and Sm being more on the latter side. Since most of these species have similar structure with open \textit{f}-shells, it is somewhat expected that they would behave similarly. As such, the competition between these elements with respect to domination of scattering/fluorescence or cooling likely comes down to details in their atomic structure. Overall, this very lanthanide-rich model showcases how difficult it may be to identify single species in a composition representative of low $Y_e$ ejecta. The atomic nature of the lanthanides, and typical lack of stand-out strong transitions, yields spectra with broad, blended absorption and emission features that often arise from many different species. The peaks formed, even at 20 days, tend to arise in escape windows within the line blocking, rather than corresponding to any intrinsically important emission lines.

\subsection{Time-Dependent Effects on Spectra}
\label{subsec:timedep_spectra}

\begin{figure}
    \includegraphics[trim={0.cm 0.1cm 0.4cm 0.3cm},clip,width = 0.47\textwidth]{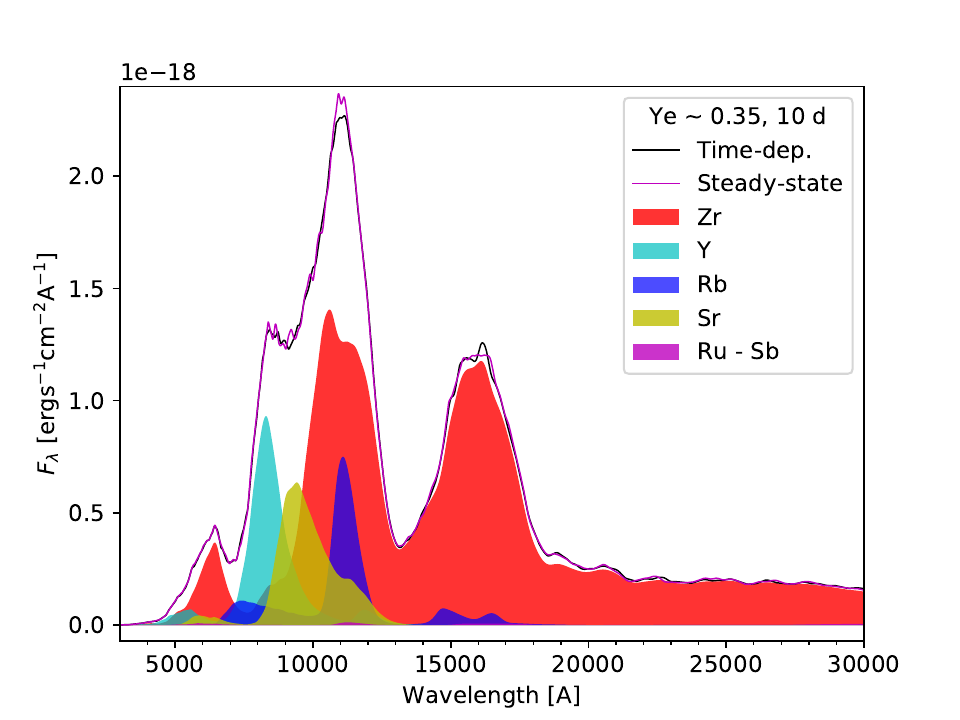} 
    \includegraphics[trim={0.cm 0.1cm 0.4cm 0.3cm},clip,width = 0.47\textwidth]{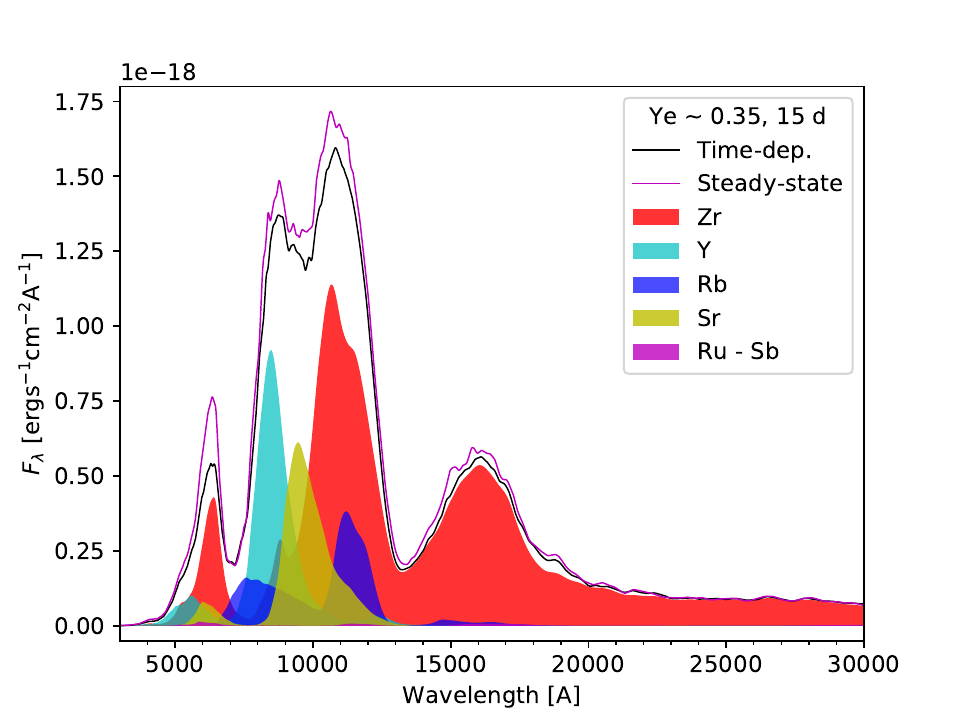}
    \includegraphics[trim={0.cm 0.1cm 0.4cm 0.3cm},clip,width = 0.47\textwidth]{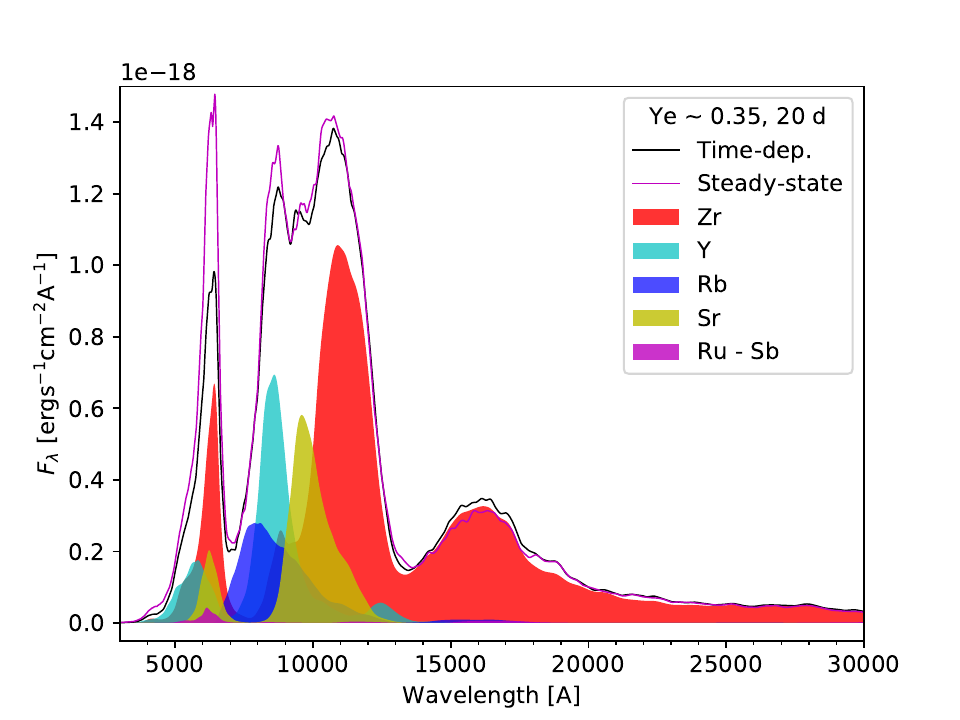}
    \caption{The key emitting species in the $Y_e \sim 0.35$ model at 10, 15 and 20 days. The black spectra are the time-dependent solutions, whilst the magenta spectra are the steady-state solutions. The filled area showing elemental contributions are with respect to the time-dependent solution.}
    \label{fig:Ye035_species_time}
\end{figure}

We now take a closer look at the effects of using the full time-dependent equations on the emergent spectra. As shown previously in Section \ref{sec:thermo_results}, only the $Y_e \sim 0.35$ model appears to show significant effects in the thermodynamic quantities, and this is also reflected in the spectral output. Conversely, the $Y_e \sim 0.25, 0.15$ models show little to no difference in their emergent spectra. The comparison of the total spectral output in steady-state and time-dependent modes from 10 to 20 days for the $Y_e \sim 0.35$ model are shown in Fig. \ref{fig:Ye035_species_time} along with the elemental contributions of the same key elements as identified above. 

As expected, we find that time-dependent effects increase as time goes on, following the same trend as seen in Section \ref{sec:thermo_results}. These changes are too small to affect the emergent spectrum at 10 days after merger, but differences become noticeable from 15 days onwards. The time-dependent solution has a smaller emission peak at $\sim 6000 \ang$, and at 20 days, more emission around $1.7 \mu$m. As the main effects of the time-dependent results are to lower temperature and the degree of ionisation, we must consider both in tandem to explain the effects on the emergent spectrum. Since the strongest emitting species in the model are neutral and singly ionised species, a lower ionisation degree will increase the abundance of these species, which would, to first order, imply increasing their emission. 

However, the ejecta temperature is also cooler, such that transitions excited by thermal collisions are weaker. For many of the emission lines, we therefore have a competing effect, where although we have a higher abundance of an emitting species, the total emission may actually decrease due to lower emissivity from cooler temperatures (see Fig. \ref{fig:Ye035_20d_species_detail} for a detailed look of time-dependent effects on the key species). These competing effects leading to changes in the emergent spectrum, highlight the complex nature of time-dependent effects not only on the thermodynamic state of the ejecta, but also on the emergent spectra. These results suggest that accurate spectral analysis of low density, lanthanide-free ejecta may require time-dependent, NLTE modelling in order to be properly interpreted.

\subsection{The Effects of Strontium on Lanthanide-Free Ejecta}
\label{subsec:SrII_focus}

\begin{figure}
    \includegraphics[trim={0.0cm 0.1cm 0.4cm 0.3cm},clip,width = 0.47\textwidth]{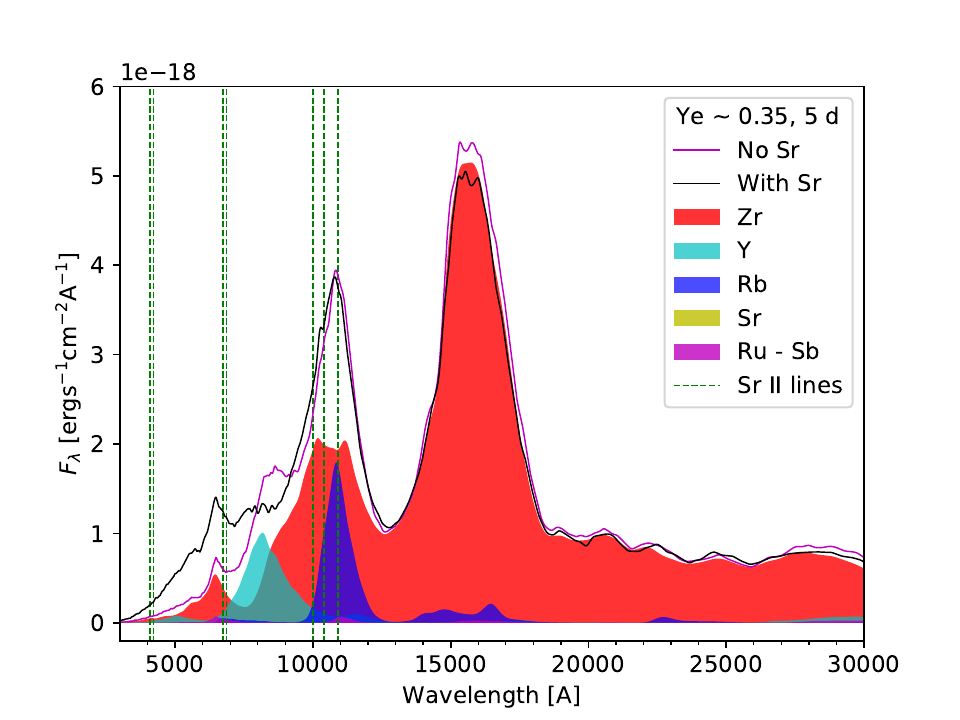} 
    \includegraphics[trim={0.0cm 0.1cm 0.4cm 0.3cm},clip,width = 0.47\textwidth]{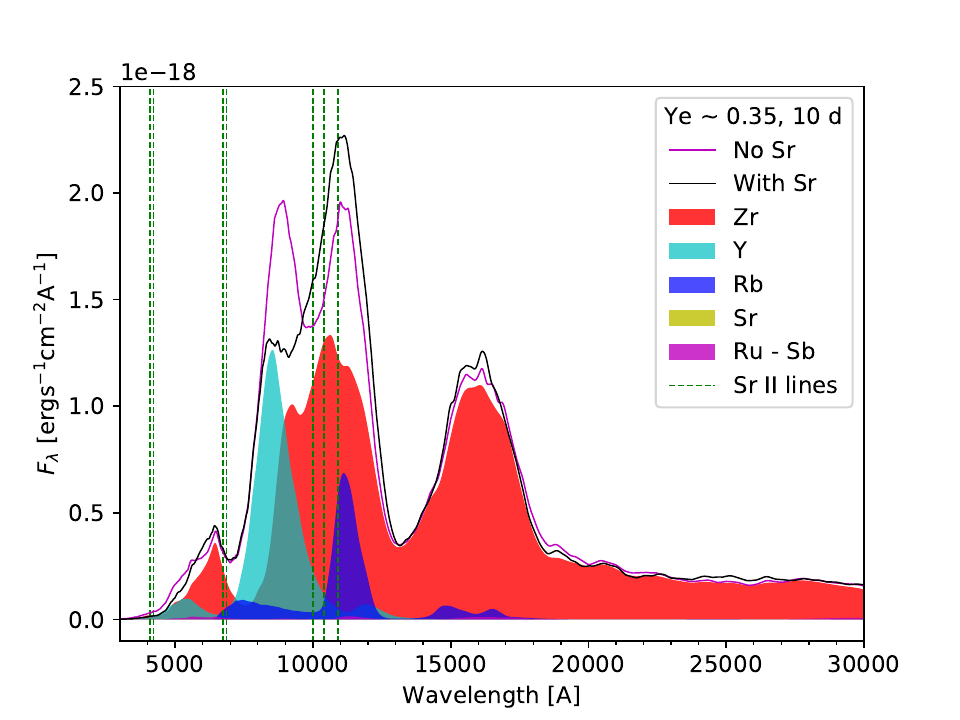}
    \includegraphics[trim={0.0cm 0.1cm 0.4cm 0.3cm},clip,width = 0.47\textwidth]{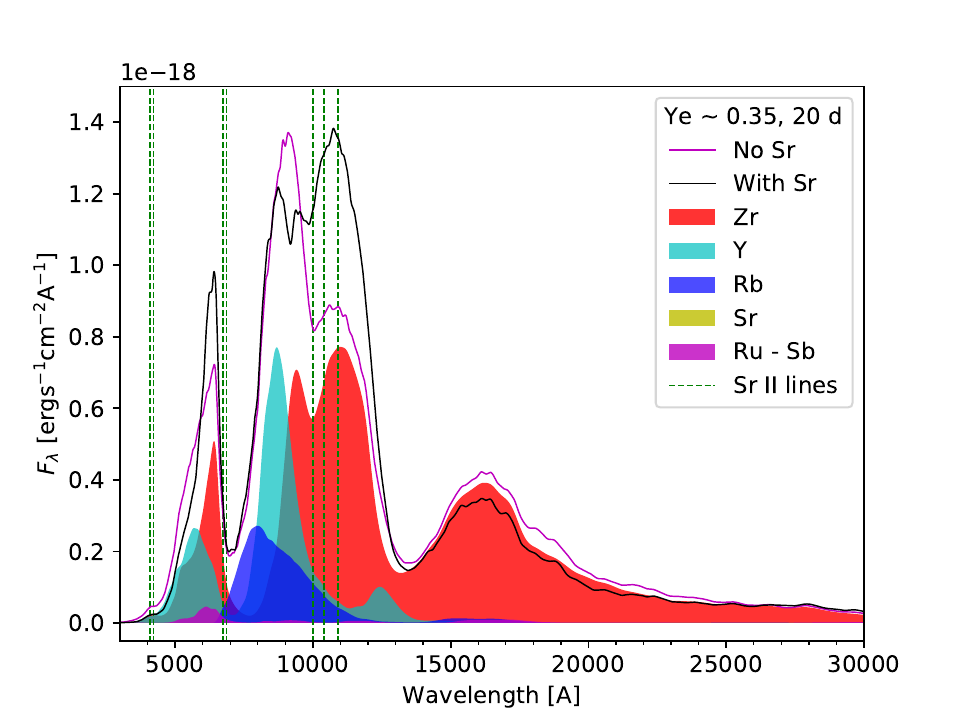} 
    \caption{The effect of including (black) and excluding (yellow) Sr on the $Y_e \sim 0.35$ model spectrum. The elemental contributions of Rb, Y and Zr are marked out by the fills, and correspond to the Sr-free model. The vertical dashed green lines mark out the rest wavelengths of the Sr\,\textsc{ii} $4000~\ang$, $6800~\ang$ doublets and the $10000~\ang$ triplet respectively.}
    \label{fig:Ye035_noSr_detail}
\end{figure}

Sr has been proposed as responsible for a spectral feature in AT2017gfo seen up to about a week after merger \citep{Watson.etal:2019,Domoto.etal:22}, though an alternative He origin has also been suggested \citep[see e.g.][]{Perego.etal:22,Tarumi.etal:23}. To study in more detail how our models are affected by Sr, we run again the $Y_e \sim 0.35$ model with Sr removed. In order to maintain a constant total ejecta mass, the mass fractions of the other elements were correspondingly adjusted (see Table \ref{tab:compositions}), while the energy deposition remained as before.

Considering first the thermodynamic state of the ejecta, we generally find higher temperatures when removing Sr from the model, in the range of $\sim$100--1500~K from the innermost to outermost ejecta layers respectively, the difference typically becoming more important as time progresses. Sr has a particularly strong effect on the ejecta's temperature due to the efficient cooling in the Sr\,\textsc{ii} $\sim 4000~ \ang$ channel. This transition is found to provide up to $\sim 40$ per cent of the total cooling for ejecta layers where Sr\,\textsc{ii} is highly abundant (e.g. Fig. \ref{fig:coolvflux}). As such, removing Sr from the model significantly reduces the cooling capacity of the ejecta, thereby increasing the temperature.

Although most individual species become slightly more ionised, the overall electron fraction $x_e$ of the medium decreases. This is because Sr\,\textsc{i} and\,\textsc{ii} are quite easily ionised, and the typically dominant species is therefore Sr\,\textsc{iii}. Conversely, most other species in the ejecta are more abundant in their singly ionised states. As such, removing Sr from the model leads to an overall decrease of electron fraction $x_e$, although other elements are slightly more ionised. A smaller number of free electrons in the ejecta further aids in increasing the temperature.

The emergent spectra of the $Y_e \sim 0.35$ model with and without Sr are shown in Fig. \ref{fig:Ye035_noSr_detail}. We focus on the changes arising from Y and Zr, as we find that the spectral features of the other elements in the model do not appear to be strongly affected. The presence of Sr in the ejecta mainly causes changes in the wavelength range $\lambda \sim$ 8000--12000~$\ang$ by effect of the $10000~ \ang$ triplet. Certain effects can also be seen at shorter wavelengths, but line formation there is extremely complex, while there is also an effect at longer wavelengths, though it is relatively minor. Thus, we focus on the 8000--12000~$\ang$ range in our analysis. 

At 5 and 10 days, the presence of Sr gives a more extended and deeper absorption trough at $\sim 8500~ \ang$ in the spectrum compared to a composition without it, suppressing flux levels specifically around this wavelength. The effect on the $1.1~ \mu$m peak is small however, with Sr slightly decreasing flux levels at 5 days and slightly increasing them at 10 days. Thus, the situation appears far from a single-line P-Cygni limit within 10 days after merger. Only at 20 days does Sr noticeably increase the strength of that feature, then by a factor of roughly 1.4, while slightly decreasing the flux levels at $\sim 8500~ \ang$ by absorption from the $10000~ \ang$ triplet.

In section \ref{subsec:highYe_species}, we identified the 1.1 $\mu$m spectral peak as arising due to an escape window opening up around this wavelength. Its persistence also for compositions without Sr would then be explained by several other lines from Y, Zr, and Rb also providing optically thick lines up to, and around this wavelength. Sr helps out in this chain of reprocessing radiation towards the escape window, but it is not crucial. However, we see a marked change in the spectral shape around $10000~ \ang$ from 10 days onwards when Sr is added. Therefore, these models support Sr as being active in KN spectral formation, in particular through its $10000~ \ang$ triplet. However, the specific impact of Sr may be complex, with the effect of the $10000~ \ang$ feature being not easily distinguishable from other line scattering processes. We note again that our analysis here is limited to epochs $\geq5$ days, so the situation may be different at earlier phases.

\section{Comparison to AT2017gfo}
\label{sec:AT2017gfo}

In this section we compare the evolution of our models to that of AT2017gfo with respect to bolometric luminosity (Fig. \ref{fig:lbol}) and colour evolution in \textit{grizJHK} colours (Fig. \ref{fig:colours}). Given the limited accuracy of wavelengths in our data set, the models generally do not yield accurate predictions for specific features. However, we expect that the models are able to reasonably capture the general SED. For these reasons we here focus our comparison with AT2017gfo mostly to colours, which test the general shape and evolution of the SED. 

\subsection{Bolometric and Optical Light Curves}
\label{subsec:LC_evolution}
\begin{figure}
    \includegraphics[trim={0.0cm 0.cm 0.4cm 0.3cm},width = 0.475\textwidth]{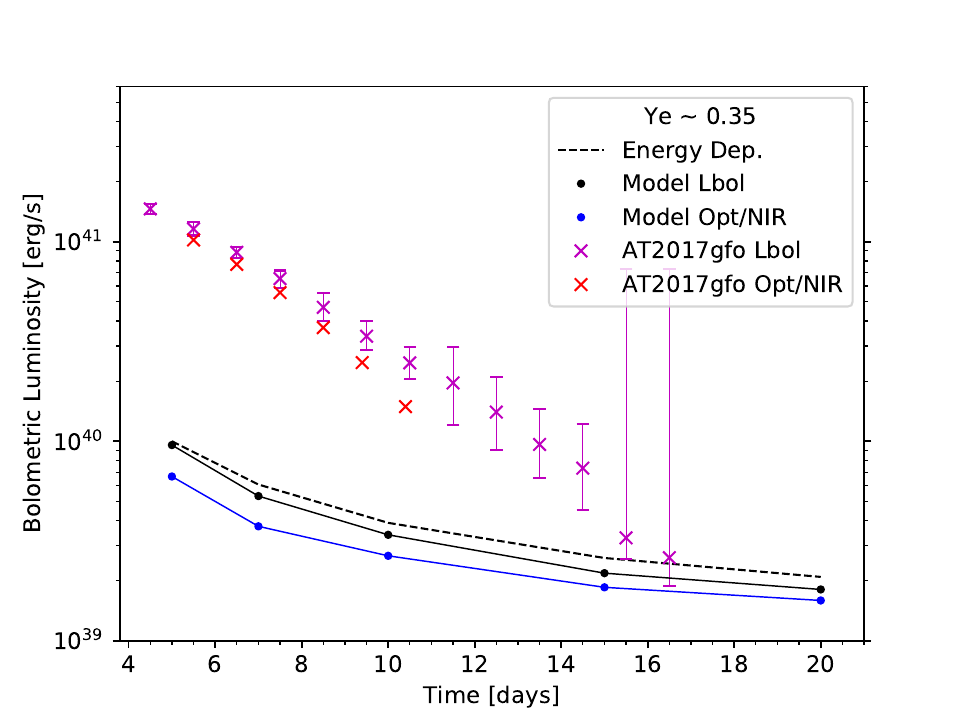} 
    \includegraphics[trim={0.0cm 0.cm 0.4cm 0.3cm},width = 0.475\textwidth]{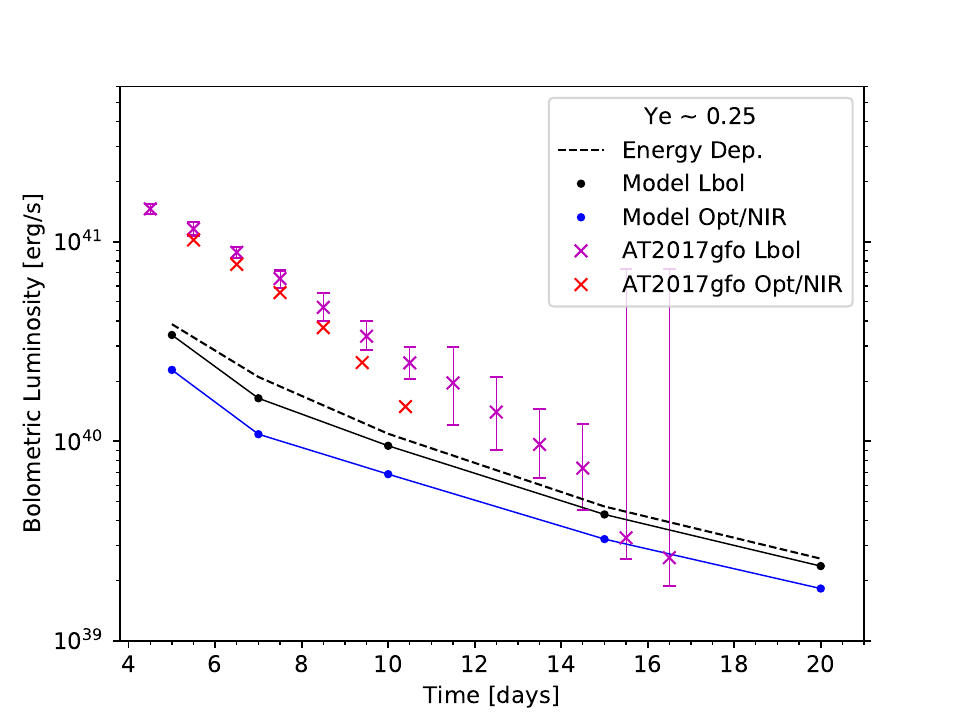}
    \includegraphics[trim={0.0cm 0.cm 0.4cm 0.3cm},width = 0.475\textwidth]{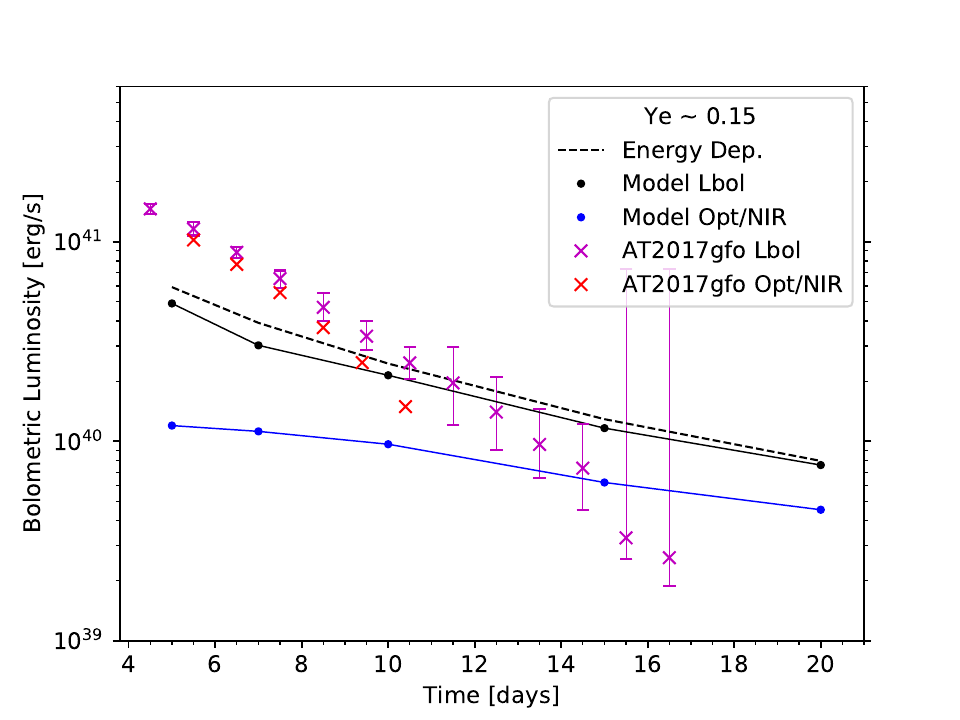}
    \caption{The bolometric lightcurves of each model compared to those of AT2017gfo as calculated by \citet{Waxman.etal:18}. The black points arise from integration of photometric data, while the red points are luminosity estimates in the 0.3--2.4~$\mu$m range calculated by integrating a blackbody fit. We compare our model to the latter by integrating our spectra in the same range, corresponding to the wavelengths of the X-shooter spectra of AT2017gfo. We note that no uncertainties are provided for the luminosity calculated from the blackbody fit. The energy deposition in our models (including thermalisation effects) are shown by dashed black lines; the emergent bolometric luminosity is somewhat lower than this due to adiabatic losses as photons scatter.}
    \label{fig:lbol}
\end{figure}

\begin{figure*}
    \centering
    \includegraphics[trim={0.0cm 0.1cm 0.4cm 0.3cm},clip,width = 0.47\textwidth]{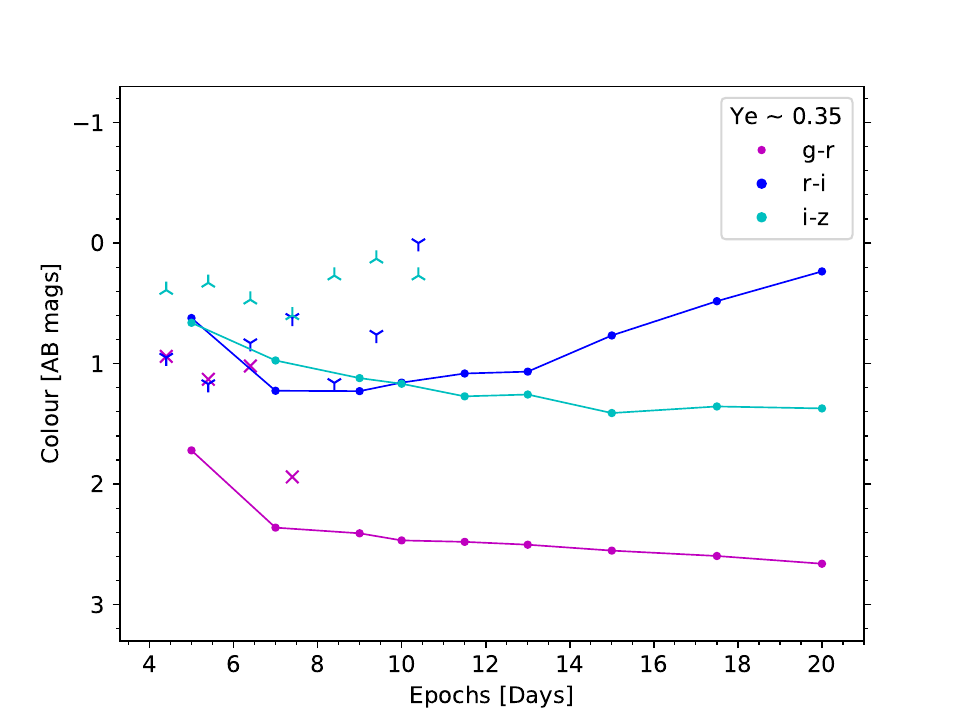} 
    \includegraphics[trim={0.0cm 0.1cm 0.4cm 0.3cm},clip,width = 0.47\textwidth]{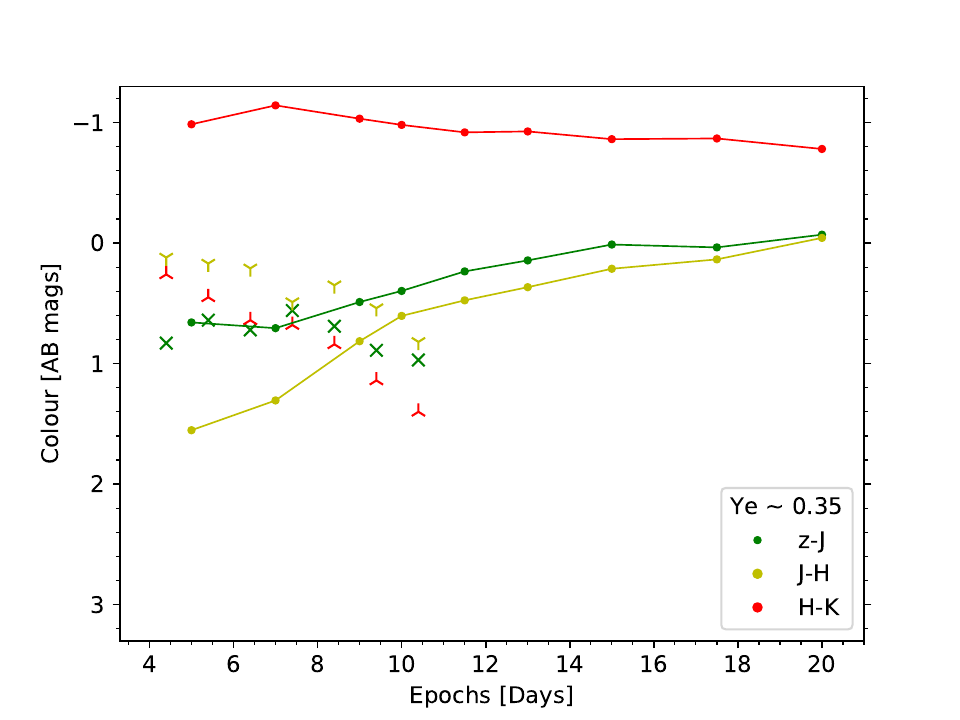}
    \includegraphics[trim={0.0cm 0.1cm 0.4cm 0.3cm},clip,width = 0.47\textwidth]{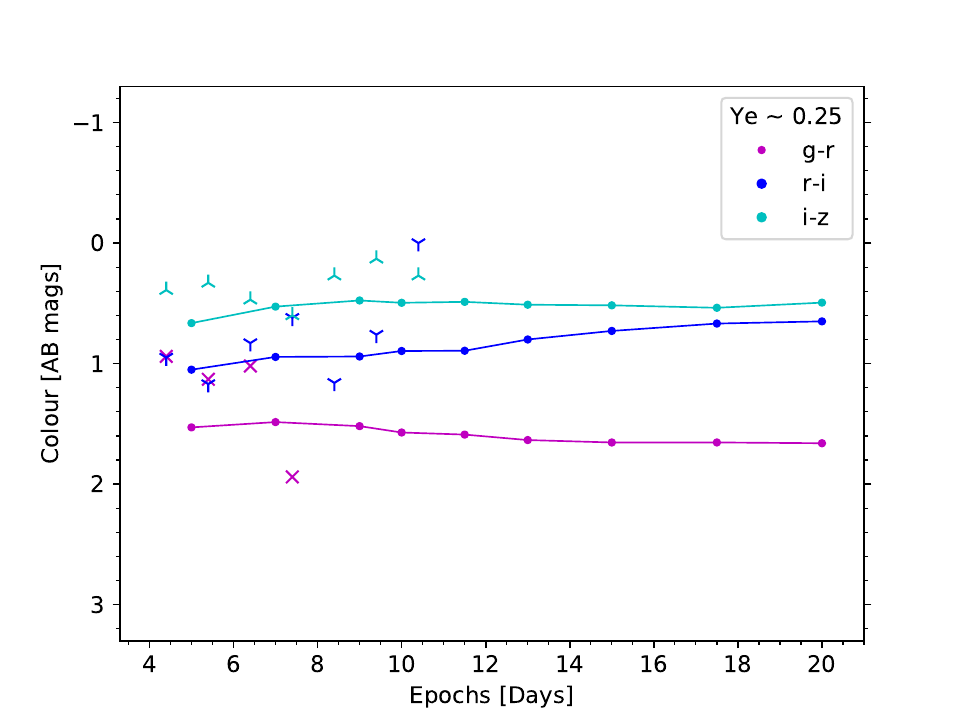}
    \includegraphics[trim={0.0cm 0.1cm 0.4cm 0.3cm},clip,width = 0.47\textwidth]{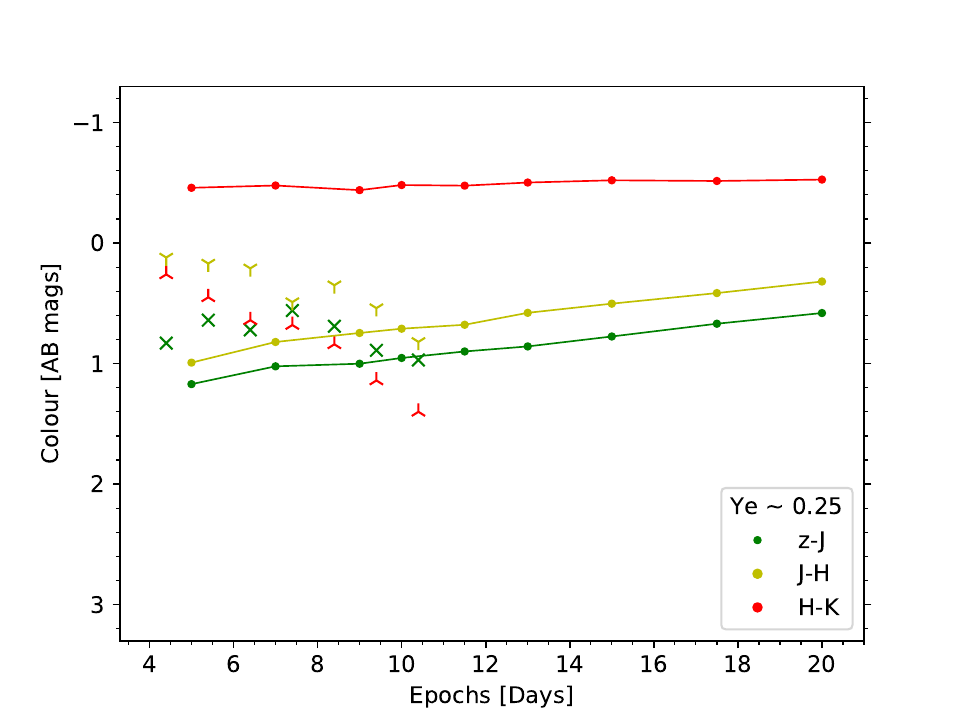} 
    \includegraphics[trim={0.0cm 0.1cm 0.4cm 0.3cm},clip,width = 0.47\textwidth]{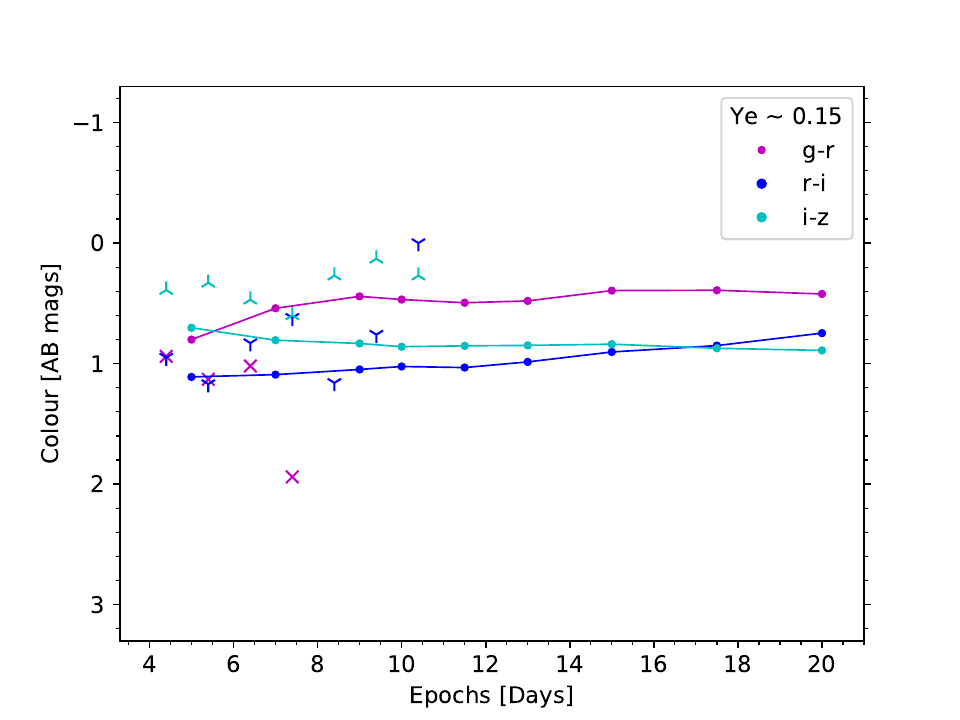}
    \includegraphics[trim={0.0cm 0.1cm 0.4cm 0.3cm},clip,width = 0.47\textwidth]{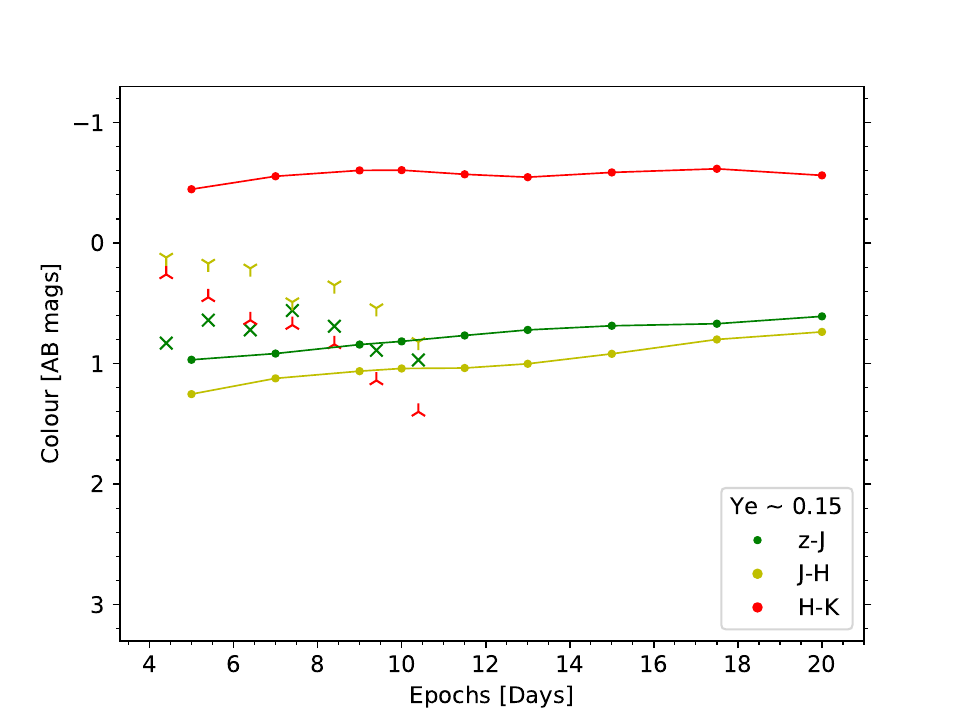}   
    \caption{Colour evolution of our models (points) in optical and NIR colours on the left and right hand sides respectively, compared to the observed colour evolution of AT2017gfo (crosses and Y-shapes) in the first 20 days after merger. We note that AT2017gfo is missing observations in several photometric bands after 10 days, so we limit ourselves to the first 10 days which have reliable photometric measurements in most bands, aside from g-band which stops after 7.4 days.} 
    \label{fig:colours}
\end{figure*}

Considering first the bolometric light curve (LC) in Fig. \ref{fig:lbol}, we see that our models are typically fainter than AT2017gfo at early times, until about 10 days after, with only the $Y_e \sim 0.15$ model reaching comparable and greater luminosities past 10 days. The disparity is worse at earlier times, with the $Y_e \sim 0.35$ model being more than an order of magnitude dimmer than AT2017gfo at 5 days after merger. Conversely, the $Y_e \sim 0.15$ model has a bolometric luminosity mostly consistent with AT2017gfo from 10 days onwards, though it should be noted from the error bars that the observed luminosity is poorly constrained at these late epochs. This higher luminosity is consistent with the greater power of the model, which can be seen in Figure \ref{fig:energy_dep}. There, we see that at these epochs, this model has a spontaneous fission contribution to the energy deposition that is roughly equal to the $\beta$-decay contribution, with $\alpha$-decay also adding a further $\sim$ 20 per cent.

There are several possible reasons for such a dim bolometric luminosity at early times. These include too little ejecta mass, particularities in the raw decay power for the chosen $Y_e$, or too inefficient thermalisation, leading to reduced energy deposition to the ejecta. Another significant factor may be time-dependent photon diffusion effects, which are not taken into account by \texttt{SUMO}. Diffusion was previously discussed in Section \ref{subsec:excitation}, where previous LC studies including diffusion effects have found that these play a role until 5--20 days, depending on composition \citep[e.g.][]{tanaka:opacities:2020,Bulla:2023}.

As our ejecta mass of $0.05~\rm{\Msol}$ lies well within the range of most estimates for AT2017gfo \citep[e.g.][]{Kasen.etal:17,Pian2017,Tanaka.etal:17,Smartt.etal:17,Waxman.etal:18}, and our thermalisation physics comes from well accepted semi-analytical fits \citep{Barnes.etal:16,Waxman.etal:19,Kasen.Barnes:19}, which were also well reproduced numerically by \citet{Hotokezaka.etal:20}, we do not believe these two explanations to be the main cause. With respect to the raw radioactive power arising from the composition, we can see from the top panel of Fig. \ref{fig:lbol} that our $Y_e \sim 0.35$ model is naturally a low power model. This is a particularity of this $Y_e$, as the nuclear power of this model between 1--20d is quite small owing to the relatively few isotopes that substantially contribute to radioactive heating \citep[see fig. 5 in][]{Wanajo.etal:14}. It should be noted however, that nuclear power at slightly higher $Y_e \sim 0.4$ is larger due to the important contributions from the $\beta$-decay chains of $^{66}$Ni and $^{72}$Zn \citep{Wanajo:18}. Furthermore, AT2017gfo is expected to synthesise heavier elements past the second r-process peak \citep[e.g.][]{Kasen.etal:17,Tanaka.etal:17,Rosswog.etal:18,Waxman.etal:18}, and potentially trans-lead elements which may provide additional power by efficiently thermalising $\alpha$-decay particles \citep[e.g.][]{Wanajo.etal:14,Wanajo:18}. As such, it is expected that the bolometric luminosity of this light composition model be lower than that of AT2017gfo.

The other two models, with heavier elemental compositions, do not have intrinsically low power, and their dimmer luminosity for $\leq$10 days thus likely arises from the omission of time-dependent photon diffusion in \texttt{SUMO}. In \citet{tanaka:opacities:2020} and \citet{Hotokezaka.etal:20}, emergent bolometric light curves are plotted compared to the instantaneous deposition after thermalisation. Is is seen there that diffusion is going on in the models up to 10--30 days, depending on the ejecta. On the other hand, the bolometric light curve is never more than a factor $\sim$2 brighter than the deposition. \citet{Bulla:2023} shows that the ratio can depend strongly on viewing angle in multi-D model, obtaining up to a factor 3 difference for polar viewing angles, at 5 days, but a negligible ($\lesssim$ 10 per cent) effect for equatorial angles. For the angle inferred for AT2017fgo of $\theta \sim 30 \degree$ \citep[e.g.][]{Pian2017,Troja.etal:17,Finstad.etal:18,Bulla:19} the factor is about 2, similar to the 1D results. One should note that LTE models with complete thermalisation of photon absorptions, such as these, may overestimate the duration of the diffusion phase as fluorescence to longer, more optically thin wavelength regions is not allowed to occur. However, by what extent cannot currently be addressed, and the AT2017gfo data itself also does not allow this to be determined. In contrast to observed supernova light curves, for which the diffusion phase is clearly identifiable, the fact that all KN ejecta are radioactive produces a rather indistinct difference between diffusion and steady-state light curves. As such, we may take a factor 2 as an upper limit to the luminosity factor.

Additionally, the factor between emergent luminosity and deposited energy is not exactly unity in models with stationary radiation fields such as ours. This is due to adiabatic degradation of the radiation field (note distinction to the adiabatic cooling of the thermal electrons as the ejecta expand) as photons interact on their way out. We find in our models that this adiabatic loss factor is $\sim$ 5--25 per cent, depending on model and epoch. As such, the factor 2 discussed above gets compounded due to adiabatic degradation, giving a total factor $\lesssim $3. We assess this is likely the driving factor behind the significantly too low luminosities of our $\rm{Ye} \sim 0.25$ and $Y_e \sim 0.15$ models between 5--10 days. Studies of the effects of photon diffusion in the context of SN Type Ia have found that while the bolometric luminosity may be reduced by a factor of 2--3 when omitted, the spectra remain very similar \citep{Kasen.etal:06,Shen.etal:21}. As such, we do not make any predictions for KN lightcurves at early times when diffusion may still be playing a role, but expect that the general SED shape, and thus the colours of our models, remain accurate. 

\subsection{Colours}
\label{subsec:colours}
We compare the colour evolution of our models to that of AT2017gfo in Fig. \ref{fig:colours}. Optical colours $g-r$, $r-i$ and $i-z$ are shown in the left hand panels, while NIR colours $z-J$, $J-H$ and $H-K$ are shown in the right hand panels. We note that the observed g-band photometry for AT2017gfo was limited to upper limits past 7.4 days \citep[e.g.][]{Villar.etal:17}, and so the $g-r$ colour for AT2017gfo is only shown up to that epoch. We see a relatively flat colour evolution for our models, with the exception of the $Y_e \sim 0.35$ model in the first few epochs. While the NIR colours of AT2017gfo show a marked reddening with time, our models instead tend to get slightly bluer over time in the NIR colours.

\begin{figure}
    \includegraphics[trim={0.4cm 0cm 0.4cm 0.3cm},width = 0.49\textwidth]{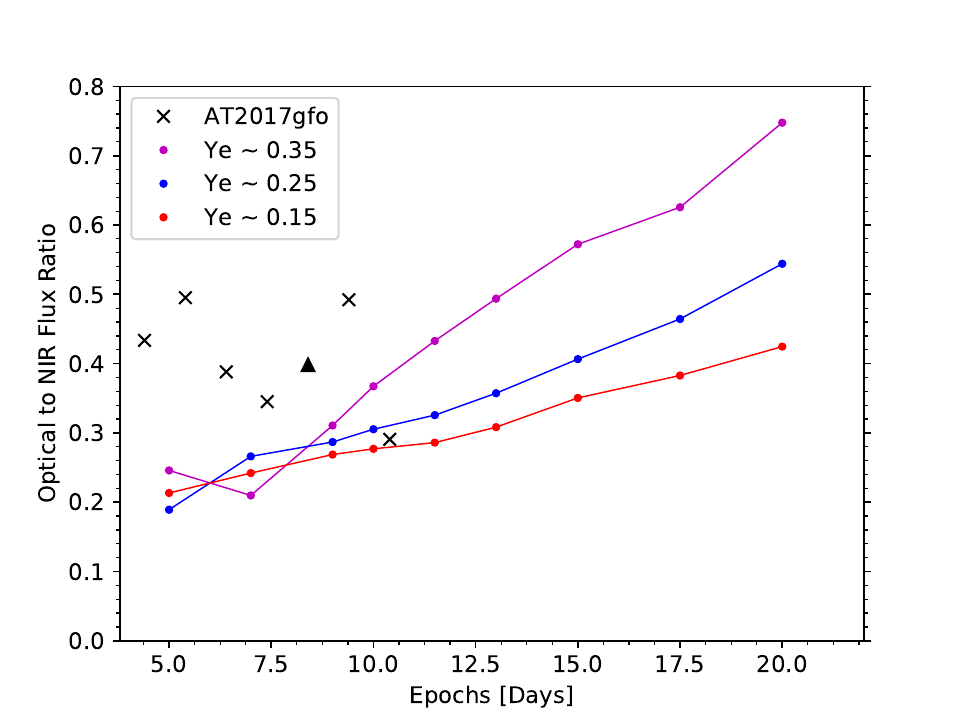} 
    \caption{The ratio of flux in the optical ($3000 \geq \lambda \geq 10000~\ang$) to the NIR ($10000 \geq \lambda \geq 24000~\ang$) for our models (solid lines) and from the spectra of AT2017gfo (crosses). The triangle at 8.4 days for AT2017gfo is a lower limit due to the X-Shooter spectrum lacking data below $\lambda = 6000~\ang$.}
    \label{fig:fluxratio}
\end{figure}

Although the colour evolution of both our models and AT2017gfo do not immediately appear to present a marked blue to red evolution in the 5--10 day regime, these colours are taken between adjacent bands in which the evolution may depend on the (dis)-appearance of individual features. In order to provide a 'colour' evolution with a larger perspective, we calculate the ratio of flux in the optical (3000--10000~$\ang$) to that in the NIR (10000--24000~$\ang$), shown in Fig. \ref{fig:fluxratio}. There, we see a clear trend for our models, where more flux is present in the optical regime as time goes on, i.e. the models get 'bluer'. The lightest model ($Y_e \sim 0.35$) evolves the fastest, while the $Y_e \sim 0.15$ model, with the highest lanthanide abundance, evolves the slowest. AT2017gfo, conversely, shows a large spread in its flux ratio values, consistent with a constant ratio over time, with the exception of the 10.4 day value which indicates transfer from optical to NIR (note that the point at 8.4 days is missing flux below $6000~\ang$ and so is a lower limit).

The increasing optical flux of our models is explained by the increasing temperatures (Figs. \ref{fig:thermo_evolution}) combined with the decreasing optical depths in the optical regime (Figs. \ref{fig:Ye035_lines},\ref{fig:Ye025_lines},\ref{fig:Ye015_lines}). As time progresses, the radiation field therefore generally becomes bluer, and is also able to escape at these wavelengths. At early times, blue photons are forced to scatter/fluoresce to redder wavelengths in order to escape, particularly for our heavier composition models. Considering models at 5 days (top panels of Figs. \ref{fig:Ye035_lines},\ref{fig:Ye025_lines},\ref{fig:Ye015_lines}), we see many optically thick lines throughout the whole ejecta at wavelengths $\lambda \lesssim 10000~\ang$. Taking the lowest temperature of our ejecta yielding the reddest photons, $T \sim 2500~$K in the innermost layer at the earliest epoch, to be representative of characteristic photon energies, we find, using Wien's law, a peak photon wavelength of $\lambda \sim 11500 \ang$, well within the optically thick regime for this inner layer. As outer layers have higher temperatures, the representative photon wavelength is pushed to bluer, and more optically thick wavelengths. Therefore, we find that scattering/fluorescence play a critical role in the spectral formation of KNe, and are expected to continue doing so until the optical depth has dropped sufficiently such that the KN enters a truly optically thin regime. 

The only exception to the increasing optical flux trend in our models arises in the transition of the $Y_e \sim 0.35$ model from 5 to 7 days after merger, where we see comparatively more flux in the NIR. This does not arise due to increased flux in the NIR, but rather a decrease in optical flux. This particular evolution highlights that the SED evolution of KNe in this regime also depends highly on specific features, and not only general temperature and optical depth. 

Alongside the complex colour evolution shown in Fig. \ref{fig:colours}, our results imply that the temperature of the ejecta cannot be reliably inferred from the shape of the SED. As an example, the best fitting blackbody to the 5 days spectrum of the $Y_e \sim 0.25$ model gives a temperature of $\sim$2000 K, whereas the ejecta temperature ranges from $\sim$ 4000--12000 K. Furthermore, we see that the $Y_e \sim 0.15$ model, which typically has the lowest optical to NIR flux ratio, is actually \textit{hotter} than the other models, as seen in Fig. \ref{fig:thermo_evolution}. Since the temperature solution of the gas in NLTE is a balance of radioactive heating and predominantly line cooling, both of which depend heavily on composition, it is not surprising that both the temperature solution and emergent spectra are highly different between these models. 

In general, the model spectra generated in this study are not expected or intended to be particularly similar to AT2017gfo given the simplicity of the morphology (1D) and homogeneous compositions, as well as limited accuracy of our atomic data and relatively simplified treatment of several NLTE processes. With respect to the ejecta model, AT2017gfo is believed to have had multiple components with different compositions and complex 3D morphology \citep[see e.g.][for reviews]{Perego.etal:17,Metzger:19,Shibata.Hotokezaka:19}. The apparent lack of a 'continuum'\footnote{The 'continuum' in line-emission dominated KNe is technically an overlap of many weak lines, and does not arise from continuous processes.} in our spectra may arise from the 1D, homogeneous nature of our models. For a given composition, the same optically thick lines will be blocking similar regions of the spectrum, further blueshifted from their rest wavelength the farther out they are in the ejecta. As such, blue photons are continuously line blocked and cannot escape without relying on scattering and fluorescence.

In a more realistic 3D inhomogeneous model, photons will have access to more escape routes where differing compositions provide different escape windows in the line blocking. The recent study of \citet{Shingles2023} modelled the early time emission from a purely dynamical ejecta component ($M_{ej} = 0.005 \Msol$) also with fluorescence fully considered. There, they found significant dimensional effects on the emergent spectrum, notably that the 1D spectra are less 'continuum' dominated than the 3D spectra. At this point however, it is not fully known to what extent geometrical effects play a role in the total emergent spectrum, since the dominant wind component (typically $\gtrsim 90\%$ of the ejecta mass) is expected to be somewhat more spherically symmetric than the dynamical component \citep[e.g.][]{Kawaguchi.etal:21,Kawaguchi.etal:23,Neuweiler.etal:23}. As such further studies taking into account key transfer processes such as fluorescence, and comparing geometries and compositions in 3D while also including the disc wind component are required to thoroughly establish the effects of dimensionality.

\section{Discussion and Conclusion}
\label{sec:discussion}

We have conducted 1D NLTE radiative transfer simulations with the spectral synthesis code \textsc{sumo} in order to generate KN spectra from 5 to 20 days after merger. We study three uniform-composition, multi-zone models of varying characteristic $Y_e \sim 0.35, 0.25, 0.15$, with compositions including up to 30 different elements in each model, and nuclear decay power from \citet{Wanajo.etal:14}.

We determine the temperature profiles of the ejecta, showing that the temperature for the most part increases monotonically outwards, although in some cases the peak is reached at an interior mass coordinate. As has been shown also in previous works \citep{Hotokezaka.etal:21,Pognan.etal:22a}, KN temperatures increase with time in the post-diffusion phase. Our model temperatures are 2500--5000~K in the innermost ejecta layers, while outermost layers have a larger range of 5000--35000~K, depending on model and epoch. The ionisation degree correspondingly spans $x_e \sim$ 0.7--2.2 (innermost layer) and $x_e \sim $1.6--2.9 (outermost layer). Thus, neutral, singly ionised, and doubly ionised ions all play a role for 5--20 days KN spectral formation. 

We show that KN ejecta are to a large extent still opaque in the optical/NIR due to line blocking, even up to 20d. Much of the spectral shape and features of our models are determined by the location of optically thin, or reduced optical depth windows in this line blocking. We typically find that wavelengths of $\lambda \lesssim 7000~\ang$ are completely line-blocked, whereas at longer wavelengths the blocking is partial. The emergent peaks tend to arise in windows of reduced optical depth rather than at the location of e.g. important cooling lines.

We find that for a lanthanide-free composition ($Y_e \sim 0.35$), the neutral and single ionised species of group I--IV elements of the first r-process peak, Rb, Sr, Y and Zr, dominate the spectral formation. These elements have few valence electrons, with strong transitions between low lying states giving contributions at distinct wavelengths. This makes these species promising candidates for identification in current and future KN observations. Zr\,\textsc{i}, which has many closely packed multiplets of the same parity at low energies, also has many (semi)-forbidden transitions at red wavelengths alongside the allowed transitions. This leads to Zr\,\textsc{i} dominating the NIR regime of the lanthanide-free model; our models indicate that at this $Y_e$ almost the entire NIR spectrum is formed by Zr.

We test in greater detail the effect of Sr on the emergent spectrum by running the $Y_e \sim 0.35$ model with Sr omitted. We find that the presence or absence of Sr significantly impacts the 8000--12000~$\ang$ spectral region, through the Sr\,\textsc{ii} $\lambda_0 = 10039, 100330, 10918~\ang$ triplet. Secondary spectral effects also arise from the impact of Sr on the thermodynamic state of the ejecta, which become somewhat hotter and more ionised. Notably, we find that the Sr\,\textsc{ii} doublet at $\sim 4000~\ang$ is a particularly efficient cooling transition, and its removal leads to higher temperatures. However, we also find that inferring the presence of Sr directly from the spectral shape around $10000~\ang$ may in general not be straightforward, as we find many optically thick lines from other species at similar wavelengths, and quite complex spectral formation at those wavelengths. Our model gives a peak around 1.1 $\mu$m also without any Sr (or He).

We establish that Rb\,\textsc{i} has an important and active doublet transition at experimentally measured rest wavelength $\lambda_0 = 7802, 7950~\ang$ ($\sim 8900$ \AA\  in our model atom). In our lanthanide-free model ($Y_e \sim 0.35$), we see strong scattering in this transition as it remains optically thick throughout the entire ejecta even to 20 days. This transition may be an alternative to the proposed Y\,\textsc{ii} transitions \citep{Sneppen.Watson:23Y} for the $\sim 7600~\ang$ P-Cygni like feature in the spectrum of AT2017gfo. 

In the lanthanide-bearing $Y_e \sim 0.25$ and $0.15$ models, we find that the lanthanide species dominate the spectral formation. We identify several specific lanthanides -- neodymium (Nd), samarium (Sm) and dysprosium (Dy) -- which appear to play particularly important roles, and are thus promising candidates for diagnosis in 5--20 day observations of KNe characterised by low $Y_e$. Our current model atom wavelengths are unfortunately not accurate enough that we can robustly predict specific features from these, but by identifying that these particular elements are highly active in the spectral formation, we provide impetus for further efforts in better determining their atomic properties. The two actinides we consider, Th and U, make up only 0.003 of the composition by mass fraction even at $Y_e \sim 0.15$ (see Table \ref{tab:compositions}) and have little impact on the spectra in our models; detection of these may plausibly occur only for yet more neutron-rich KNe ($Y_e < 0.15$). In the $Y_e \sim 0.25$ model, we also see the continued presence of the Sr\,\textsc{ii} $10000~\ang$ triplet, providing support that such a feature may also be observable in lanthanide-bearing ejecta.

We do not see (in our $Y_e \sim 0.15$ model) the emergence of the [Te\,\textsc{iii}] $2.1~\mu$m emission line proposed as the explanation for the feature seen in the spectrum of AT2017gfo at 10 days \citep{Hotokezaka2023}. We find two reasons for this; the first is that our prescription used to calculate collision strengths for forbidden lines \citep{Axelrod:80} in this case gives a much lower value than the dedicated calculation by \citet{Madonna2018}, giving significantly lower emissivity. Second, we do not obtain optically thin conditions at 10 days around 2 microns, but instead lines from other species absorb and reprocess much of the emission from [Te\,\textsc{iii}]. The striking difference between the collision strength values, and their key role in accurate spectral modelling of KN is well illustrated by this example, which highlights the need for further high quality atomic data, in particular pertaining to collision strengths and recombination rates. As the line blocking at low $Y_e$ is dominated by lanthanides, it is also important to develop accurate model atoms for these in terms of energy levels and transitions probabilities. 

Recent works on electron impact excitation have found that the \citet{Axelrod:80} treatment may systematically underestimate collision strengths, while the accuracy of the Van Regemorter \citep{Regemorter:62} approximation varies depending on ion and electron temperature \citep[][in the context of Pt]{Bromley.etal:23}. While this may lead to underestimated forbidden line emissivities in our models, as for the [Te III] line above, we believe that these lines do not play a dominant role in determining the overall temperature solution of the models and epochs studied here. In our models, we find that allowed transitions dominate both as emission and cooling channels, and thus the accuracy of our temperature solutions will depend mainly on the Van Regemorter approximation, and hence on our line wavelengths and transition strengths. When measured values are available, we find that our calculated A-values are within an order of magnitude of these. The accuracy of the transition wavelengths is more variable, but we believe our theoretical values to be broadly accurate enough to yield reasonable temperature results. It is likely that forbidden lines play larger roles at later times when the ejecta become more optically thin, and such lines are expected to emit strongly \citep[see e.g.][]{Hotokezaka.etal:22,Hotokezaka2023}.

The increasing temperature, and the diminishing line blocking, leads to an overall 'red-to- blue' SED time evolution in the models during the 5--20d period. However, the SED changes are quite mild, and the regular photometric colors stay relatively constant over this time period. We find that the AT2017gfo color curves are similarly quite flat between 5--10d (the last epoch at which colors are available), with only the NIR colours showing a noticeable reddening trend. Considering the broader spectral evolution by comparing the model fluxes in the optical to the NIR, we find this ratio to be monotonically increasing by a factor of $\sim$ 2--3 in the 5--20d range depending on the model. The observations of AT2017gfo show a flat evolution between 5--10d, and possibly a hint of decreasing trend from the last 10.4d observation. Discrepancy in this trend may imply that single composition models (as used here) are not suitable of AT2017gfo, as many light curve analyses have previously indicated \citep[e.g.][]{Perego.etal:17,Rosswog.etal:18,Waxman.etal:18,Tanaka.etal:18}. However, verification of the accuracy of stationary radiation field models, as used here, at 5--10 days still needs to be firmly demonstrated, as diffusion may not have fully ended yet at these epochs.

We establish that time-dependent recombination and adiabatic cooling effects on our models are relatively minor, both with respect to the thermodynamic state, as well as the emergent spectra. The exception is the $Y_e \sim 0.35$ model, which obtains lower temperatures and a less ionised gas from 10 days onwards, with impact on the emergent spectra arising from 15 days onwards. The combination of low power and low density conditions previously identified in \citet{Pognan.etal:22a} as maximising time-dependent effects is thus confirmed here in a spectral context. Therefore, accurate modelling of an early, fast moving, lanthanide-free ejecta component likely requires the inclusion of time-dependent effects in the NLTE calculations. We find that a few key spectral features may be significantly affected in the case of strong time-dependent effects, which may lead to incorrect deductions on the abundance of certain prominent species.

By combining several arguments and model properties, we can conclude that KN ejecta are not optically thin at least up to 20 days after merger, and that resonance scattering and fluorescence play key roles in KN spectral formation throughout the first weeks. The temperatures, which increase with time, and the high degree of line blocking in the blue, signify that much of the cooling emission cannot directly escape. Instead, the cooling emission experiences resonance scattering and fluorescence. One consequence of this is that the KN SED bears little relation to the gas temperature, implying that blackbody fits to the emergent spectrum have little physical meaning. These results also suggest that KNe evolve qualitatively differently to SNe, which have a clear transition to a thermal emission dominated nebular phase, followed by a much later fluorescence dominated phase. It is not yet completely clear how KNe evolve, whether they move directly to this fluorescence phase, or reach the thermal nebular phase later on. Further theoretical studies combined with late time observations of nebular phase KNe are still required in order to fully elucidate the evolution of these transients.

\section*{Acknowledgements}

We acknowledge funding from the European Research Council (ERC) under the European Union's Horizon 2020 Research and Innovation Program (ERC Starting Grant 803189  -- SUPERSPEC, PI Jerkstrand). JG thanks the Swedish Research Council for the individual starting grant with contract No. 2020-05467. The computations were enabled by resources provided by the Swedish National Infrastructure for Computing (SNIC), at the PDC Center for High Performance Computing, KTH Royal Institute of Technology, partially funded by the Swedish Research Council through grant agreement no. 2018-05973. We acknowledge also computing funding by the Swedish National Research Council (grant 2018-03799, PI Jerkstrand).

\section*{Data Availability}

The data underlying this article will be shared on reasonable request to the corresponding author.



\bibliographystyle{mnras}
\bibliography{paper} 

\begin{thebibliography}{}
\makeatletter
\relax
\def\mn@urlcharsother{\let\do\@makeother \do\$\do\&\do\#\do\^\do\_\do\%\do\~}
\def\mn@doi{\begingroup\mn@urlcharsother \@ifnextchar [ {\mn@doi@}
  {\mn@doi@[]}}
\def\mn@doi@[#1]#2{\def\@tempa{#1}\ifx\@tempa\@empty \href
  {http://dx.doi.org/#2} {doi:#2}\else \href {http://dx.doi.org/#2} {#1}\fi
  \endgroup}
\def\mn@eprint#1#2{\mn@eprint@#1:#2::\@nil}
\def\mn@eprint@arXiv#1{\href {http://arxiv.org/abs/#1} {{\tt arXiv:#1}}}
\def\mn@eprint@dblp#1{\href {http://dblp.uni-trier.de/rec/bibtex/#1.xml}
  {dblp:#1}}
\def\mn@eprint@#1:#2:#3:#4\@nil{\def\@tempa {#1}\def\@tempb {#2}\def\@tempc
  {#3}\ifx \@tempc \@empty \let \@tempc \@tempb \let \@tempb \@tempa \fi \ifx
  \@tempb \@empty \def\@tempb {arXiv}\fi \@ifundefined
  {mn@eprint@\@tempb}{\@tempb:\@tempc}{\expandafter \expandafter \csname
  mn@eprint@\@tempb\endcsname \expandafter{\@tempc}}}

\bibitem[\protect\citeauthoryear{{Abbott}, {Abbott}, {Abbott}, {Acernese},
  {Ackley}, {Adams}, {Adams}  \& {et al.}}{{Abbott}
  et~al.}{2017}]{Abbott.etal:17}
{Abbott} B.~P.,  {Abbott} R.,  {Abbott} T.~D.,  {Acernese} F.,  {Ackley} K.,
  {Adams} C.,  {Adams} T.,   {et al.} 2017, \mn@doi [\apjl]
  {10.3847/2041-8213/aa91c9}, \href
  {https://ui.adsabs.harvard.edu/abs/2017ApJ...848L..12A} {848, L12}

\bibitem[\protect\citeauthoryear{{Arcones} \& {Thielemann}}{{Arcones} \&
  {Thielemann}}{2023}]{Arcones.Thielemann:2023}
{Arcones} A.,  {Thielemann} F.-K.,  2023, \mn@doi [\aapr]
  {10.1007/s00159-022-00146-x}, \href
  {https://ui.adsabs.harvard.edu/abs/2023A&ARv..31....1A} {31, 1}

\bibitem[\protect\citeauthoryear{{Axelrod}}{{Axelrod}}{1980}]{Axelrod:80}
{Axelrod} T.~S.,  1980, PhD thesis, California Univ., Santa Cruz.

\bibitem[\protect\citeauthoryear{{Banerjee}, {Tanaka}, {Kawaguchi}, {Kato}  \&
  {Gaigalas}}{{Banerjee} et~al.}{2020}]{Banerjee.etal:20}
{Banerjee} S.,  {Tanaka} M.,  {Kawaguchi} K.,  {Kato} D.,   {Gaigalas} G.,
  2020, \mn@doi [\apj] {10.3847/1538-4357/abae61}, \href
  {https://ui.adsabs.harvard.edu/abs/2020ApJ...901...29B} {901, 29}

\bibitem[\protect\citeauthoryear{{Banerjee}, {Tanaka}, {Kato}, {Gaigalas},
  {Kawaguchi}  \& {Domoto}}{{Banerjee} et~al.}{2022}]{Banerjee2022}
{Banerjee} S.,  {Tanaka} M.,  {Kato} D.,  {Gaigalas} G.,  {Kawaguchi} K.,
  {Domoto} N.,  2022, \mn@doi [\apj] {10.3847/1538-4357/ac7565}, \href
  {https://ui.adsabs.harvard.edu/abs/2022ApJ...934..117B} {934, 117}

\bibitem[\protect\citeauthoryear{{Barnes}, {Kasen}, {Wu}  \&
  {Mart{\'\i}nez-Pinedo}}{{Barnes} et~al.}{2016}]{Barnes.etal:16}
{Barnes} J.,  {Kasen} D.,  {Wu} M.-R.,   {Mart{\'\i}nez-Pinedo} G.,  2016,
  \mn@doi [\apj] {10.3847/0004-637X/829/2/110}, \href
  {https://ui.adsabs.harvard.edu/abs/2016ApJ...829..110B} {829, 110}

\bibitem[\protect\citeauthoryear{{Barnes}, {Zhu}, {Lund}, {Sprouse}, {Vassh},
  {McLaughlin}, {Mumpower}  \& {Surman}}{{Barnes} et~al.}{2021}]{Barnes2021}
{Barnes} J.,  {Zhu} Y.~L.,  {Lund} K.~A.,  {Sprouse} T.~M.,  {Vassh} N.,
  {McLaughlin} G.~C.,  {Mumpower} M.~R.,   {Surman} R.,  2021, \mn@doi [\apj]
  {10.3847/1538-4357/ac0aec}, \href
  {https://ui.adsabs.harvard.edu/abs/2021ApJ...918...44B} {918, 44}

\bibitem[\protect\citeauthoryear{Biémont, Palmeri  \& Quinet}{Biémont
  et~al.}{1999}]{DREAM_1999}
Biémont E.,  Palmeri P.,   Quinet P.,  1999, \mn@doi [Ap&SS]
  {10.1023/A:1017049314691}, 269, 635

\bibitem[\protect\citeauthoryear{{Blondin} et~al.,}{{Blondin}
  et~al.}{2022}]{Blondin2022}
{Blondin} S.,  et~al., 2022, \mn@doi [\aap] {10.1051/0004-6361/202244134},
  \href {https://ui.adsabs.harvard.edu/abs/2022A&A...668A.163B} {668, A163}

\bibitem[\protect\citeauthoryear{{Boty{\'a}nszki} \& {Kasen}}{{Boty{\'a}nszki}
  \& {Kasen}}{2017}]{Botyanski.Kasen:17}
{Boty{\'a}nszki} J.,  {Kasen} D.,  2017, \mn@doi [\apj]
  {10.3847/1538-4357/aa81d8}, \href
  {https://ui.adsabs.harvard.edu/abs/2017ApJ...845..176B} {845, 176}

\bibitem[\protect\citeauthoryear{{Bromley} et~al.,}{{Bromley}
  et~al.}{2020}]{Bromley2020}
{Bromley} S.~J.,  et~al., 2020, \mn@doi [\apjs] {10.3847/1538-4365/abaa4d},
  \href {https://ui.adsabs.harvard.edu/abs/2020ApJS..250...19B} {250, 19}

\bibitem[\protect\citeauthoryear{{Bromley}, {McCann}, {Loch}  \&
  {Ballance}}{{Bromley} et~al.}{2023}]{Bromley.etal:23}
{Bromley} S.~J.,  {McCann} M.,  {Loch} S.~D.,   {Ballance} C.~P.,  2023,
  \mn@doi [\apjs] {10.3847/1538-4365/ace5a1}, \href
  {https://ui.adsabs.harvard.edu/abs/2023ApJS..268...22B} {268, 22}

\bibitem[\protect\citeauthoryear{{Bulla}}{{Bulla}}{2019}]{Bulla:19}
{Bulla} M.,  2019, \mn@doi [\mnras] {10.1093/mnras/stz2495}, \href
  {https://ui.adsabs.harvard.edu/abs/2019MNRAS.489.5037B} {489, 5037}

\bibitem[\protect\citeauthoryear{{Bulla}}{{Bulla}}{2023}]{Bulla:2023}
{Bulla} M.,  2023, \mn@doi [\mnras] {10.1093/mnras/stad232}, \href
  {https://ui.adsabs.harvard.edu/abs/2023MNRAS.520.2558B} {520, 2558}

\bibitem[\protect\citeauthoryear{{Carvajal Gallego}, {Palmeri}  \&
  {Quinet}}{{Carvajal Gallego} et~al.}{2021}]{Carvajal2021}
{Carvajal Gallego} H.,  {Palmeri} P.,   {Quinet} P.,  2021, \mn@doi [\mnras]
  {10.1093/mnras/staa3729}, \href
  {https://ui.adsabs.harvard.edu/abs/2021MNRAS.501.1440C} {501, 1440}

\bibitem[\protect\citeauthoryear{{Carvajal Gallego}, {Deprince}, {Palmeri}  \&
  {Quinet}}{{Carvajal Gallego} et~al.}{2023}]{CarjavalGallego2023}
{Carvajal Gallego} H.,  {Deprince} J.,  {Palmeri} P.,   {Quinet} P.,  2023,
  \mn@doi [\mnras] {10.1093/mnras/stad990}, \href
  {https://ui.adsabs.harvard.edu/abs/2023MNRAS.522..312C} {522, 312}

\bibitem[\protect\citeauthoryear{{Collins}, {Bauswein}, {Sim}, {Vijayan},
  {Mart{\'\i}nez-Pinedo}, {Just}, {Shingles}  \& {Kromer}}{{Collins}
  et~al.}{2023}]{Collins2023}
{Collins} C.~E.,  {Bauswein} A.,  {Sim} S.~A.,  {Vijayan} V.,
  {Mart{\'\i}nez-Pinedo} G.,  {Just} O.,  {Shingles} L.~J.,   {Kromer} M.,
  2023, \mn@doi [\mnras] {10.1093/mnras/stad606}, \href
  {https://ui.adsabs.harvard.edu/abs/2023MNRAS.521.1858C} {521, 1858}

\bibitem[\protect\citeauthoryear{{C{\^o}t{\'e}} et~al.,}{{C{\^o}t{\'e}}
  et~al.}{2019}]{Cote.etal:19}
{C{\^o}t{\'e}} B.,  et~al., 2019, \mn@doi [\apj] {10.3847/1538-4357/ab10db},
  \href {https://ui.adsabs.harvard.edu/abs/2019ApJ...875..106C} {875, 106}

\bibitem[\protect\citeauthoryear{{Deprince}, {Carvajal Gallego}, {Godefroid},
  {Goriely}, {Palmeri}  \& {Quinet}}{{Deprince} et~al.}{2023}]{Deprince2023}
{Deprince} J.,  {Carvajal Gallego} H.,  {Godefroid} M.,  {Goriely} S.,
  {Palmeri} P.,   {Quinet} P.,  2023, \mn@doi [European Phys. J. D]
  {10.1140/epjd/s10053-023-00671-z}, \href
  {https://ui.adsabs.harvard.edu/abs/2023EPJD...77...93D} {77, 93}

\bibitem[\protect\citeauthoryear{{Domoto}, {Tanaka}, {Wanajo}  \&
  {Kawaguchi}}{{Domoto} et~al.}{2021}]{Domoto.etal:21}
{Domoto} N.,  {Tanaka} M.,  {Wanajo} S.,   {Kawaguchi} K.,  2021, \mn@doi
  [\apj] {10.3847/1538-4357/abf358}, \href
  {https://ui.adsabs.harvard.edu/abs/2021ApJ...913...26D} {913, 26}

\bibitem[\protect\citeauthoryear{{Domoto}, {Tanaka}, {Kato}, {Kawaguchi},
  {Hotokezaka}  \& {Wanajo}}{{Domoto} et~al.}{2022}]{Domoto.etal:22}
{Domoto} N.,  {Tanaka} M.,  {Kato} D.,  {Kawaguchi} K.,  {Hotokezaka} K.,
  {Wanajo} S.,  2022, \mn@doi [\apj] {10.3847/1538-4357/ac8c36}, \href
  {https://ui.adsabs.harvard.edu/abs/2022ApJ...939....8D} {939, 8}

\bibitem[\protect\citeauthoryear{{Eichler}, {Livio}, {Piran}  \&
  {Schramm}}{{Eichler} et~al.}{1989}]{Eichler.etal:89}
{Eichler} D.,  {Livio} M.,  {Piran} T.,   {Schramm} D.~N.,  1989, \mn@doi
  [\nat] {10.1038/340126a0}, \href
  {https://ui.adsabs.harvard.edu/abs/1989Natur.340..126E} {340, 126}

\bibitem[\protect\citeauthoryear{{Finstad}, {De}, {Brown}, {Berger}  \&
  {Biwer}}{{Finstad} et~al.}{2018}]{Finstad.etal:18}
{Finstad} D.,  {De} S.,  {Brown} D.~A.,  {Berger} E.,   {Biwer} C.~M.,  2018,
  \mn@doi [\apjl] {10.3847/2041-8213/aac6c1}, \href
  {https://ui.adsabs.harvard.edu/abs/2018ApJ...860L...2F} {860, L2}

\bibitem[\protect\citeauthoryear{{Fl{\"o}rs} et~al.,}{{Fl{\"o}rs}
  et~al.}{2023}]{Flors.etal:23}
{Fl{\"o}rs} A.,  et~al., 2023, \mn@doi [\mnras] {10.1093/mnras/stad2053}, \href
  {https://ui.adsabs.harvard.edu/abs/2023MNRAS.tmp.2015F} {}

\bibitem[\protect\citeauthoryear{{Fontes}, {Fryer}, {Hungerford}, {Wollaeger}
  \& {Korobkin}}{{Fontes} et~al.}{2020}]{Fontes.etal:20}
{Fontes} C.~J.,  {Fryer} C.~L.,  {Hungerford} A.~L.,  {Wollaeger} R.~T.,
  {Korobkin} O.,  2020, \mn@doi [\mnras] {10.1093/mnras/staa485}, \href
  {https://ui.adsabs.harvard.edu/abs/2020MNRAS.493.4143F} {493, 4143}

\bibitem[\protect\citeauthoryear{{Fontes}, {Fryer}, {Wollaeger}, {Mumpower}  \&
  {Sprouse}}{{Fontes} et~al.}{2023}]{Fontes2023}
{Fontes} C.~J.,  {Fryer} C.~L.,  {Wollaeger} R.~T.,  {Mumpower} M.~R.,
  {Sprouse} T.~M.,  2023, \mn@doi [\mnras] {10.1093/mnras/stac2792}, \href
  {https://ui.adsabs.harvard.edu/abs/2023MNRAS.519.2862F} {519, 2862}

\bibitem[\protect\citeauthoryear{{Freiburghaus}, {Rosswog}  \&
  {Thielemann}}{{Freiburghaus} et~al.}{1999}]{Freiburghaus.etal:99}
{Freiburghaus} C.,  {Rosswog} S.,   {Thielemann} F.~K.,  1999, \mn@doi [\apjl]
  {10.1086/312343}, \href
  {https://ui.adsabs.harvard.edu/abs/1999ApJ...525L.121F} {525, L121}

\bibitem[\protect\citeauthoryear{{Gaigalas}, {Kato}, {Rynkun},
  {Rad{\v{z}}i{\={u}}t{\.{e}}}  \& {Tanaka}}{{Gaigalas}
  et~al.}{2019}]{Gaigalas.etal:19}
{Gaigalas} G.,  {Kato} D.,  {Rynkun} P.,  {Rad{\v{z}}i{\={u}}t{\.{e}}} L.,
  {Tanaka} M.,  2019, \mn@doi [\apjs] {10.3847/1538-4365/aaf9b8}, \href
  {https://ui.adsabs.harvard.edu/abs/2019ApJS..240...29G} {240, 29}

\bibitem[\protect\citeauthoryear{{Gaigalas}, {Rynkun},
  {Rad{\v{z}}i{\={u}}t{\.{e}}}, {Kato}, {Tanaka}  \& {J{\"o}nsson}}{{Gaigalas}
  et~al.}{2020}]{Gaigalas2020}
{Gaigalas} G.,  {Rynkun} P.,  {Rad{\v{z}}i{\={u}}t{\.{e}}} L.,  {Kato} D.,
  {Tanaka} M.,   {J{\"o}nsson} P.,  2020, \mn@doi [\apjs]
  {10.3847/1538-4365/ab881a}, \href
  {https://ui.adsabs.harvard.edu/abs/2020ApJS..248...13G} {248, 13}

\bibitem[\protect\citeauthoryear{{Gillanders}, {McCann}, {Sim}, {Smartt}  \&
  {Ballance}}{{Gillanders} et~al.}{2021}]{Gillanders.etal:21}
{Gillanders} J.~H.,  {McCann} M.,  {Sim} S.~A.,  {Smartt} S.~J.,   {Ballance}
  C.~P.,  2021, \mn@doi [\mnras] {10.1093/mnras/stab1861}, \href
  {https://ui.adsabs.harvard.edu/abs/2021MNRAS.506.3560G} {506, 3560}

\bibitem[\protect\citeauthoryear{{Gillanders}, {Smartt}, {Sim}, {Bauswein}  \&
  {Goriely}}{{Gillanders} et~al.}{2022}]{Gillanders.etal:22}
{Gillanders} J.~H.,  {Smartt} S.~J.,  {Sim} S.~A.,  {Bauswein} A.,   {Goriely}
  S.,  2022, \mn@doi [\mnras] {10.1093/mnras/stac1258}, \href
  {https://ui.adsabs.harvard.edu/abs/2022MNRAS.515..631G} {515, 631}

\bibitem[\protect\citeauthoryear{{Gillanders}, {Sim}, {Smartt}, {Goriely}  \&
  {Bauswein}}{{Gillanders} et~al.}{2023}]{Gillanders.etal:23}
{Gillanders} J.~H.,  {Sim} S.~A.,  {Smartt} S.~J.,  {Goriely} S.,   {Bauswein}
  A.,  2023, arXiv e-prints, \href
  {https://ui.adsabs.harvard.edu/abs/2023arXiv230615055G} {p. arXiv:2306.15055}

\bibitem[\protect\citeauthoryear{{Gu}}{{Gu}}{2008}]{Gu:08}
{Gu} M.~F.,  2008, \mn@doi [Canadian Journal of Physics] {10.1139/P07-197},
  \href {https://ui.adsabs.harvard.edu/abs/2008CaJPh..86..675G} {86, 675}

\bibitem[\protect\citeauthoryear{{Hotokezaka} \& {Nakar}}{{Hotokezaka} \&
  {Nakar}}{2020}]{Hotokezaka.etal:20}
{Hotokezaka} K.,  {Nakar} E.,  2020, \mn@doi [\apj] {10.3847/1538-4357/ab6a98},
  \href {https://ui.adsabs.harvard.edu/abs/2020ApJ...891..152H} {891, 152}

\bibitem[\protect\citeauthoryear{{Hotokezaka}, {Tanaka}, {Kato}  \&
  {Gaigalas}}{{Hotokezaka} et~al.}{2021}]{Hotokezaka.etal:21}
{Hotokezaka} K.,  {Tanaka} M.,  {Kato} D.,   {Gaigalas} G.,  2021, \mn@doi
  [\mnras] {10.1093/mnras/stab1975}, \href
  {https://ui.adsabs.harvard.edu/abs/2021MNRAS.506.5863H} {506, 5863}

\bibitem[\protect\citeauthoryear{{Hotokezaka}, {Tanaka}, {Kato}  \&
  {Gaigalas}}{{Hotokezaka} et~al.}{2022}]{Hotokezaka.etal:22}
{Hotokezaka} K.,  {Tanaka} M.,  {Kato} D.,   {Gaigalas} G.,  2022, \mn@doi
  [\mnras] {10.1093/mnrasl/slac071}, \href
  {https://ui.adsabs.harvard.edu/abs/2022MNRAS.515L..89H} {515, L89}

\bibitem[\protect\citeauthoryear{{Hotokezaka}, {Tanaka}, {Kato}  \&
  {Gaigalas}}{{Hotokezaka} et~al.}{2023}]{Hotokezaka2023}
{Hotokezaka} K.,  {Tanaka} M.,  {Kato} D.,   {Gaigalas} G.,  2023, \mn@doi
  [\mnras] {10.1093/mnrasl/slad128}, \href
  {https://ui.adsabs.harvard.edu/abs/2023MNRAS.526L.155H} {526, L155}

\bibitem[\protect\citeauthoryear{{Jerkstrand}}{{Jerkstrand}}{2011}]{Jerkstrand:11}
{Jerkstrand} A.,  2011, PhD thesis, Department of Astronomy, Stockholm
  University

\bibitem[\protect\citeauthoryear{{Jerkstrand}}{{Jerkstrand}}{2017}]{Jerkstrand:17}
{Jerkstrand} A.,  2017, {Spectra of Supernovae in the Nebular Phase}.
Springer, Cham, p.~795, \mn@doi{10.1007/978-3-319-21846-5\_29}

\bibitem[\protect\citeauthoryear{{Jerkstrand}, {Fransson}  \&
  {Kozma}}{{Jerkstrand} et~al.}{2011}]{Jerkstrand.etal:11}
{Jerkstrand} A.,  {Fransson} C.,   {Kozma} C.,  2011, \mn@doi [\aap]
  {10.1051/0004-6361/201015937}, \href
  {https://ui.adsabs.harvard.edu/abs/2011A&A...530A..45J} {530, A45}

\bibitem[\protect\citeauthoryear{{Jerkstrand}, {Fransson}, {Maguire}, {Smartt},
  {Ergon}  \& {Spyromilio}}{{Jerkstrand} et~al.}{2012}]{Jerkstrand.etal:12}
{Jerkstrand} A.,  {Fransson} C.,  {Maguire} K.,  {Smartt} S.,  {Ergon} M.,
  {Spyromilio} J.,  2012, \mn@doi [\aap] {10.1051/0004-6361/201219528}, \href
  {https://ui.adsabs.harvard.edu/abs/2012A&A...546A..28J} {546, A28}

\bibitem[\protect\citeauthoryear{{Jerkstrand}, {Smartt}  \&
  {Heger}}{{Jerkstrand} et~al.}{2016}]{Jerkstrand.etal:16}
{Jerkstrand} A.,  {Smartt} S.~J.,   {Heger} A.,  2016, \mn@doi [\mnras]
  {10.1093/mnras/stv2369}, \href
  {https://ui.adsabs.harvard.edu/abs/2016MNRAS.455.3207J} {455, 3207}

\bibitem[\protect\citeauthoryear{{Just}, {Kullmann}, {Goriely}, {Bauswein},
  {Janka}  \& {Collins}}{{Just} et~al.}{2022}]{Just2022}
{Just} O.,  {Kullmann} I.,  {Goriely} S.,  {Bauswein} A.,  {Janka} H.~T.,
  {Collins} C.~E.,  2022, \mn@doi [\mnras] {10.1093/mnras/stab3327}, \href
  {https://ui.adsabs.harvard.edu/abs/2022MNRAS.510.2820J} {510, 2820}

\bibitem[\protect\citeauthoryear{{Kasen} \& {Barnes}}{{Kasen} \&
  {Barnes}}{2019}]{Kasen.Barnes:19}
{Kasen} D.,  {Barnes} J.,  2019, \mn@doi [\apj] {10.3847/1538-4357/ab06c2},
  \href {https://ui.adsabs.harvard.edu/abs/2019ApJ...876..128K} {876, 128}

\bibitem[\protect\citeauthoryear{{Kasen}, {Thomas}  \& {Nugent}}{{Kasen}
  et~al.}{2006}]{Kasen.etal:06}
{Kasen} D.,  {Thomas} R.~C.,   {Nugent} P.,  2006, \mn@doi [\apj]
  {10.1086/506190}, \href
  {https://ui.adsabs.harvard.edu/abs/2006ApJ...651..366K} {651, 366}

\bibitem[\protect\citeauthoryear{{Kasen}, {Badnell}  \& {Barnes}}{{Kasen}
  et~al.}{2013}]{Kasen.etal:13}
{Kasen} D.,  {Badnell} N.~R.,   {Barnes} J.,  2013, \mn@doi [\apj]
  {10.1088/0004-637X/774/1/25}, \href
  {https://ui.adsabs.harvard.edu/abs/2013ApJ...774...25K} {774, 25}

\bibitem[\protect\citeauthoryear{{Kasen}, {Metzger}, {Barnes}, {Quataert}  \&
  {Ramirez-Ruiz}}{{Kasen} et~al.}{2017}]{Kasen.etal:17}
{Kasen} D.,  {Metzger} B.,  {Barnes} J.,  {Quataert} E.,   {Ramirez-Ruiz} E.,
  2017, \mn@doi [\nat] {10.1038/nature24453}, \href
  {https://ui.adsabs.harvard.edu/abs/2017Natur.551...80K} {551, 80}

\bibitem[\protect\citeauthoryear{{Kawaguchi}, {Fujibayashi}, {Shibata},
  {Tanaka}  \& {Wanajo}}{{Kawaguchi} et~al.}{2021}]{Kawaguchi.etal:21}
{Kawaguchi} K.,  {Fujibayashi} S.,  {Shibata} M.,  {Tanaka} M.,   {Wanajo} S.,
  2021, \mn@doi [\apj] {10.3847/1538-4357/abf3bc}, \href
  {https://ui.adsabs.harvard.edu/abs/2021ApJ...913..100K} {913, 100}

\bibitem[\protect\citeauthoryear{{Kawaguchi}, {Fujibayashi}, {Domoto},
  {Kiuchi}, {Shibata}  \& {Wanajo}}{{Kawaguchi}
  et~al.}{2023}]{Kawaguchi.etal:23}
{Kawaguchi} K.,  {Fujibayashi} S.,  {Domoto} N.,  {Kiuchi} K.,  {Shibata} M.,
  {Wanajo} S.,  2023, \mn@doi [\mnras] {10.1093/mnras/stad2430}, \href
  {https://ui.adsabs.harvard.edu/abs/2023MNRAS.525.3384K} {525, 3384}

\bibitem[\protect\citeauthoryear{{Kozma} \& {Fransson}}{{Kozma} \&
  {Fransson}}{1992}]{Kozma.Fransson:92}
{Kozma} C.,  {Fransson} C.,  1992, \mn@doi [\apj] {10.1086/171311}, \href
  {https://ui.adsabs.harvard.edu/abs/1992ApJ...390..602K} {390, 602}

\bibitem[\protect\citeauthoryear{Kramida, {Yu.~Ralchenko}, Reader  \& {and NIST
  ASD Team}}{Kramida et~al.}{2020}]{NIST_ASD}
Kramida A.,  {Yu.~Ralchenko} Reader J.,   {and NIST ASD Team} 2020, {NIST
  Atomic Spectra Database (ver. 5.8), [Online]. Available:
  {\tt{https://physics.nist.gov/asd}} [2021, September 22]. National Institute
  of Standards and Technology, Gaithersburg, MD.}

\bibitem[\protect\citeauthoryear{{Lawler}, {Schmidt}  \& {Hartog}}{{Lawler}
  et~al.}{2022}]{Lawler.etal:22}
{Lawler} J.~E.,  {Schmidt} J.~R.,   {Hartog} E.~A.~D.,  2022, \mn@doi [\jqsrt]
  {10.1016/j.jqsrt.2022.108283}, \href
  {https://ui.adsabs.harvard.edu/abs/2022JQSRT.28908283L} {289, 108283}

\bibitem[\protect\citeauthoryear{{Li} \& {Paczy{\'n}ski}}{{Li} \&
  {Paczy{\'n}ski}}{1998}]{Li.Paczynski:98}
{Li} L.-X.,  {Paczy{\'n}ski} B.,  1998, \mn@doi [\apjl] {10.1086/311680}, \href
  {https://ui.adsabs.harvard.edu/abs/1998ApJ...507L..59L} {507, L59}

\bibitem[\protect\citeauthoryear{{Lodders}}{{Lodders}}{2021}]{Lodders:21}
{Lodders} K.,  2021, \mn@doi [\ssr] {10.1007/s11214-021-00825-8}, \href
  {https://ui.adsabs.harvard.edu/abs/2021SSRv..217...44L} {217, 44}

\bibitem[\protect\citeauthoryear{{Lodders}, {Palme}  \& {Gail}}{{Lodders}
  et~al.}{2009}]{Lodders.etal:09}
{Lodders} K.,  {Palme} H.,   {Gail} H.~P.,  2009, \mn@doi [Landolt
  B\"{o}rnstein] {10.1007/978-3-540-88055-4\_34}, \href
  {https://ui.adsabs.harvard.edu/abs/2009LanB...4B..712L} {4B, 712}

\bibitem[\protect\citeauthoryear{{Longair}}{{Longair}}{2011}]{Longair2011}
{Longair} M.~S.,  2011, {High Energy Astrophysics}.
{Cambridge University Press}

\bibitem[\protect\citeauthoryear{{Lotz}}{{Lotz}}{1967}]{Lotz:67}
{Lotz} W.,  1967, \mn@doi [Zeitschrift fur Physik] {10.1007/BF01325928}, \href
  {https://ui.adsabs.harvard.edu/abs/1967ZPhy..206..205L} {206, 205}

\bibitem[\protect\citeauthoryear{{Madonna} et~al.,}{{Madonna}
  et~al.}{2018}]{Madonna2018}
{Madonna} S.,  et~al., 2018, \mn@doi [\apjl] {10.3847/2041-8213/aaccef}, \href
  {https://ui.adsabs.harvard.edu/abs/2018ApJ...861L...8M} {861, L8}

\bibitem[\protect\citeauthoryear{{Margutti} \& {Chornock}}{{Margutti} \&
  {Chornock}}{2021}]{Margutti.Chornock:21}
{Margutti} R.,  {Chornock} R.,  2021, \mn@doi [\araa]
  {10.1146/annurev-astro-112420-030742}, \href
  {https://ui.adsabs.harvard.edu/abs/2021ARA&A..59..155M} {59, 155}

\bibitem[\protect\citeauthoryear{{Mart{\'\i}nez-Pinedo}
  et~al.,}{{Mart{\'\i}nez-Pinedo} et~al.}{2007}]{Martinez.etal:07}
{Mart{\'\i}nez-Pinedo} G.,  et~al., 2007, \mn@doi [Progress in Particle and
  Nuclear Physics] {10.1016/j.ppnp.2007.01.018}, \href
  {https://ui.adsabs.harvard.edu/abs/2007PrPNP..59..199M} {59, 199}

\bibitem[\protect\citeauthoryear{{McCann}, {Bromley}, {Loch}  \&
  {Ballance}}{{McCann} et~al.}{2022}]{McCann2022}
{McCann} M.,  {Bromley} S.,  {Loch} S.~D.,   {Ballance} C.~P.,  2022, \mn@doi
  [\mnras] {10.1093/mnras/stab3285}, \href
  {https://ui.adsabs.harvard.edu/abs/2022MNRAS.509.4723M} {509, 4723}

\bibitem[\protect\citeauthoryear{{Metzger}}{{Metzger}}{2019}]{Metzger:19}
{Metzger} B.~D.,  2019, \mn@doi [Living Reviews in Relativity]
  {10.1007/s41114-019-0024-0}, \href
  {https://ui.adsabs.harvard.edu/abs/2019LRR....23....1M} {23, 1}

\bibitem[\protect\citeauthoryear{{Metzger} et~al.,}{{Metzger}
  et~al.}{2010}]{Metzger.etal:10}
{Metzger} B.~D.,  et~al., 2010, \mn@doi [\mnras]
  {10.1111/j.1365-2966.2010.16864.x}, \href
  {https://ui.adsabs.harvard.edu/abs/2010MNRAS.406.2650M} {406, 2650}

\bibitem[\protect\citeauthoryear{{Moore}}{{Moore}}{1971}]{Moore:71a}
{Moore} C.~E.,  1971, {Atomic Energy Levels as Derived from the Analysis of
  Optical Spectra – Chromium through Niobium}, unknown,
  \mn@doi{10.6028/NBS.NSRDS.35v2}

\bibitem[\protect\citeauthoryear{{Nedora} et~al.,}{{Nedora}
  et~al.}{2021}]{Nedora.etal:21}
{Nedora} V.,  et~al., 2021, \mn@doi [\apj] {10.3847/1538-4357/abc9be}, \href
  {https://ui.adsabs.harvard.edu/abs/2021ApJ...906...98N} {906, 98}

\bibitem[\protect\citeauthoryear{{Neuweiler}, {Dietrich}, {Bulla}, {Chaurasia},
  {Rosswog}  \& {Ujevic}}{{Neuweiler} et~al.}{2023}]{Neuweiler.etal:23}
{Neuweiler} A.,  {Dietrich} T.,  {Bulla} M.,  {Chaurasia} S.~V.,  {Rosswog} S.,
    {Ujevic} M.,  2023, \mn@doi [\prd] {10.1103/PhysRevD.107.023016}, \href
  {https://ui.adsabs.harvard.edu/abs/2023PhRvD.107b3016N} {107, 023016}

\bibitem[\protect\citeauthoryear{{Palmer}}{{Palmer}}{1977}]{Palmer:77}
{Palmer} B.~A.,  1977, PhD thesis, Purdue University, Indiana

\bibitem[\protect\citeauthoryear{{Perego}, {Radice}  \& {Bernuzzi}}{{Perego}
  et~al.}{2017}]{Perego.etal:17}
{Perego} A.,  {Radice} D.,   {Bernuzzi} S.,  2017, \mn@doi [\apjl]
  {10.3847/2041-8213/aa9ab9}, \href
  {https://ui.adsabs.harvard.edu/abs/2017ApJ...850L..37P} {850, L37}

\bibitem[\protect\citeauthoryear{{Perego} et~al.,}{{Perego}
  et~al.}{2022}]{Perego.etal:22}
{Perego} A.,  et~al., 2022, \mn@doi [\apj] {10.3847/1538-4357/ac3751}, \href
  {https://ui.adsabs.harvard.edu/abs/2022ApJ...925...22P} {925, 22}

\bibitem[\protect\citeauthoryear{{Pian} et~al.,}{{Pian}
  et~al.}{2017}]{Pian2017}
{Pian} E.,  et~al., 2017, \mn@doi [\nat] {10.1038/nature24298}, \href
  {https://ui.adsabs.harvard.edu/abs/2017Natur.551...67P} {551, 67}

\bibitem[\protect\citeauthoryear{{Pognan}, {Jerkstrand}  \& {Grumer}}{{Pognan}
  et~al.}{2022a}]{Pognan.etal:22a}
{Pognan} Q.,  {Jerkstrand} A.,   {Grumer} J.,  2022a, \mn@doi [\mnras]
  {10.1093/mnras/stab3674}, \href
  {https://ui.adsabs.harvard.edu/abs/2022MNRAS.510.3806P} {510, 3806}

\bibitem[\protect\citeauthoryear{{Pognan}, {Jerkstrand}  \& {Grumer}}{{Pognan}
  et~al.}{2022b}]{Pognan.etal:22b}
{Pognan} Q.,  {Jerkstrand} A.,   {Grumer} J.,  2022b, \mn@doi [\mnras]
  {10.1093/mnras/stac1253}, \href
  {https://ui.adsabs.harvard.edu/abs/2022MNRAS.513.5174P} {513, 5174}

\bibitem[\protect\citeauthoryear{{Prantzos}, {Abia}, {Cristallo}, {Limongi}  \&
  {Chieffi}}{{Prantzos} et~al.}{2020}]{Prantzos.etal:20}
{Prantzos} N.,  {Abia} C.,  {Cristallo} S.,  {Limongi} M.,   {Chieffi} A.,
  2020, \mn@doi [\mnras] {10.1093/mnras/stz3154}, \href
  {https://ui.adsabs.harvard.edu/abs/2020MNRAS.491.1832P} {491, 1832}

\bibitem[\protect\citeauthoryear{{Rad{\v{z}}i{\={u}}t{\.{e}}}, {Gaigalas},
  {Kato}, {Rynkun}  \& {Tanaka}}{{Rad{\v{z}}i{\={u}}t{\.{e}}}
  et~al.}{2020}]{Radziute.etal:20}
{Rad{\v{z}}i{\={u}}t{\.{e}}} L.,  {Gaigalas} G.,  {Kato} D.,  {Rynkun} P.,
  {Tanaka} M.,  2020, \mn@doi [\apjs] {10.3847/1538-4365/ab8312}, \href
  {https://ui.adsabs.harvard.edu/abs/2020ApJS..248...17R} {248, 17}

\bibitem[\protect\citeauthoryear{{Rosswog}, {Liebend{\"o}rfer}, {Thielemann},
  {Davies}, {Benz}  \& {Piran}}{{Rosswog} et~al.}{1999}]{Rosswog.etal:99}
{Rosswog} S.,  {Liebend{\"o}rfer} M.,  {Thielemann} F.~K.,  {Davies} M.~B.,
  {Benz} W.,   {Piran} T.,  1999, \mn@doi [\aap]
  {10.48550/arXiv.astro-ph/9811367}, \href
  {https://ui.adsabs.harvard.edu/abs/1999A&A...341..499R} {341, 499}

\bibitem[\protect\citeauthoryear{{Rosswog}, {Sollerman}, {Feindt}, {Goobar},
  {Korobkin}, {Wollaeger}, {Fremling}  \& {Kasliwal}}{{Rosswog}
  et~al.}{2018}]{Rosswog.etal:18}
{Rosswog} S.,  {Sollerman} J.,  {Feindt} U.,  {Goobar} A.,  {Korobkin} O.,
  {Wollaeger} R.,  {Fremling} C.,   {Kasliwal} M.~M.,  2018, \mn@doi [\aap]
  {10.1051/0004-6361/201732117}, \href
  {https://ui.adsabs.harvard.edu/abs/2018A&A...615A.132R} {615, A132}

\bibitem[\protect\citeauthoryear{{Rutten}}{{Rutten}}{2003}]{Rutten:03}
{Rutten} R.~J.,  2003, {Radiative Transfer in Stellar Atmospheres}.
{Sterrekindig Instituut Utrecht}

\bibitem[\protect\citeauthoryear{Ryabchikova, Piskunov, Kurucz, Stempels,
  Heiter, Pakhomov  \& Barklem}{Ryabchikova et~al.}{2015}]{VALD3_2015}
Ryabchikova T.,  Piskunov N.,  Kurucz R.~L.,  Stempels H.~C.,  Heiter U.,
  Pakhomov Y.,   Barklem P.~S.,  2015, \mn@doi [Physica Scripta]
  {10.1088/0031-8949/90/5/054005}, 90, 054005

\bibitem[\protect\citeauthoryear{{Rybicki} \& {Lightman}}{{Rybicki} \&
  {Lightman}}{1979}]{Rybicki.Lightman:79}
{Rybicki} G.~B.,  {Lightman} A.~P.,  1979, {Radiative processes in
  astrophysics}.
{Wiley VCH}

\bibitem[\protect\citeauthoryear{{Rynkun}, {Banerjee}, {Gaigalas}, {Tanaka},
  {Rad{\v{z}}i{\={u}}t{\.{e}}}  \& {Kato}}{{Rynkun} et~al.}{2022}]{Rynkun2022}
{Rynkun} P.,  {Banerjee} S.,  {Gaigalas} G.,  {Tanaka} M.,
  {Rad{\v{z}}i{\={u}}t{\.{e}}} L.,   {Kato} D.,  2022, \mn@doi [\aap]
  {10.1051/0004-6361/202141513}, \href
  {https://ui.adsabs.harvard.edu/abs/2022A&A...658A..82R} {658, A82}

\bibitem[\protect\citeauthoryear{{Shen}, {Blondin}, {Kasen}, {Dessart},
  {Townsley}, {Boos}  \& {Hillier}}{{Shen} et~al.}{2021}]{Shen.etal:21}
{Shen} K.~J.,  {Blondin} S.,  {Kasen} D.,  {Dessart} L.,  {Townsley} D.~M.,
  {Boos} S.,   {Hillier} D.~J.,  2021, \mn@doi [\apjl]
  {10.3847/2041-8213/abe69b}, \href
  {https://ui.adsabs.harvard.edu/abs/2021ApJ...909L..18S} {909, L18}

\bibitem[\protect\citeauthoryear{{Shibata} \& {Hotokezaka}}{{Shibata} \&
  {Hotokezaka}}{2019}]{Shibata.Hotokezaka:19}
{Shibata} M.,  {Hotokezaka} K.,  2019, \mn@doi [Annual Review of Nuclear and
  Particle Science] {10.1146/annurev-nucl-101918-023625}, \href
  {https://ui.adsabs.harvard.edu/abs/2019ARNPS..69...41S} {69, 41}

\bibitem[\protect\citeauthoryear{{Shingles} et~al.,}{{Shingles}
  et~al.}{2020}]{Shingles.etal:20}
{Shingles} L.~J.,  et~al., 2020, \mn@doi [\mnras] {10.1093/mnras/stz3412},
  \href {https://ui.adsabs.harvard.edu/abs/2020MNRAS.492.2029S} {492, 2029}

\bibitem[\protect\citeauthoryear{{Shingles} et~al.,}{{Shingles}
  et~al.}{2023}]{Shingles2023}
{Shingles} L.~J.,  et~al., 2023, \mn@doi [\apjl] {10.3847/2041-8213/acf29a},
  \href {https://ui.adsabs.harvard.edu/abs/2023ApJ...954L..41S} {954, L41}

\bibitem[\protect\citeauthoryear{{Shull} \& {van Steenberg}}{{Shull} \& {van
  Steenberg}}{1982}]{Shull.Steenberg:82}
{Shull} J.~M.,  {van Steenberg} M.,  1982, \mn@doi [\apjs] {10.1086/190769},
  \href {https://ui.adsabs.harvard.edu/abs/1982ApJS...48...95S} {48, 95}

\bibitem[\protect\citeauthoryear{{Simsarian}, {Orozco}, {Sprouse}  \&
  {Zhao}}{{Simsarian} et~al.}{1998}]{Simsarian.etal:98}
{Simsarian} J.~E.,  {Orozco} L.~A.,  {Sprouse} G.~D.,   {Zhao} W.~Z.,  1998,
  \mn@doi [\pra] {10.1103/PhysRevA.57.2448}, \href
  {https://ui.adsabs.harvard.edu/abs/1998PhRvA..57.2448S} {57, 2448}

\bibitem[\protect\citeauthoryear{{Smartt} et~al.,}{{Smartt}
  et~al.}{2017}]{Smartt.etal:17}
{Smartt} S.~J.,  et~al., 2017, \mn@doi [\nat] {10.1038/nature24303}, \href
  {https://ui.adsabs.harvard.edu/abs/2017Natur.551...75S} {551, 75}

\bibitem[\protect\citeauthoryear{{Sneppen} \& {Watson}}{{Sneppen} \&
  {Watson}}{2023}]{Sneppen.Watson:23Y}
{Sneppen} A.,  {Watson} D.,  2023, \mn@doi [\aap]
  {10.1051/0004-6361/202346421}, \href
  {https://ui.adsabs.harvard.edu/abs/2023A&A...675A.194S} {675, A194}

\bibitem[\protect\citeauthoryear{{Spencer} \& {Fano}}{{Spencer} \&
  {Fano}}{1954}]{Spencer.Fano:54}
{Spencer} L.~V.,  {Fano} U.,  1954, \mn@doi [Physical Review]
  {10.1103/PhysRev.93.1172}, \href
  {https://ui.adsabs.harvard.edu/abs/1954PhRv...93.1172S} {93, 1172}

\bibitem[\protect\citeauthoryear{{Symbalisty} \& {Schramm}}{{Symbalisty} \&
  {Schramm}}{1982}]{Symbalisty.Schramm:82}
{Symbalisty} E.,  {Schramm} D.~N.,  1982, \aplett, \href
  {https://ui.adsabs.harvard.edu/abs/1982ApL....22..143S} {22, 143}

\bibitem[\protect\citeauthoryear{{Tanaka} \& {Hotokezaka}}{{Tanaka} \&
  {Hotokezaka}}{2013}]{Tanaka.Hotokezaka:13}
{Tanaka} M.,  {Hotokezaka} K.,  2013, \mn@doi [\apj]
  {10.1088/0004-637X/775/2/113}, \href
  {https://ui.adsabs.harvard.edu/abs/2013ApJ...775..113T} {775, 113}

\bibitem[\protect\citeauthoryear{{Tanaka} et~al.,}{{Tanaka}
  et~al.}{2017}]{Tanaka.etal:17}
{Tanaka} M.,  et~al., 2017, \mn@doi [\pasj] {10.1093/pasj/psx121}, \href
  {https://ui.adsabs.harvard.edu/abs/2017PASJ...69..102T} {69, 102}

\bibitem[\protect\citeauthoryear{{Tanaka} et~al.,}{{Tanaka}
  et~al.}{2018}]{Tanaka.etal:18}
{Tanaka} M.,  et~al., 2018, \mn@doi [\apj] {10.3847/1538-4357/aaa0cb}, \href
  {https://ui.adsabs.harvard.edu/abs/2018ApJ...852..109T} {852, 109}

\bibitem[\protect\citeauthoryear{Tanaka, Kato, Gaigalas  \& Kawaguchi}{Tanaka
  et~al.}{2020}]{tanaka:opacities:2020}
Tanaka M.,  Kato D.,  Gaigalas G.,   Kawaguchi K.,  2020, \mn@doi [\mnras]
  {10.1093/mnras/staa1576}, 496, 1369

\bibitem[\protect\citeauthoryear{{Tanvir} et~al.,}{{Tanvir}
  et~al.}{2017}]{Tanvir.etal:17}
{Tanvir} N.~R.,  et~al., 2017, \mn@doi [\apjl] {10.3847/2041-8213/aa90b6},
  \href {https://ui.adsabs.harvard.edu/abs/2017ApJ...848L..27T} {848, L27}

\bibitem[\protect\citeauthoryear{{Tarumi}, {Hotokezaka}, {Domoto}  \&
  {Tanaka}}{{Tarumi} et~al.}{2023}]{Tarumi.etal:23}
{Tarumi} Y.,  {Hotokezaka} K.,  {Domoto} N.,   {Tanaka} M.,  2023, arXiv
  e-prints, \href {https://ui.adsabs.harvard.edu/abs/2023arXiv230213061T} {p.
  arXiv:2302.13061}

\bibitem[\protect\citeauthoryear{{Troja} et~al.,}{{Troja}
  et~al.}{2017}]{Troja.etal:17}
{Troja} E.,  et~al., 2017, \mn@doi [\nat] {10.1038/nature24290}, \href
  {https://ui.adsabs.harvard.edu/abs/2017Natur.551...71T} {551, 71}

\bibitem[\protect\citeauthoryear{{Vieira}, {Ruan}, {Haggard}, {Ford}, {Drout},
  {Fern{\'a}ndez}  \& {Badnell}}{{Vieira} et~al.}{2023}]{Viera2023}
{Vieira} N.,  {Ruan} J.~J.,  {Haggard} D.,  {Ford} N.,  {Drout} M.~R.,
  {Fern{\'a}ndez} R.,   {Badnell} N.~R.,  2023, \mn@doi [\apj]
  {10.3847/1538-4357/acae72}, \href
  {https://ui.adsabs.harvard.edu/abs/2023ApJ...944..123V} {944, 123}

\bibitem[\protect\citeauthoryear{{Villar} et~al.,}{{Villar}
  et~al.}{2017}]{Villar.etal:17}
{Villar} V.~A.,  et~al., 2017, \mn@doi [\apjl] {10.3847/2041-8213/aa9c84},
  \href {https://ui.adsabs.harvard.edu/abs/2017ApJ...851L..21V} {851, L21}

\bibitem[\protect\citeauthoryear{{Volz} \& {Schmoranzer}}{{Volz} \&
  {Schmoranzer}}{1996}]{Volz:96}
{Volz} U.,  {Schmoranzer} H.,  1996, \mn@doi [Physica Scripta Volume T]
  {10.1088/0031-8949/1996/T65/007}, \href
  {https://ui.adsabs.harvard.edu/abs/1996PhST...65...48V} {65, 48}

\bibitem[\protect\citeauthoryear{{Wanajo}}{{Wanajo}}{2018}]{Wanajo:18}
{Wanajo} S.,  2018, \mn@doi [\apj] {10.3847/1538-4357/aae0f2}, \href
  {https://ui.adsabs.harvard.edu/abs/2018ApJ...868...65W} {868, 65}

\bibitem[\protect\citeauthoryear{{Wanajo}, {Sekiguchi}, {Nishimura}, {Kiuchi},
  {Kyutoku}  \& {Shibata}}{{Wanajo} et~al.}{2014}]{Wanajo.etal:14}
{Wanajo} S.,  {Sekiguchi} Y.,  {Nishimura} N.,  {Kiuchi} K.,  {Kyutoku} K.,
  {Shibata} M.,  2014, \mn@doi [\apjl] {10.1088/2041-8205/789/2/L39}, \href
  {https://ui.adsabs.harvard.edu/abs/2014ApJ...789L..39W} {789, L39}

\bibitem[\protect\citeauthoryear{{Watson} et~al.,}{{Watson}
  et~al.}{2019}]{Watson.etal:2019}
{Watson} D.,  et~al., 2019, \mn@doi [\nat] {10.1038/s41586-019-1676-3}, \href
  {https://ui.adsabs.harvard.edu/abs/2019Natur.574..497W} {574, 497}

\bibitem[\protect\citeauthoryear{{Waxman}, {Ofek}, {Kushnir}  \&
  {Gal-Yam}}{{Waxman} et~al.}{2018}]{Waxman.etal:18}
{Waxman} E.,  {Ofek} E.~O.,  {Kushnir} D.,   {Gal-Yam} A.,  2018, \mn@doi
  [\mnras] {10.1093/mnras/sty2441}, \href
  {https://ui.adsabs.harvard.edu/abs/2018MNRAS.481.3423W} {481, 3423}

\bibitem[\protect\citeauthoryear{{Waxman}, {Ofek}  \& {Kushnir}}{{Waxman}
  et~al.}{2019}]{Waxman.etal:19}
{Waxman} E.,  {Ofek} E.~O.,   {Kushnir} D.,  2019, \mn@doi [\apj]
  {10.3847/1538-4357/ab1f71}, \href
  {https://ui.adsabs.harvard.edu/abs/2019ApJ...878...93W} {878, 93}

\bibitem[\protect\citeauthoryear{{Wollaeger} et~al.,}{{Wollaeger}
  et~al.}{2018}]{Wollaeger2018}
{Wollaeger} R.~T.,  et~al., 2018, \mn@doi [\mnras] {10.1093/mnras/sty1018},
  \href {https://ui.adsabs.harvard.edu/abs/2018MNRAS.478.3298W} {478, 3298}

\bibitem[\protect\citeauthoryear{{Wollaeger} et~al.,}{{Wollaeger}
  et~al.}{2021}]{Wollaeger.etal:21}
{Wollaeger} R.~T.,  et~al., 2021, \mn@doi [\apj] {10.3847/1538-4357/ac0d03},
  \href {https://ui.adsabs.harvard.edu/abs/2021ApJ...918...10W} {918, 10}

\bibitem[\protect\citeauthoryear{{van Regemorter}}{{van
  Regemorter}}{1962}]{Regemorter:62}
{van Regemorter} H.,  1962, \mn@doi [\apj] {10.1086/147445}, \href
  {https://ui.adsabs.harvard.edu/abs/1962ApJ...136..906V} {136, 906}

\makeatother
\end{thebibliography}



\appendix

\section{Model Composition and Energy Deposition}
\label{app:model_extra}

\begin{figure}
    \includegraphics[trim={0.0cm 0.cm 0.4cm 0.2cm},width = 0.49\textwidth]{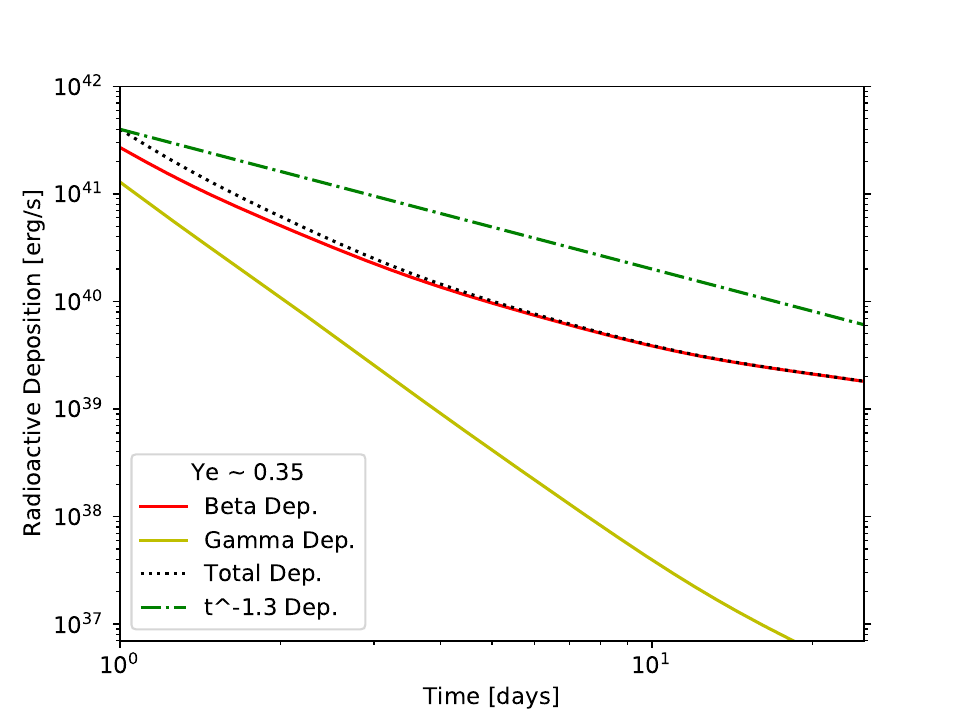} 
    \includegraphics[trim={0.0cm 0.cm 0.4cm 0.2cm},width = 0.49\textwidth]{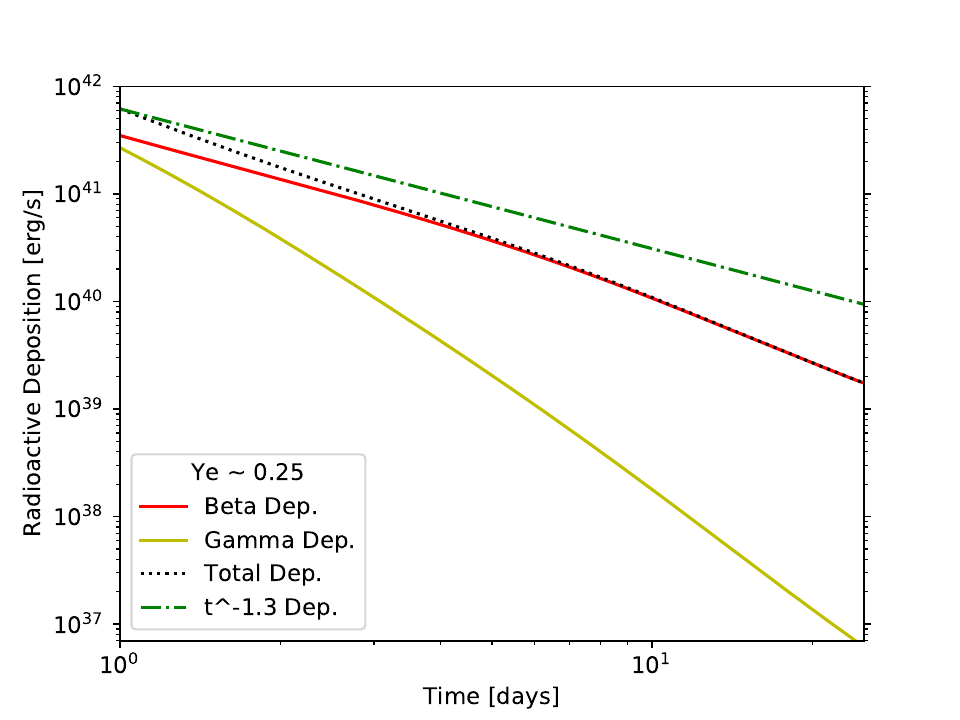}
    \includegraphics[trim={0.0cm 0.cm 0.4cm 0.2cm},width = 0.49\textwidth]{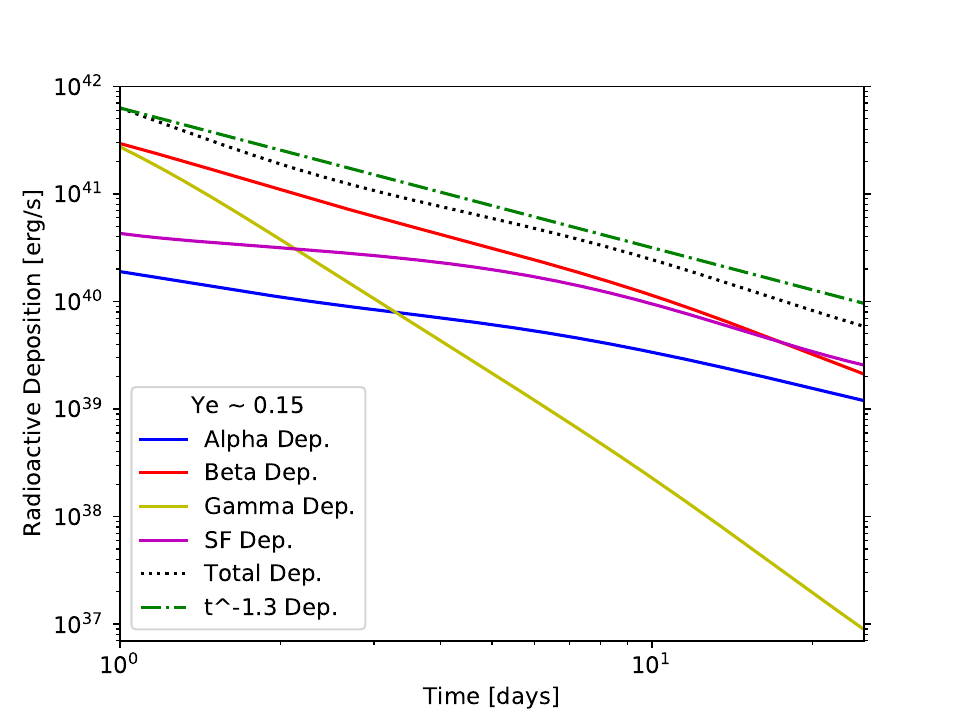}
    \caption{The energy deposition to the spectral simulation separated by decay product, and calculated consistently with the evolution of composition shown in Fig. \ref{fig:compositions}. The canonical $t^{-1.3}$ power law is shown to illustrate the importance of thermalisation and the thermalisation break. We note that only the $Y_e \sim 0.15$ model has significant alpha decay and spontaneous fission contributions.}
    \label{fig:energy_dep}
\end{figure}

In this appendix, we present the exact compositions of our three models, as well as the $Y_e \sim 0.35$ model without Sr in Table \ref{tab:compositions}. We also show the equations used for the thermalisation of the radioactive decay products in Equations \ref{eq:alpha_therm} - \ref{eq:t_sf}, taken from \citet{Barnes.etal:16,Waxman.etal:19,Kasen.Barnes:19}. The additional consideration to $\beta$-decay from \citet{Waxman.etal:19} is reflected in the exponent of $-1.5$ for the thermalisation fraction of $\beta$-decay electrons, Equation \ref{eq:beta_therm}. The total energy depositions by model, with each product's contribution can be visualised in Fig. \ref{fig:energy_dep}.

\begin{align}
    \label{eq:alpha_therm}
    & f_{\alpha} = (1 + \frac{t}{t_{\alpha}})^{-1.5}
    \\
    \label{eq:beta_therm}
    & f_{\beta} = (1 + \frac{t}{t_e})^{-1.5}
    \\
    \label{eq:gamma_therm}
    & f_{\gamma} = \left(1 - e^{-(\frac{t_{\gamma}}{t})^2} \right)
    \\
    \label{eq:sf_therm}
    & f_{\rm{SF}} = \frac{\ln\left(1 + 2(\frac{t}{t_{\rm{SF}}})^2 \right)}{2(\frac{t}{t_{\rm{SF}}})^2}
\end{align}

\begin{align}
    \label{eq:t_alpha}
    & t_{\alpha} = 38.7 \left(\frac{M_{\rm{zone}}}{0.01 \rm{\Msol}}\right)^{2/3} \left(\frac{v_{\rm{zone}}}{0.2 \rm{c}}\right)^{-2}~\mbox{d}
    \\
    \label{eq:t_beta}
    & t_{\beta} = 12.9 \left(\frac{M_{\rm{zone}}}{0.01 \rm{\Msol}}\right)^{2/3} \left(\frac{v_{\rm{zone}}}{0.2 \rm{c}}\right)^{-2}~\mbox{d}
    \\
    \label{eq:t_gamma}
    & t_{\gamma} = 0.3 \left(\frac{M_{\rm{zone}}}{0.01 \rm{\Msol}}\right)^{1/2} \left(\frac{v_{\rm{zone}}}{0.2 \rm{c}}\right)^{-1}~\mbox{d}
    \\
    \label{eq:t_sf}
    & t_{\rm{SF}} = 16.77 \left(\frac{M_{\rm{zone}}}{0.005 \rm{\Msol}}\right)^{1/2} \left(\frac{v_{\rm{zone}}}{0.2 \rm{c}}\right)^{-1.5}~\mbox{d}
\end{align}

\begin{table}
    \caption{Model compositions by mass fractions. The dashed horizontal lines indicate separations between groups of elements, e.g. light elements, first r-process peak, second r-process peak, lanthanides, third r-process peak and actinides.}
    \centering
    \begin{tabular}{c|c|c|c|c}
    Element & $Y_e \sim 0.35 $ & $Y_e \sim 0.35$, no Sr & $Y_e \sim 0.25 $ & $Y_e \sim 0.15 $ \\
    \hline 
    $^{22}$Ti & 0.0006 & 0.0007 & - & - \\
    $^{23}$V & 0.0005 & 0.0006 & - & - \\ 
    $^{24}$Cr & 0.0119 & 0.0147 & - & - \\
    $^{25}$Mn & 0.0009 & 0.0011 & - & - \\
    $^{26}$Fe & 0.0258 & 0.0319 & - & - \\
    $^{27}$Co & 0.0004 & 0.0005 & - & - \\
    $^{28}$Ni & 0.0240 & 0.0297 & - & - \\
    $^{29}$Cu & 0.0079 & 0.0098 & - & - \\
    \hdashline
    $^{30}$Zn & 0.0106 & 0.0131 & 0.0249 & - \\
    $^{31}$Ga & 0.0025 & 0.0032 & 0.0112 & - \\
    $^{32}$Ge & 0.0095 & 0.0118 & 0.0519 & - \\
    $^{33}$As & 0.0008 & 0.0010 & - & - \\
    $^{34}$Se & 0.1272 & 0.1570 & 0.1269 & - \\
    $^{35}$Br & 0.0241 & 0.0298 & 0.0207 & - \\
    $^{36}$Kr & 0.2638 & 0.3256 & 0.0909 & - \\
    $^{37}$Rb & 0.1033 & 0.1276 & 0.0304 & - \\
    $^{38}$Sr & 0.1898 & - & 0.0375 & - \\
    $^{39}$Y & 0.0261 & 0.0322 & 0.0058 & - \\
    $^{40}$Zr & 0.1105 & 0.1364 & 0.0749 & 0.0151 \\
    \hdashline
    $^{41}$Nb & 0.0007 & 0.0009 & -  & - \\
    $^{42}$Mo & 0.0220 & 0.0271 & 0.0235 & - \\
    $^{44}$Ru & 0.0252 & 0.0311 & 0.0395 & 0.0099 \\
    $^{45}$Rh & 0.0007 & 0.0008 & -  & - \\
    $^{46}$Pd & 0.0059 & 0.0073 & 0.0283 & - \\
    $^{47}$Ag & 0.0008 & 0.0010 & 0.0096 & - \\
    $^{48}$Cd & 0.0023 & 0.0029 & 0.0542 & 0.0125 \\
    $^{49}$In & 0.0003 & 0.0003 & 0.0077 & - \\
    $^{50}$Sn & 0.0013 & 0.0016 & 0.1749 & 0.0937 \\
    $^{51}$Sb & 0.0003 & 0.0003 & 0.0424 & 0.0216 \\
    $^{52}$Te & - & - & 0.0573 & 0.0730 \\
    $^{53}$I & - & - & 0.0428 & 0.0388 \\
    $^{54}$Xe & - & - & 0.0179 & 0.0551 \\
    $^{55}$Cs & - & - & 0.0092 & 0.0338 \\
    $^{56}$Ba & - & - & 0.0018 & - \\
    \hdashline
    $^{58}$Ce & - & - & 0.0029 & 0.0179 \\
    $^{60}$Nd & - & - & 0.0039 & 0.0289 \\
    $^{62}$Sm & - & - & 0.0022 & 0.0218 \\
    $^{63}$Eu & - & - & 0.0016 & 0.0204 \\
    $^{64}$Gd & - & - & 0.0023 & 0.0365 \\
    $^{65}$Tb & - & - & - & 0.0148 \\
    $^{66}$Dy & - & - & 0.0019 & 0.0430 \\
    $^{68}$Er & - & - & 0.0010 & 0.0334 \\
    $^{69}$Tm & - & - & - & 0.0132 \\
    $^{70}$Yb & - & - & - & 0.0221 \\
    \hdashline
    $^{72}$Hf & - & - & - & 0.0228 \\
    $^{73}$Ta & - & - & - & 0.0925 \\
    $^{75}$Re & - & - & - & 0.0343 \\
    $^{76}$Os & - & - & - & 0.1573 \\
    $^{77}$Ir & - & - & - & 0.0361 \\
    $^{78}$Pt & - & - & - & 0.0368 \\
    $^{79}$Au & - & - & - & 0.0088 \\
    $^{82}$Pb & - & - & - & 0.0023 \\
    $^{83}$Bi & - & - & - & 0.0008 \\
    \hdashline
    $^{90}$Th & - & - & - & 0.0013 \\
    $^{92}$U & - & - & - & 0.0015 \\
    \hline
    \end{tabular}
    \label{tab:compositions}
\end{table}

\newpage

\section{Additional Plots}
\label{app:extra_results}

\begin{figure*}
    \centering
    \includegraphics[trim={0.4cm 0.cm 0.4cm 0.3cm},width = 0.47\textwidth]{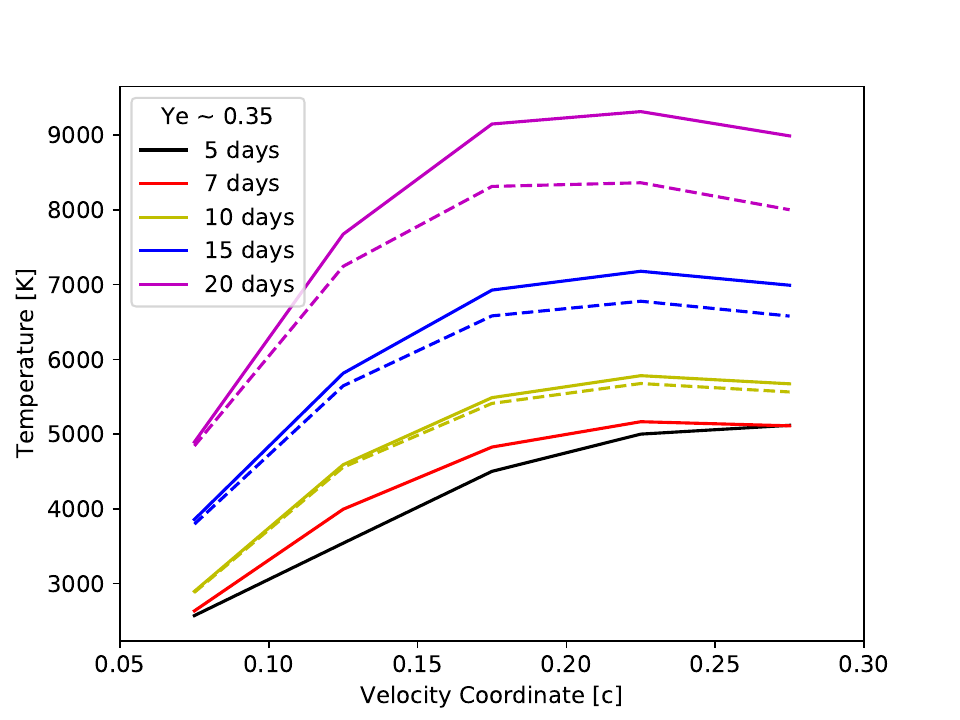} 
    \includegraphics[trim={0.4cm 0.cm 0.4cm 0.3cm},width = 0.47\textwidth]{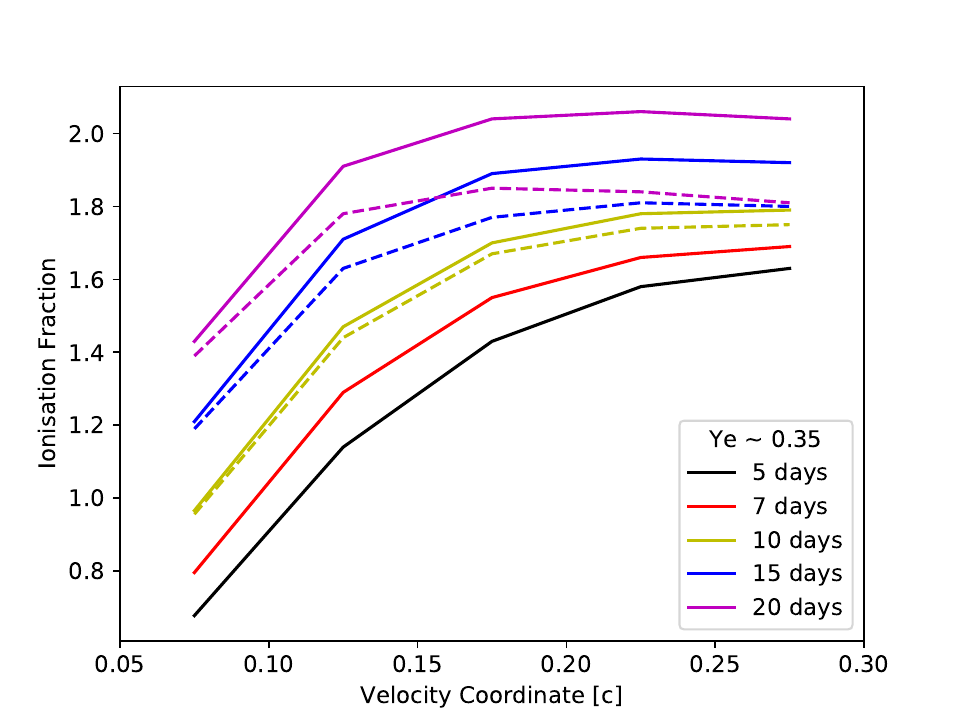} 
    \includegraphics[trim={0.4cm 0.cm 0.4cm 0.3cm},width = 0.47\textwidth]{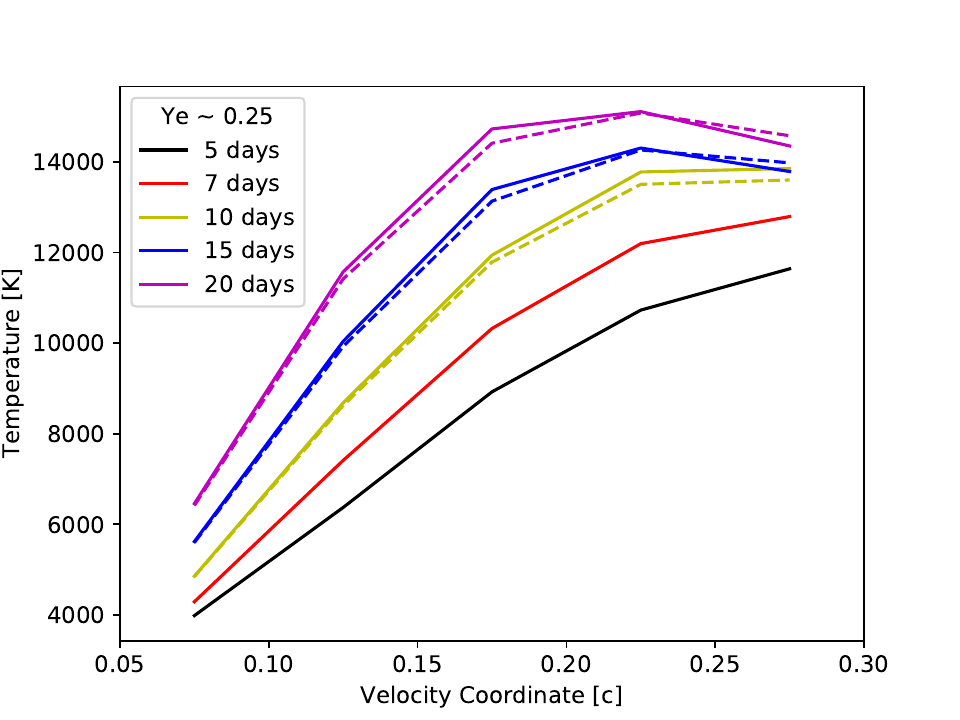}
    \includegraphics[trim={0.4cm 0.cm 0.4cm 0.3cm},width = 0.47\textwidth]{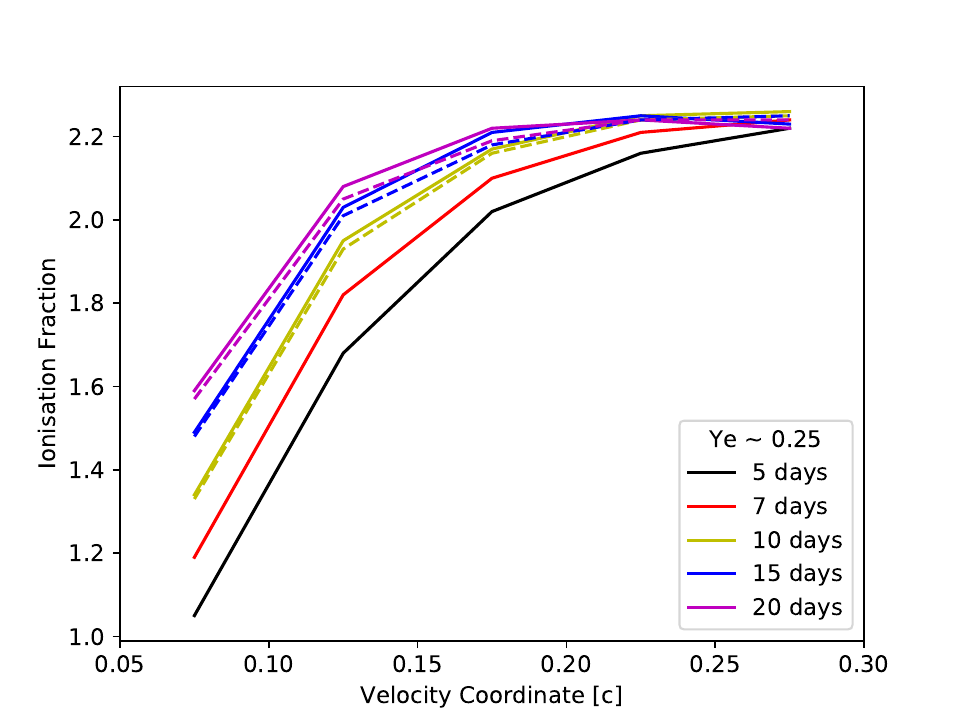}
    \includegraphics[trim={0.4cm 0.cm 0.4cm 0.3cm},width = 0.47\textwidth]{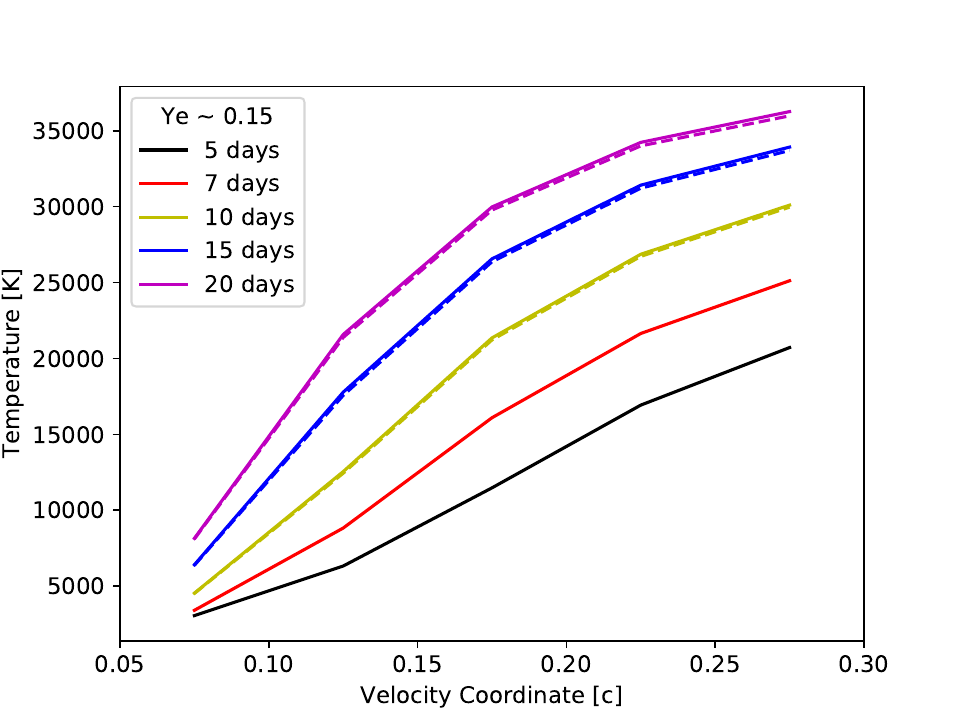}
    \includegraphics[trim={0.4cm 0.cm 0.4cm 0.3cm},width = 0.47\textwidth]{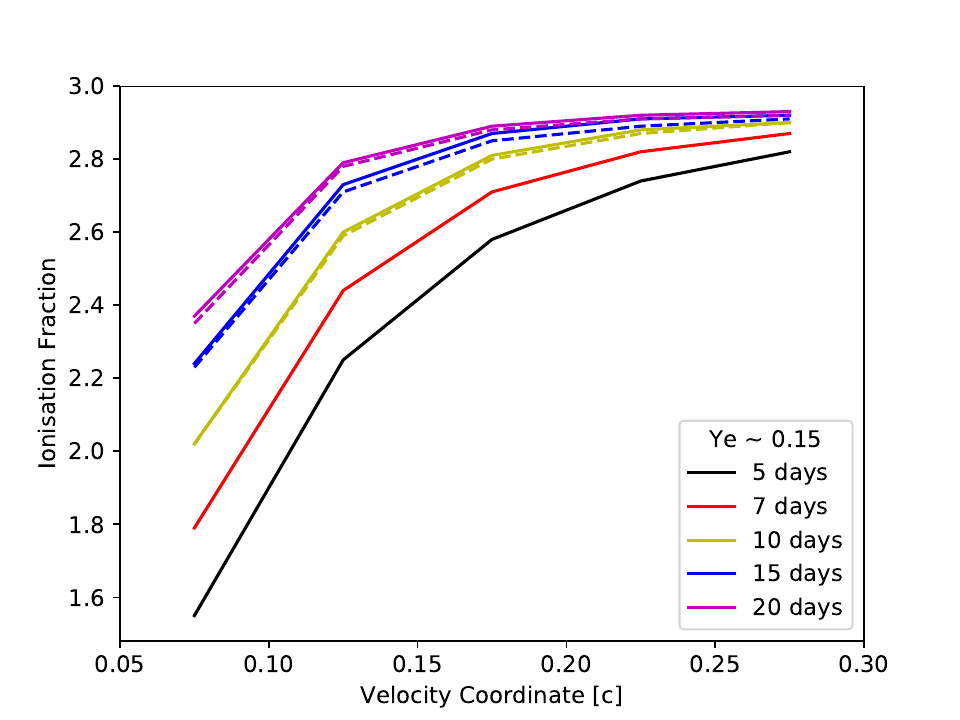}
    \caption{The temperature and ionisation structure profiles of the models. The time dependent solutions are indicated by the dashed lines. Note that time dependent mode is only run from 10 days onwards.}
    \label{fig:thermo_profile}
\end{figure*}

This appendix collects additional plots of auxiliary utility to the main body of the paper. In Fig. \ref{fig:thermo_profile}, we show the thermodynamic evolution of our models in terms of profiles, e.g. layer slices. This provides an alternative visualisation to Fig. \ref{fig:thermo_evolution}, and notably shows the changes in stratification of the ejecta in the $Y_e \sim 0.35, 0.25$ models as time progresses. 

\begin{figure*}
    \centering
    \includegraphics[trim={0.2cm 0cm 0.6cm 0.3cm},clip,width = 0.49\textwidth]{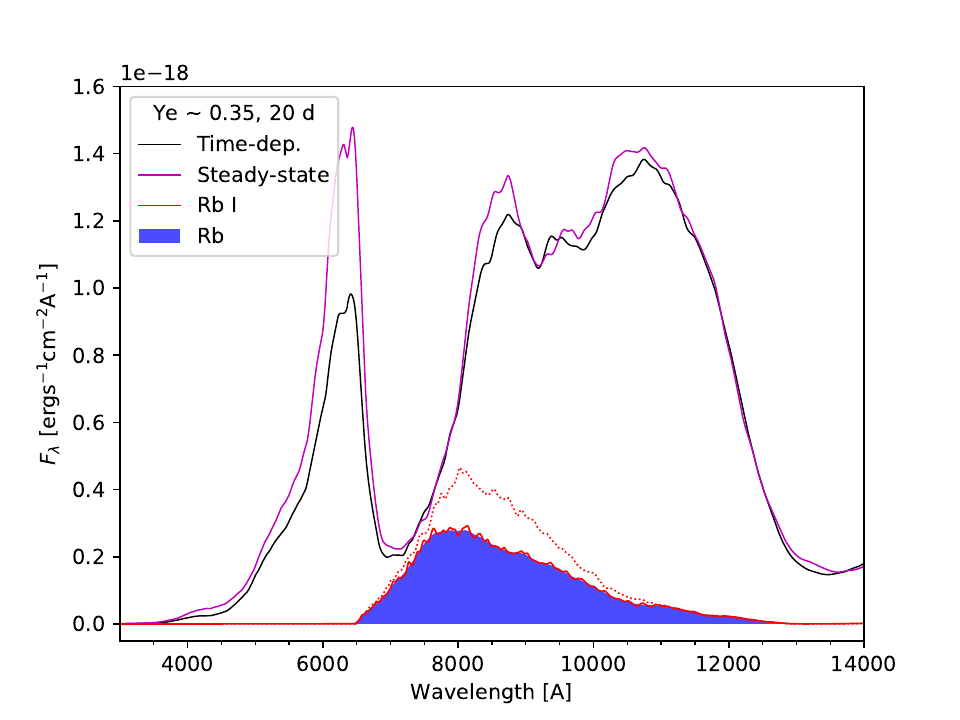} 
    \includegraphics[trim={0.2cm 0cm 0.6cm 0.3cm},clip,width = 0.49\textwidth]{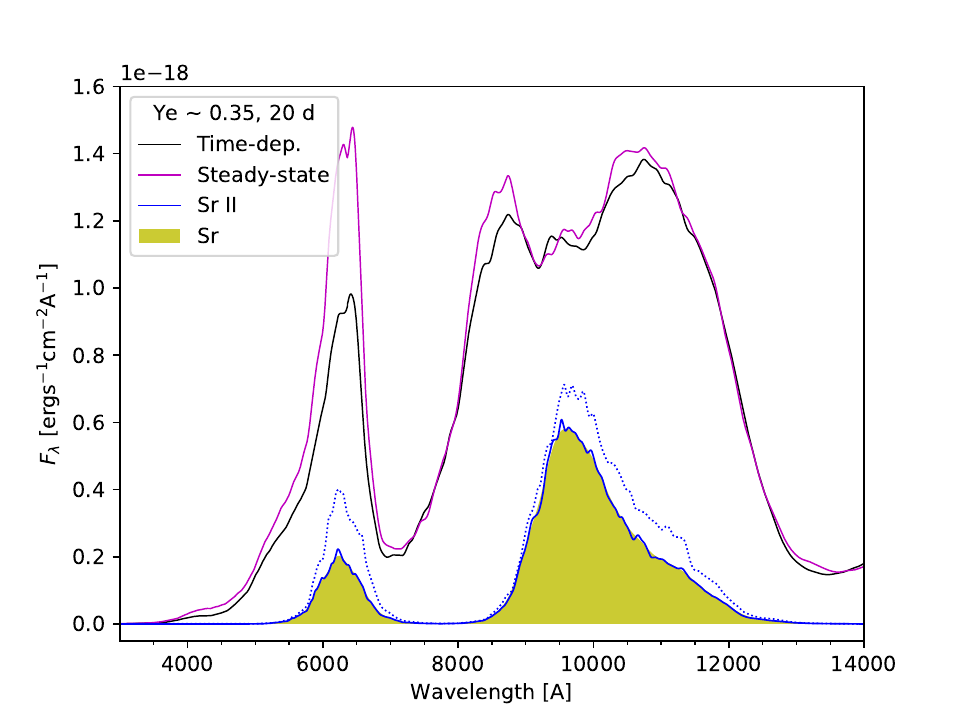}
    \includegraphics[trim={0.2cm 0cm 0.6cm 0.3cm},clip,width = 0.49\textwidth]{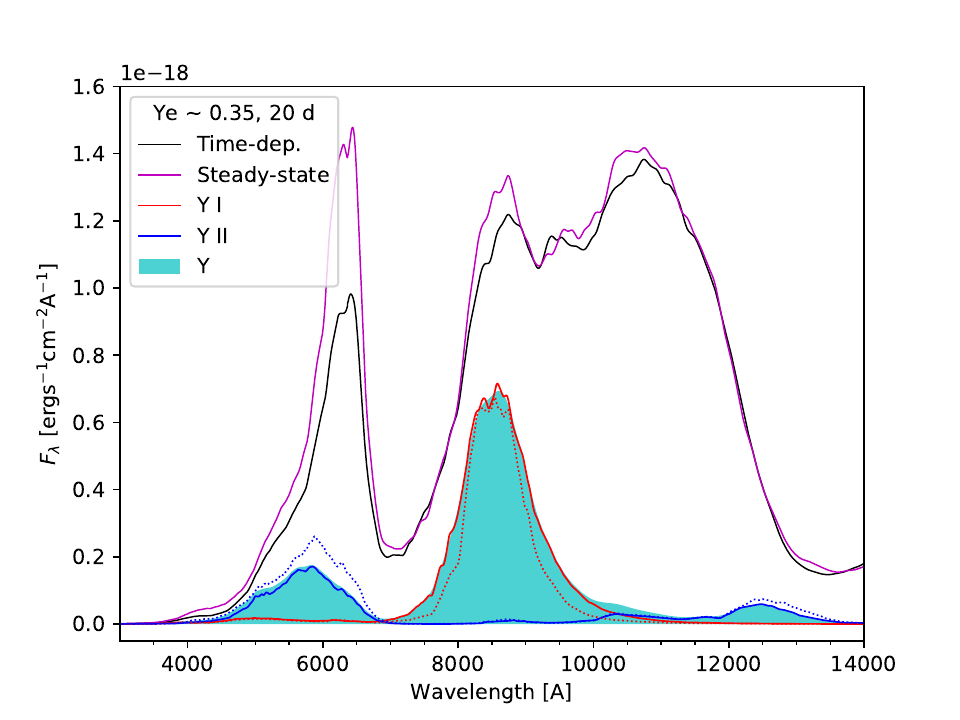}
    \includegraphics[trim={0.2cm 0cm 0.6cm 0.3cm},clip,width = 0.49\textwidth]{Spectra/FigB2a.pdf} 
    \caption{Individual contributions of Rb, Sr, Y and Zr for the high $Y_e$ model at 20 days. Time-dependent effects are shown by comparing the time-dependent spectrum (black) to the steady-state spectrum (magenta), as well as the effect on the individual species (solid for time-dependent, dotted for steady-state). The filled areas are the total elemental contribution in the time-dependent solution. Note that the x-axis is scaled differently for each element, to focus on the wavelength ranges where they contribute the most to the spectrum.}
    \label{fig:Ye035_20d_species_detail}
\end{figure*}

We also show a detailed visualisation of time-dependent effects on the individual species of the $Y_e \sim 0.35$ model at 20 days after merger, when the time-dependent effects are strongest, in Fig. \ref{fig:Ye035_20d_species_detail}.

\section{Atomic Data}
\label{app:atomic_data}
This part of the appendix gives information on the fundamental parameters used in the atomic structure calculations, most importantly the scope of included spectroscopic configurations. The information is summarized in Tab. \ref{tab:atomic_data}. The first to third columns (Z, El and Sp) define the atomic number, the element name and the ionic spectrum designation. The fourth ($E^\mathrm{NIST}_\mathrm{i}$) gives the (first) ionization energy as listed by the NIST Atomic Spectra Database \citep{NIST_ASD} in units of eV. The fifth to seventh ($n_\mathrm{lev}$, $n_\mathrm{lev}^\mathrm{red}$ and $n_\mathrm{trans}$) specifies the number of computed fine-structure levels, the number of such levels below the $E^\mathrm{NIST}_\mathrm{i}$ ionization energy and, finally, the total number of allowed and forbidden transitions between the $n_\mathrm{trans}$ levels (with different multipole contributions summed up for each line). The eighth column ($n^\mathrm{cfg}_\mathrm{opt}$) specifies how many configurations that were included in the Dirac-Fock-Slater optimization of the common central, screening potential in the order of the configurations in the following columns (i.e. a '2' implies that the ground configuration and the first excited configuration was used in the optimization). Finally, column nine and ten ('Ground Conf.' and 'Additional Configurations') contains the ground and excited configurations included in the atomic model applied to each ion. 

\begin{table*}\centering
\caption{Fundamental atomic parameters defining the atomic structure calculations. See Appendix \ref{app:atomic_data} for further information and definitions.} \label{tab:atomic_data}
\scriptsize
\begin{tabularx}{\textwidth}{lrrrrrrrlXX}
\toprule
Z &El & Sp &$E^\mathrm{NIST}_\mathrm{i}$[eV] &$n_\mathrm{lev}$ &$n_\mathrm{lev}^\mathrm{red}$ &$n_\mathrm{trans}$ &$n^\mathrm{cfg}_\mathrm{opt}$ & Ground Conf. & Additional Configurations \\ \midrule
        26 & Fe & I & 7.9024681 & 3195 & 667 & 153526 & 1 & 3d6 4s2 & 3d8, 3d7 4s1, 3d6 4s1 4p1, 3d7 4p1, 3d5 4s2 4p1, 3d6 4s1 5s1, 3d7 5s1, 3d6 4s1 5p1, 3d6 4s1 4d1, 3d7 5p1, 3d7 4d1, 3d6 4s1 6s1, 3d7 6s1, 3d6 4s1 6p1, 3d6 4s1 5d1 \\
        ~ & ~ & II & 16.1992 & 3467 & 1647 & 901922 & 2 & 3d6 4s1 & 3d7, 3d5 4s2, 3d6 4p1, 3d5 4s1 4p1, 3d6 5s1, 3d6 4d1, 3d6 5p1, 3d6 6s1, 3d6 4f1, 3d5 4s1 5s1, 3d6 5d1, 3d6 6p1, 3d5 4p2, 3d5 4s1 4d1, 3d6 7s1 \\
        ~ & ~ & III & 30.651 & 2338 & 1853 & 1108723 & 1 & 3d6 & 3d5 4s1, 3d5 4p1, 3d5 4d1, 3d5 5s1, 3d5 5p1, 3d5 4f1, 3d4 4s1 4p1, 3d5 5d1, 3d5 6s1, 3d5 6p1 \\
        ~ & ~ & IV & 54.91 & 736 & 730 & 190780 & 1 & 3d5 & 3d4 4s1, 3d4 4p1, 3d3 4s1 4p1, 3d4 5s1, 3d4 5p1 \\
        27 & Co & I & 7.88101 & 778 & 423 & 64859 & 3 & 3p6 3d7 4s2 & 3p6 3d8 4s1, 3p6 3d9, 3p6 3d7 4s1 4p1, 3p6 3d8 4p1, 3p6 3d8 5s1, 3p6 3d7 4s1 5s1, 3p6 3d8 4d1, 3p6 3d7 4s1 4d1 \\
        ~ & ~ & II & 17.0844 & 905 & 721 & 177823 & 1 & 3p6 3d8 & 3p6 3d7 4s1, 3p6 3d6 4s2, 3p6 3d7 4p1, 3p6 3d6 4s1 4p1, 3p6 3d7 5s1, 3p6 3d7 4d1, 3p6 3d7 5p1, 3p6 3d7 6s1 \\
        ~ & ~ & III & 33.5 & 601 & 596 & 122474 & 1 & 3p6 3d7 & 3p6 3d6 4s1, 3p6 3d6 4p1, 3p6 3d6 4d1, 3p6 3d6 5s1 \\
        ~ & ~ & IV & 51.27 & 1088 & 975 & 318197 & 1 & 3p6 3d6 & 3p6 3d5 4s1, 3p6 3d5 4p1, 3p6 3d4 4s1 4p1, 3p6 3d5 4d1, 3p6 3d5 5s1 \\
        28 & Ni & I & 7.639878 & 236 & 156 & 9661 & 3 & 3p6 3d8 4s2 & 3p6 3d9 4s1, 3p6 3d9 4p1, 3p6 3d10, 3p6 3d8 4s1 4p1, 3p6 3d9 5s1, 3p6 3d8 4s1 5s1, 3p6 3d9 5p1, 3p6 3d9 4d1, 3p6 3d9 6s1, 3p6 3d9 6p1, 3p6 3d9 5d1, 3p6 3d9 4f1 \\
        ~ & ~ & II & 18.168838 & 587 & 519 & 97570 & 1 & 3p6 3d9 & 3p6 3d8 4s1, 3p6 3d7 4s2, 3p6 3d8 4p1, 3p6 3d7 4s1 4p1, 3p6 3d8 5s1, 3p6 3d8 4d1, 3p6 3d8 5p1, 3p6 3d8 6s1, 3p6 3d8 4f1, 3p6 3d8 5d1 \\
        ~ & ~ & III & 35.187 & 867 & 825 & 235160 & 1 & 3p6 3d8 & 3p6 3d7 4s1, 3p6 3d7 4p1, 3p6 3d6 4s2, 3p6 3d7 4d1, 3p6 3d7 5s1, 3p6 3d7 5p1, 3p6 3d6 4s1 4p1 \\
        ~ & ~ & IV & 54.92 & 818 & 818 & 232957 & 1 & 3p6 3d7 & 3p6 3d6 4s1, 3p6 3d6 4p1, 3p6 3d5 4s2, 3p6 3d6 4d1, 3p6 3d6 5s1, 3p6 3d6 5p1 \\
        29 & Cu & I & 7.72638 & 38 & 32 & 449 & 3 & 3p6 3d10 4s1 & 3p6 3d9 4s2, 3p6 3d10 4p1, 3p6 3d9 4s1 4p1, 3p6 3d10 5s1, 3p6 3d10 5p1, 3p6 3d10 4d1, 3p6 3d10 6s1, 3p6 3d10 6p1, 3p6 3d10 5d1 \\
        ~ & ~ & II & 20.29239 & 204 & 193 & 14480 & 1 & 3p6 3d10 & 3p6 3d9 4s1, 3p6 3d9 4p1, 3p6 3d8 4s2, 3p6 3d8 4s1 4p1, 3p6 3d9 5s1, 3p6 3d9 4d1, 3p6 3d9 5p1, 3p6 3d9 6s1, 3p6 3d9 4f1, 3p6 3d9 5d1, 3p6 3d9 6p1 \\
        ~ & ~ & III & 36.841 & 587 & 565 & 116015 & 1 & 3p6 3d9 & 3p6 3d8 4s1, 3p6 3d8 4p1, 3p6 3d7 4s2, 3p6 3d8 5s1, 3p6 3d8 4d1, 3p6 3d8 5p1, 3p6 3d7 4s1 4p1, 3p6 3d8 4f1, 3p6 3d8 6s1, 3p6 3d8 5d1 \\
        ~ & ~ & IV & 57.38 & 397 & 397 & 54954 & 1 & 3p6 3d8 & 3p6 3d7 4s1, 3p6 3d7 4p1, 3p6 3d7 4d1, 3p6 3d6 4s2, 3p6 3d7 5s1 \\
        30 & Zn & I & 9.394197 & 29 & 29 & 348 & 1 & 3p6 3d10 4s2 & 3p6 3d10 4s1 4p1, 3p6 3d10 4s1 5s1, 3p6 3d10 4s1 5p1, 3p6 3d10 4s1 4d1, 3p6 3d10 4s1 6s1, 3p6 3d10 4s1 6p1, 3p6 3d10 4s1 5d1, 3p6 3d10 4s1 4f1 \\
        ~ & ~ & II & 17.96439 & 40 & 36 & 564 & 1 & 3p6 3d10 4s1 & 3p6 3d10 4p1, 3p6 3d9 4s2, 3p6 3d10 5s1, 3p6 3d10 4d1, 3p6 3d10 5p1, 3p6 3d9 4s1 4p1, 3p6 3d10 6s1, 3p6 3d10 4f1, 3p6 3d10 5d1, 3p6 3d10 6p1 \\
        ~ & ~ & III & 39.7233 & 150 & 149 & 8624 & 1 & 3p6 3d10 & 3p6 3d9 4s1, 3p6 3d9 4p1, 3p6 3d8 4s2, 3p6 3d9 4d1, 3p6 3d9 5s1, 3p6 3d9 5p1, 3p6 3d8 4s1 4p1 \\
        ~ & ~ & IV & 59.573 & 382 & 382 & 52015 & 1 & 3p6 3d9 & 3p6 3d8 4s1, 3p6 3d8 4p1, 3p6 3d8 4d1, 3p6 3d8 5p1, 3p6 3d8 4f1, 3p6 3d8 6p1, 3p6 3d8 5f1 \\
        31 & Ga & I & 5.999302 & 22 & 17 & 130 & 1 & 3d10 4s2 4p1 & 3d10 4s2 5s1, 3d10 4s2 5p1, 3d10 4s2 4d1, 3d10 4s2 6s1, 3d10 4s1 4p2, 3d10 4s2 6p1, 3d10 4s2 5d1, 3d10 4s2 4f1 \\
        ~ & ~ & II & 20.51514 & 26 & 26 & 290 & 1 & 3d10 4s2 & 3d10 4s1 4p1, 3d10 4s1 5s1, 3d10 4p2, 3d10 4s1 4d1, 3d10 4s1 5p1, 3d10 4s1 6s1, 3d10 4s1 5d1 \\
        ~ & ~ & III & 30.72576 & 12 & 12 & 62 & 1 & 3d10 4s1 & 3d10 4p1, 3d9 4s2, 3d10 5s1, 3d10 4d1, 3d10 5p1, 3d10 4f1 \\
        ~ & ~ & IV & 63.241 & 51 & 51 & 1087 & 1 & 3d10 & 3d9 4s1, 3d9 4p1, 3d9 4d1, 3d9 5s1, 3d9 5p1 \\
        32 & Ge & I & 7.899435 & 79 & 72 & 2182 & 2 & 3d10 4s2 4p2 & 3d10 4s1 4p3, 3d10 4s2 4p1 5s1, 3d10 4s2 4p1 5p1, 3d10 4s2 4p1 4d1, 3d10 4s2 4p1 6s1, 3d10 4s2 4p1 6p1, 3d10 4s2 4p1 5d1, 3d10 4s2 4p1 4f1 \\
        ~ & ~ & II & 15.93461 & 22 & 22 & 223 & 1 & 3d10 4s2 4p1 & 3d10 4s1 4p2, 3d10 4s2 5s1, 3d10 4s2 5p1, 3d10 4s2 4d1, 3d10 4s2 6s1, 3d10 4s2 5d1, 3d10 4s2 4f1, 3d10 4s2 6p1 \\
        ~ & ~ & III & 34.0576 & 32 & 32 & 412 & 1 & 3d10 4s2 & 3d10 4s1 4p1, 3d10 4p2, 3d10 4s1 5s1, 3d10 4s1 4d1, 3d10 4s1 5p1, 3d10 4s1 4f1, 3d10 4s1 6s1, 3d10 4s1 5d1, 3d10 4s1 7s1 \\
        ~ & ~ & IV & 45.7155 & 15 & 15 & 100 & 1 & 3d10 4s1 & 3d10 4p1, 3d10 4d1, 3d10 5s1, 3d10 5p1, 3d10 4f1, 3d9 4s2, 3d10 5d1, 3d10 6s1 \\
        33 & As & I & 9.78855 & 99 & 78 & 2757 & 1 & 3d10 4s2 4p3 & 3d10 4s2 4p2 5s1, 3d10 4s1 4p4, 3d10 4s2 4p2 5p1, 3d10 4s2 4p2 4d1, 3d10 4s2 4p2 6s1, 3d10 4s2 4p2 6p1 \\
        ~ & ~ & II & 18.5892 & 67 & 67 & 1989 & 1 & 3d10 4s2 4p2 & 3d10 4s1 4p3, 3d10 4s2 4p1 5s1, 3d10 4s2 4p1 4d1, 3d10 4s2 4p1 5p1, 3d10 4s2 4p1 6s1, 3d10 4s2 4p1 5d1, 3d10 4s2 4p1 6p1 \\
        ~ & ~ & III & 28.349 & 25 & 25 & 290 & 1 & 3d10 4s2 4p1 & 3d10 4s1 4p2, 3d10 4s2 5s1, 3d10 4s2 4d1, 3d10 4s2 5p1, 3d10 4p3, 3d10 4s2 6s1, 3d10 4s2 4f1, 3d10 4s2 5d1 \\
        ~ & ~ & IV & 50.15 & 30 & 30 & 365 & 1 & 3d10 4s2 & 3d10 4s1 4p1, 3d10 4s1 4d1, 3d10 4p2, 3d10 4s1 5s1, 3d10 4s1 5p1, 3d10 4s1 4f1, 3d10 4s1 5d1, 3d10 4s1 6s1 \\
        34 & Se & I & 9.752392 & 157 & 120 & 6030 & 1 & 4s2 4p4 & 4s2 4p3 5s1, 4s2 4p3 5p1, 4s2 4p3 4d1, 4s2 4p3 6s1, 4s2 4p3 6p1, 4s2 4p3 5d1 \\
        ~ & ~ & II & 21.196 & 78 & 78 & 2758 & 1 & 4s2 4p3 & 4s1 4p4, 4s2 4p2 5s1, 4s2 4p2 5p1, 4s2 4p2 6s1, 4s2 4p2 5d1 \\
        ~ & ~ & III & 31.697 & 57 & 57 & 1432 & 1 & 4s2 4p2 & 4s1 4p3, 4s2 4p1 4d1, 4s2 4p1 5s1, 4s2 4p1 5p1, 4s2 4p1 6s1, 4s2 4p1 5d1 \\
        ~ & ~ & IV & 42.947 & 20 & 20 & 190 & 1 & 4s2 4p1 & 4s1 4p2, 4s2 4d1, 4s2 5s1, 4p3, 4s2 5p1 \\
        35 & Br & I & 11.81381 & 117 & 109 & 5256 & 1 & 4s2 4p5 & 4s2 4p4 5s1, 4s2 4p4 5p1, 4s2 4p4 4d1, 4s2 4p4 6s1, 4s1 4p6, 4s2 4p4 6p1, 4s2 4p4 5d1 \\
        ~ & ~ & II & 21.591 & 173 & 172 & 11816 & 1 & 4s2 4p4 & 4s2 4p3 5s1, 4s1 4p5, 4s2 4p3 4d1, 4s2 4p3 5p1, 4s2 4p3 6s1, 4s2 4p3 5d1, 4s2 4p3 4f1 \\
        ~ & ~ & III & 34.871 & 106 & 106 & 5005 & 1 & 4s2 4p3 & 4s1 4p4, 4s2 4p2 4d1, 4s2 4p2 5s1, 4s2 4p2 5p1, 4s2 4p2 5d1, 4s2 4p2 6s1 \\
        ~ & ~ & IV & 47.782 & 45 & 45 & 898 & 1 & 4s2 4p2 & 4s1 4p3, 4s2 4p1 4d1, 4s2 4p1 5s1, 4s2 4p1 5p1, 4s2 4p1 6s1 \\
        36 & Kr & I & 13.9996053 & 53 & 53 & 1216 & 1 & 4s2 4p6 & 4s2 4p5 5s1, 4s2 4p5 5p1, 4s2 4p5 4d1, 4s2 4p5 6s1, 4s2 4p5 6p1, 4s2 4p5 5d1 \\
        ~ & ~ & II & 24.35984 & 147 & 147 & 9115 & 1 & 4s2 4p5 & 4s1 4p6, 4s2 4p4 5s1, 4s2 4p4 4d1, 4s2 4p4 5p1, 4s2 4p4 6s1, 4s2 4p4 5d1, 4s2 4p4 6p1, 4s2 4p4 4f1 \\
        ~ & ~ & III & 35.838 & 172 & 171 & 12169 & 1 & 4s2 4p4 & 4s1 4p5, 4s2 4p3 4d1, 4s2 4p3 5s1, 4s2 4p3 5p1, 4s2 4p3 6s1, 4s2 4p3 5d1, 4s2 4p3 6d1, 4p6 \\
        ~ & ~ & IV & 50.85 & 106 & 106 & 5005 & 1 & 4s2 4p3 & 4s1 4p4, 4s2 4p2 4d1, 4s2 4p2 5s1, 4s2 4p2 5p1, 4s2 4p2 5d1, 4s2 4p2 6s1 \\
        37 & Rb & I & 4.177128 & 17 & 17 & 130 & 1 & 4s2 4p6 5s1 & 4s2 4p6 5p1, 4s2 4p6 4d1, 4s2 4p6 6s1, 4s2 4p6 6p1, 4s2 4p6 5d1, 4s2 4p6 7s1, 4s2 4p6 4f1, 4s2 4p6 7p1, 4s2 4p6 6d1 \\
        ~ & ~ & II & 27.28954 & 93 & 93 & 3564 & 1 & 4s2 4p6 & 4s2 4p5 5s1, 4s2 4p5 4d1, 4s2 4p5 5p1, 4s2 4p5 6s1, 4s2 4p5 5d1, 4s2 4p5 6p1, 4s2 4p5 4f1, 4s2 4p5 7s1, 4s2 4p5 6d1, 4s2 4p5 5f1 \\
        ~ & ~ & III & 39.247 & 96 & 96 & 4052 & 1 & 4s2 4p5 & 4s1 4p6, 4s2 4p4 4d1, 4s2 4p4 5s1, 4s2 4p4 5p1, 4s2 4p4 6s1, 4s2 4p4 5d1 \\
        ~ & ~ & IV & 52.2 & 133 & 133 & 7422 & 1 & 4s2 4p4 & 4s1 4p5, 4s2 4p3 4d1, 4s2 4p3 5s1, 4s2 4p3 5p1, 4s2 4p3 5d1, 4s2 4p3 6s1 \\
\bottomrule
\end{tabularx}
\end{table*}

\begin{table*}\centering
\scriptsize
\caption{Table \ref{tab:atomic_data} continued.}
\begin{tabularx}{\textwidth}{lrrrrrrrlXX}
\toprule
Z &El & Sp &$E^\mathrm{NIST}_\mathrm{i}$[eV] &$n_\mathrm{lev}$ &$n_\mathrm{lev}^\mathrm{red}$ &$n_\mathrm{trans}$ &$n^\mathrm{cfg}_\mathrm{opt}$ & Ground Conf. & Additional Configurations \\ \midrule
        38 & Sr & I & 5.6948674 & 50 & 46 & 907 & 1 & 4s2 4p6 5s2 & 4s2 4p6 5s1 5p1, 4s2 4p6 5s1 4d1, 4s2 4p6 5s1 6s1, 4s2 4p6 4d1 5p1, 4s2 4p6 5s1 6p1, 4s2 4p6 5s1 5d1, 4s2 4p6 5p2, 4s2 4p6 5s1 7s1, 4s2 4p6 5s1 4f1, 4s2 4p6 5s1 7p1, 4s2 4p6 5s1 6d1 \\
        ~ & ~ & II & 11.0302764 & 27 & 27 & 190 & 1 & 4s2 4p6 5s1 & 4s2 4p6 4d1, 4s2 4p6 5p1, 4s2 4p6 6s1, 4s2 4p6 5d1, 4s2 4p6 6p1, 4s2 4p6 4f1, 4s2 4p6 7s1, 4s2 4p6 6d1, 4s2 4p6 7p1, 4s2 4p6 5f1, 4s2 4p6 5g1 \\
        ~ & ~ & III & 42.88353 & 65 & 65 & 1749 & 1 & 4s2 4p6 & 4s2 4p5 4d1, 4s2 4p5 5s1, 4s2 4p5 5p1, 4s2 4p5 5d1, 4s2 4p5 6s1, 4s2 4p5 4f1, 4s2 4p5 6p1 \\
        ~ & ~ & IV & 56.28 & 118 & 118 & 5765 & 1 & 4s2 4p5 & 4s1 4p6, 4s2 4p4 4d1, 4s2 4p4 5s1, 4s2 4p4 5p1, 4s2 4p4 5d1, 4s2 4p4 4f1 \\
        39 & Y & I & 6.21726 & 128 & 111 & 5423 & 1 & 4p6 4d1 5s2 & 4p6 5s2 5p1, 4p6 4d2 5s1, 4p6 4d1 5s1 5p1, 4p6 4d2 5p1, 4p6 4d3, 4p6 5s2 6s1, 4p6 4d1 5s1 6s1, 5p2 4p6 5s1, 4p6 5s2 5d1, 4p6 5s2 6p1 \\
        ~ & ~ & II & 12.2236 & 99 & 99 & 3800 & 1 & 4p6 5s2 & 4p6 4d1 5s1, 4p6 4d2, 4p6 5s1 5p1, 4p6 4d1 5p1, 4p6 4d1 6s1, 4p6 5s1 6s1, 4p6 4d1 5d1, 4p6 5p2, 4p6 4d1 6p1, 4p6 5s1 5d1, 4p6 5s1 6p1, 4p6 4d1 4f1 \\
        ~ & ~ & III & 20.52441 & 17 & 17 & 126 & 1 & 4p6 4d1 & 4p6 5s1, 4p6 5p1, 4p6 6s1, 4p6 5d1, 4p6 6p1, 4p6 4f1, 4p6 7s1, 4p6 6d1, 4p6 5f1 \\
        ~ & ~ & IV & 60.6072 & 65 & 65 & 1749 & 1 & 4p6 & 4p5 4d1, 4p5 5s1, 4p5 5p1, 4p5 5d1, 4p5 4f1, 4p5 6s1, 4p5 6p1 \\
        40 & Zr & I & 6.63412 & 788 & 599 & 131460 & 1 & 4p6 4d2 5s2 & 4p6 4d3 5s1, 4p6 4d2 5s1 5p1, 4p6 4d1 5s2 5p1, 4p6 4d4, 4p6 4d3 5p1, 4p6 4d2 5s1 6s1, 4p6 4d2 5s1 6p1, 4p6 4d2 5s1 5d1, 4p6 4d3 6s1, 4p6 4d2 5p2, 4p6 4d2 5s1 7p1 \\
        ~ & ~ & II & 13.13 & 188 & 188 & 13747 & 1 & 4p6 4d2 5s1 & 4p6 4d3, 4p6 4d1 5s2, 4p6 4d2 5p1, 4p6 4d1 5s1 5p1, 4p6 4d2 6s1, 4p6 4d2 5d1 \\
        ~ & ~ & III & 23.17 & 84 & 84 & 2728 & 1 & 4p6 4d2 & 4p6 4d1 5s1, 4p6 5s2, 4p6 4d1 5p1, 4p6 5s1 5p1, 4p6 4d1 5d1, 4p6 4d1 6s1, 4p6 4d1 6p1, 4p6 4d1 4f1 \\
        ~ & ~ & IV & 34.41836 & 14 & 14 & 87 & 1 & 4p6 4d1 & 4p6 5s1, 4p6 5p1, 4p6 5d1, 4p6 6s1, 4p6 4f1, 4p6 6p1, 4p6 6d1 \\
        41 & Nb & I & 6.75885 & 649 & 513 & 93849 & 3 & 4d4 5s1 & 4d3 5s2, 4d5, 4d3 5s1 5p1, 4d4 5p1, 4d4 6s1, 4d3 5s1 6s1 \\
        ~ & ~ & II & 14.32 & 487 & 486 & 83641 & 3 & 4d4 & 4d3 5s1, 4d2 5s2, 4d3 5p1, 4d2 5s1 5p1, 4d3 6s1, 4d3 5d1 \\
        ~ & ~ & III & 25.04 & 188 & 188 & 13747 & 1 & 4d3 & 4d2 5s1, 4d2 5p1, 4d1 5s2, 4d1 5s1 5p1, 4d2 5d1, 4d2 6s1 \\
        ~ & ~ & IV & 37.611 & 52 & 52 & 1086 & 1 & 4d2 & 4d1 5s1, 4d1 5p1, 5s2, 5s1 5p1, 4d1 5d1, 4d1 6s1 \\
        42 & Mo & I & 7.09243 & 1654 & 401 & 54542 & 1 & 4d5 5s1 & 4d4 5s2, 4d6, 4d5 5p1, 4d4 5s1 5p1, 4d5 6s1, 4d5 5d1, 4d5 7s1, 4d4 5s1 6s1, 4d5 6d1 \\
        ~ & ~ & II & 16.16 & 851 & 832 & 242182 & 1 & 4d5 & 4d4 5s1, 4d4 5p1, 4d3 5s2, 4d3 5s1 5p1, 4d4 6s1, 4d4 5d1 \\
        ~ & ~ & III & 27.13 & 487 & 487 & 83833 & 1 & 4d4 & 4d3 5s1, 4d3 5p1, 4d2 5s2, 4d2 5s1 5p1, 4d3 6s1, 4d3 5d1 \\
        ~ & ~ & IV & 40.33 & 188 & 188 & 13747 & 1 & 4d3 & 4d2 5s1, 4d2 5p1, 4d1 5s2, 4d1 5s1 5p1, 4d2 6s1, 4d2 5d1 \\
        43 & Tc & I & 7.11938 & 2026 & 486 & 81950 & 1 & 4d5 5s2 & 4d6 5s1, 4d7, 4d5 5s1 5p1, 4d6 5p1, 4d5 5s1 6s1, 4d6 6s1, 4d5 5s1 5d1, 4d6 5d1, 4d6 6p1 \\
        ~ & ~ & II & 15.26 & 1122 & 993 & 331124 & 2 & 4d5 5s1 & 4d6, 4d5 5p1, 4d4 5s2, 4d4 5s1 5p1, 4d5 6s1, 4d5 5d1 \\
        ~ & ~ & III & 29.55 & 851 & 851 & 253888 & 1 & 4d5 & 4d4 5s1, 4d4 5p1, 4d3 5s2, 4d3 5s1 5p1, 4d4 6s1, 4d4 5d1 \\
        ~ & ~ & IV & 41 & 487 & 487 & 83833 & 1 & 4d4 & 4d3 5s1, 4d3 5p1, 4d2 5s2, 4d2 5s1 5p1, 4d3 6s1, 4d3 5d1 \\
        44 & Ru & I & 7.3605 & 1545 & 630 & 133998 & 3 & 4d7 5s1 & 4d6 5s2, 4d8, 4d6 5s1 5p1, 4d7 5p1, 4d7 6s1, 4d7 6p1, 4d7 5d1, 4d6 5s1 6s1, 4d6 5s1 5d1 \\
        ~ & ~ & II & 16.76 & 1472 & 1006 & 350226 & 1 & 4d7 & 4d6 5s1, 4d5 5s2, 4d6 5p1, 4d5 5s1 5p1, 4d6 6s1, 4d6 5d1, 4d5 5s1 6p1 \\
        ~ & ~ & III & 28.47 & 728 & 727 & 177431 & 1 & 4d6 & 4d5 5s1, 4d5 5p1, 4d5 5d1, 4d5 6s1 \\
        ~ & ~ & IV & 45 & 851 & 851 & 253887 & 1 & 4d5 & 4d4 5s1, 4d4 5p1, 4d3 5s2, 4d3 5s1 5p1, 4d4 6s1, 4d4 5d1 \\
        45 & Rh & I & 7.4589 & 98 & 94 & 3494 & 3 & 4d8 5s1 & 4d9, 4d8 5p1, 4d7 5s2, 4d8 6s1 \\
        ~ & ~ & II & 18.08 & 339 & 339 & 41465 & 1 & 4d8 & 4d7 5s1, 4d7 5p1, 4d6 5s2, 4d7 6s1, 4d7 6p1 \\
        ~ & ~ & III & 31.06 & 818 & 816 & 231952 & 1 & 4d7 & 4d6 5s1, 4d6 5p1, 4d5 5s2, 4d6 5d1, 4d6 6s1, 4d6 6p1 \\
        ~ & ~ & IV & 42 & 976 & 976 & 321490 & 1 & 4d6 & 4d5 5s1, 4d5 5p1, 4d4 5s2, 4d5 5d1, 4d5 6s1, 4d5 6p1 \\
        46 & Pd & I & 8.336839 & 150 & 114 & 5098 & 2 & 4d10 & 4d9 5s1, 4d8 5s2, 4d9 5p1, 4d9 6s1, 4d8 5s1 5p1, 4d9 6p1, 4d9 5d1 \\
        ~ & ~ & II & 19.43 & 423 & 402 & 59830 & 1 & 4d9 & 4d8 5s1, 4d8 5p1, 4d7 5s2, 4d7 5s1 5p1, 4d8 6s1, 4d8 5d1, 4d8 6p1 \\
        ~ & ~ & III & 32.93 & 555 & 548 & 104876 & 1 & 4d8 & 4d7 5s1, 4d7 5p1, 4d7 6s1, 4d6 5s1 5p1 \\
        ~ & ~ & IV & 46 & 781 & 781 & 211678 & 1 & 4d7 & 4d6 5s1, 4d6 5p1, 4d6 5d1, 4d6 6s1, 4d6 6p1 \\
        47 & Ag & I & 7.576234 & 18 & 18 & 146 & 3 & 4d10 5s1 & 4d10 5p1, 4d9 5s2, 4d10 6s1, 4d10 6p1, 4d10 5d1, 4d10 7s1, 4d10 7p1, 4d10 6d1, 4d10 4f1, 4d10 8s1 \\
        ~ & ~ & II & 21.4844 & 150 & 143 & 7956 & 1 & 4d10 & 4d9 5s1, 4d9 5p1, 4d8 5s2, 4d9 6s1, 4d9 5d1, 4d8 5s1 5p1, 4d9 6p1 \\
        ~ & ~ & III & 34.8 & 210 & 210 & 17001 & 1 & 4d9 & 4d8 5s1, 4d8 5p1, 4d7 5s2, 4d8 6s1, 4d8 5d1, 4d8 6p1 \\
        ~ & ~ & IV & 49 & 507 & 507 & 90263 & 1 & 4d8 & 4d7 5s1, 4d7 5p1, 4d6 5s2, 4d7 6s1, 4d7 5d1, 4d7 6p1 \\
        48 & Cd & I & 8.99382 & 29 & 29 & 348 & 1 & 4d10 5s2 & 4d10 5s1 5p1, 4d10 5s1 6s1, 4d10 5s1 6p1, 4d10 5s1 5d1, 4d10 5s1 7s1, 4d10 5s1 7p1, 4d10 5s1 6d1, 4d10 5s1 4f1 \\
        ~ & ~ & II & 16.908313 & 40 & 34 & 505 & 1 & 4d10 5s1 & 4d10 5p1, 4d9 5s2, 4d10 6s1, 4d10 5d1, 4d10 6p1, 4d9 5s1 5p1, 4d10 7s1, 4d10 4f1, 4d10 6d1, 4d10 7p1 \\
        ~ & ~ & III & 37.468 & 48 & 48 & 926 & 1 & 4d10 & 4d9 5s1, 4d9 5p1, 4d8 5s2, 4d9 5d1, 4d9 6s1 \\
        ~ & ~ & IV & 51 & 165 & 165 & 10423 & 1 & 4d9 & 4d8 5s1, 4d8 5p1, 4d7 5s2, 4d8 5d1, 4d8 6s1 \\
        49 & In & I & 5.7863556 & 22 & 17 & 130 & 1 & 4d10 5s2 5p1 & 4d10 5s2 6s1, 4d10 5s2 6p1, 4d10 5s2 5d1, 4d10 5s1 5p2, 4d10 5s2 7s1, 4d10 5s2 7p1, 4d10 5s2 6d1, 4d10 5s2 4f1 \\
        ~ & ~ & II & 18.87041 & 34 & 34 & 477 & 1 & 4d10 5s2 & 4d10 5s1 5p1, 4d10 5s1 6s1, 4d10 5s1 5d1, 4d10 5p2, 4d10 5s1 6p1, 4d10 5s1 7s1, 4d10 5s1 4f1, 4d10 5s1 6d1, 4d10 5s1 7p1 \\
        ~ & ~ & III & 28.04415 & 35 & 35 & 536 & 1 & 4d10 5s1 & 4d10 5p1, 4d9 5s2, 4d10 6s1, 4d10 5d1, 4d10 6p1, 4d9 5s1 5p1, 4d10 4f1 \\
        ~ & ~ & IV & 55.45 & 48 & 48 & 926 & 1 & 4d10 & 4d9 5s1, 4d9 5p1, 4d8 5s2, 4d9 5d1, 4d9 6s1 \\
        50 & Sn & I & 7.343918 & 95 & 86 & 3123 & 1 & 4d10 5s2 5p2 & 4d10 5s2 5p1 6s1, 4d10 5s1 5p3, 4d10 5s2 5p1 6p1, 4d10 5s2 5p1 5d1, 4d10 5s2 5p1 7s1, 4d10 5s2 5p1 7p1, 4d10 5s2 5p1 6d1, 4d10 5s2 5p1 4f1, 4d10 5s2 5p1 8s1, 4d10 5s2 5p1 7d1 \\
        ~ & ~ & II & 14.63307 & 25 & 25 & 282 & 1 & 4d10 5s2 5p1 & 4d10 5s1 5p2, 4d10 5s2 6s1, 4d10 5s2 5d1, 4d10 5s2 6p1, 4d10 5s2 7s1, 4d10 5s2 4f1, 4d10 5s2 6d1, 4d10 5s2 7p1, 4d10 5s2 8s1, 4d10 5s2 5f1 \\
        ~ & ~ & III & 30.506 & 34 & 34 & 477 & 1 & 4d10 5s2 & 4d10 5s1 5p1, 4d10 5p2, 4d10 5s1 6s1, 4d10 5s1 5d1, 4d10 5s1 6p1, 4d10 4f1 5s1, 4d10 5s1 7s1, 4d10 5s1 6d1, 4d10 5s1 7p1 \\
        ~ & ~ & IV & 40.74 & 40 & 40 & 682 & 1 & 4d10 5s1 & 4d10 5p1, 4d10 5d1, 4d9 5s2, 4d10 6s1, 4d10 6p1, 4d10 4f1, 4d10 6d1, 4d10 7s1, 4d10 5g1, 4d9 5s1 5p1 \\
        51 & Sb & I & 8.608389 & 206 & 114 & 5865 & 1 & 5s2 5p3 & 5s2 5p2 6s1, 5s2 5p2 6p1, 5s2 5p2 5d1, 5s2 5p2 7s1, 5s2 5p2 6d1, 5s2 5p2 7p1, 5s2 5p2 4f1, 5s2 5p2 8s1, 5s2 5p2 8p1, 5s2 5p2 7d1 \\
        ~ & ~ & II & 16.626 & 74 & 69 & 2006 & 1 & 5s2 5p2 & 5s1 5p3, 5s2 5p1 6s1, 5s2 5p1 5d1, 5s2 5p1 6p1, 5s2 5p1 7s1, 5s2 5p1 6d1, 5s2 5p1 4f1, 5p4 \\
        ~ & ~ & III & 25.3235 & 21 & 21 & 202 & 1 & 5s2 5p1 & 5s1 5p2, 5s2 6s1, 5s2 5d1, 5s2 6p1, 5s2 4f1, 5s2 7s1, 5s2 6d1, 5s2 8s1 \\
        ~ & ~ & IV & 43.804 & 30 & 30 & 365 & 1 & 5s2 & 5s1 5p1, 5p2, 5s1 5d1, 5s1 6s1, 5s1 6p1, 5s1 4f1, 5s1 6d1, 5s1 7s1 \\
        52 & Te & I & 9.00966 & 245 & 99 & 4077 & 1 & 5s2 5p4 & 5s2 5p3 6s1, 5s2 5p3 6p1, 5s2 5p3 5d1, 5s2 5p3 7s1, 5s2 5p3 7p1, 5s2 5p3 6d1, 5s2 5p3 4f1, 5s2 5p3 8s1, 5s2 5p3 7d1 \\
        ~ & ~ & II & 18.6 & 165 & 158 & 10651 & 1 & 5s2 5p3 & 5s1 5p4, 5s2 5p2 6s1, 5s2 5p2 5d1, 5s2 5p2 6p1, 5s2 5p2 7s1, 5s2 5p2 6d1, 5s2 5p2 4f1, 5s2 5p2 7p1, 5s2 5p2 8s1 \\
        ~ & ~ & III & 27.84 & 57 & 57 & 1432 & 1 & 5s2 5p2 & 5s1 5p3, 5s2 5p1 5d1, 5s2 5p1 6s1, 5s2 5p1 6p1, 5s2 5p1 6d1, 5s2 5p1 7s1 \\
        ~ & ~ & IV & 37.4155 & 18 & 18 & 153 & 1 & 5s2 5p1 & 5s1 5p2, 5s2 5d1, 5s2 6s1, 5s2 6p1, 5s2 6d1, 5s2 7s1 \\
\bottomrule
\end{tabularx}
\end{table*}

\begin{table*}\centering
\scriptsize
\caption{Table \ref{tab:atomic_data} continued.}
\begin{tabularx}{\textwidth}{lrrrrrrrlXX}
\toprule
Z &El & Sp &$E^\mathrm{NIST}_\mathrm{i}$[eV] &$n_\mathrm{lev}$ &$n_\mathrm{lev}^\mathrm{red}$ &$n_\mathrm{trans}$ &$n^\mathrm{cfg}_\mathrm{opt}$ & Ground Conf. & Additional Configurations \\ \midrule
        53 & I & I & 10.45126 & 203 & 141 & 8605 & 1 & 5s2 5p5 & 5s2 5p4 6s1, 5s2 5p4 6p1, 5s2 5p4 5d1, 5s2 5p4 7s1, 5s2 5p4 7p1, 5s2 5p4 6d1, 5s2 5p4 4f1, 5s2 5p4 8s1, 5s2 5p4 8p1, 5s2 5p4 7d1 \\
        ~ & ~ & II & 19.13126 & 289 & 247 & 24237 & 1 & 5s2 5p4 & 5s2 5p3 6s1, 5s1 5p5, 5s2 5p3 5d1, 5s2 5p3 6p1, 5s2 5p3 7s1, 5s2 5p3 6d1, 5s2 5p3 4f1, 5s2 5p3 7p1, 5s2 5p3 8s1, 5s2 5p3 7d1, 5s2 5p3 5f1 \\
        ~ & ~ & III & 29.57 & 121 & 121 & 6438 & 1 & 5s2 5p3 & 5s1 5p4, 5s2 5p2 5d1, 5s2 5p2 6s1, 5s2 5p2 7s1, 5s2 5p2 6d1, 5s2 5p2 8s1, 5s2 5p2 7d1 \\
        ~ & ~ & IV & 40.357 & 63 & 63 & 1740 & 1 & 5s2 5p2 & 5s1 5p3, 5s2 5p1 5d1, 5s2 5p1 6s1, 5s2 5p1 6d1, 5s2 5p1 7s1, 5s2 5p1 7d1, 5s2 5p1 8s1 \\
        54 & Xe & I & 12.1298436 & 81 & 70 & 2043 & 1 & 5s2 5p6 & 5s2 5p5 6s1, 5s2 5p5 6p1, 5s2 5p5 5d1, 5s2 5p5 7s1, 5s2 5p5 7p1, 5s2 5p5 6d1, 5s2 5p5 8s1, 5s2 5p5 4f1, 5s2 5p5 7d1 \\
        ~ & ~ & II & 20.975 & 155 & 148 & 9251 & 1 & 5s2 5p5 & 5s1 5p6, 5s2 5p4 6s1, 5s2 5p4 5d1, 5s2 5p4 6p1, 5s2 5p4 7s1, 5s2 5p4 6d1, 5s2 5p4 4f1, 5s2 5p4 7p1, 5s2 5p4 8s1 \\
        ~ & ~ & III & 31.05 & 214 & 213 & 17769 & 1 & 5s2 5p4 & 5s1 5p5, 5s2 5p3 5d1, 5s2 5p3 6s1, 5s2 5p3 6p1, 5s2 5p3 4f1, 5s2 5p3 6d1, 5s2 5p3 7s1, 5s2 5p3 5f1, 5p6 \\
        ~ & ~ & IV & 42.2 & 100 & 100 & 4148 & 1 & 5s2 5p3 & 5s1 5p4, 5s2 5p2 5d1, 5s2 5p2 6s1, 5s2 5p2 4f1, 5s2 5p2 6p1 \\
        55 & Cs & I & 3.893905695 & 17 & 17 & 130 & 1 & 5s2 5p6 6s1 & 5s2 5p6 6p1, 5s2 5p6 5d1, 5s2 5p6 7s1, 5s2 5p6 7p1, 5s2 5p6 6d1, 5s2 5p6 8s1, 5s2 5p6 4f1, 5s2 5p6 8p1, 5s2 5p6 7d1 \\
        ~ & ~ & II & 23.15745 & 103 & 103 & 4393 & 1 & 5s2 5p6 & 5s2 5p5 6s1, 5s2 5p5 5d1, 5s2 5p5 6p1, 5s2 5p5 7s1, 5s2 5p5 6d1, 5s2 5p5 4f1, 5s2 5p5 7p1, 5s2 5p5 8s1, 5s2 5p5 5f1, 5s2 5p5 7d1, 5s2 5p5 8p1 \\
        ~ & ~ & III & 33.195 & 185 & 184 & 13980 & 1 & 5s2 5p5 & 5s1 5p6, 5s2 5p4 5d1, 5s2 5p4 6s1, 5s2 5p4 6p1, 5s2 5p4 4f1, 5s2 5p4 6d1, 5s2 5p4 7s1, 5s2 5p4 7p1, 5s2 5p4 5f1, 5s2 5p4 8s1 \\
        ~ & ~ & IV & 43 & 153 & 153 & 9720 & 1 & 5s2 5p4 & 5s1 5p5, 5s2 5p3 5d1, 5s2 5p3 6s1, 5s2 5p3 6d1, 5s2 5p3 7s1, 5s2 5p3 7d1, 5s2 5p3 8s1 \\
        56 & Ba & I & 5.2116646 & 59 & 58 & 1431 & 1 & 5s2 5p6 6s2 & 5s2 5p6 6s1 5d1, 5s2 5p6 6s1 6p1, 5s2 5p6 5d2, 5s2 5p6 5d1 6p1, 5s2 5p6 6s1 7s1, 5s2 5p6 6s1 6d1, 5s2 5p6 6s1 7p1, 5s2 5p6 5d1 7s1, 5s2 5p6 6s1 8s1, 5s2 5p6 6p2, 5s2 5p6 6s1 4f1, 5s2 5p6 6s1 7d1 \\
        ~ & ~ & II & 10.003826 & 19 & 19 & 159 & 1 & 5s2 5p6 6s1 & 5s2 5p6 5d1, 5s2 5p6 6p1, 5s2 5p6 7s1, 5s2 5p6 6d1, 5s2 5p6 4f1, 5s2 5p6 7p1, 5s2 5p6 5f1, 5s2 5p6 8s1, 5s2 5p6 7d1, 5s2 5p6 8p1 \\
        ~ & ~ & III & 35.8438 & 93 & 93 & 3564 & 1 & 5s2 5p6 & 5s2 5p5 5d1, 5s2 5p5 6s1, 5s2 5p5 4f1, 5s2 5p5 6p1, 5s2 5p5 6d1, 5s2 5p5 7s1, 5s2 5p5 5f1, 5s2 5p5 7p1, 5s2 5p5 7d1, 5s2 5p5 8s1 \\
        ~ & ~ & IV & 47 & 60 & 60 & 1586 & 1 & 5s2 5p5 & 5s1 5p6, 5s2 5p4 5d1, 5s2 5p4 6s1, 5s2 5p4 6p1 \\
        57 & La & I & 5.5769 & 414 & 286 & 31674 & 1 & 5p6 5d1 6s2 & 5p6 5d2 6s1, 5p6 5d3, 5p6 5d1 6s1 6p1, 5p6 4f1 6s2, 5p6 6s2 6p1, 5p6 5d2 6p1, 5p6 4f1 5d1 6s1, 5p6 4f1 6s1 6p1, 5p6 5d2 7s1, 5p6 5d1 6s1 7s1, 5p6 5d2 6d1, 5p6 5d2 7p1, 5p6 4f1 5d2, 5p6 5d1 6s1 7p1, 5p6 6s2 8p1 \\
        ~ & ~ & II & 11.18496 & 66 & 66 & 1663 & 2 & 5p6 5d2 & 5p6 5d1 6s1, 5p6 4f1 6s1, 5p6 4f1 5d1, 5p6 6s2, 5p6 5d1 6p1, 5p6 6s1 6p1, 5p6 4f1 6p1 \\
        ~ & ~ & III & 19.1773 & 15 & 15 & 95 & 2 & 5p6 5d1 & 5p6 4f1, 5p6 6s1, 5p6 6p1, 5p6 7s1, 5p6 6d1, 5p6 5f1, 5p6 7p1, 5p6 8s1 \\
        ~ & ~ & IV & 49.95 & 55 & 55 & 1239 & 1 & 5p6 & 5p5 4f1, 5p5 5d1, 5p5 6s1, 5p5 6p1, 5p5 6d1, 5p5 7s1 \\
        58 & Ce & I & 5.5386 & 1920 & 1236 & 478223 & 1 & 5p6 4f1 5d1 6s2 & 5p6 4f1 5d2 6s1, 5p6 4f2 6s2, 5p6 4f2 5d1 6s1, 5p6 4f1 5d1 6s1 6p1, 5p6 4f1 5d3, 5p6 4f1 6s2 6p1, 5p6 4f2 6s1 6p1, 5p6 4f1 5d2 6p1, 5p6 4f2 5d2 \\
        ~ & ~ & II & 10.956 & 459 & 459 & 69999 & 2 & 5p6 5d2 4f1 & 5p6 4f1 5d1 6s1, 5p6 4f2 6s1, 5p6 4f2 5d1, 5p6 4f1 6s2, 5p6 4f1 5d1 6p1, 5p6 4f2 6p1, 5p6 4f1 6s1 6p1 \\
        ~ & ~ & III & 20.1974 & 237 & 235 & 18710 & 1 & 5p6 4f2 & 5p6 4f1 5d1, 5p6 4f1 6s1, 5p6 5d2, 5p6 4f1 6p1, 5p6 5d1 6s1, 5p6 4f1 6d1, 5p6 4f1 7s1, 5p6 5d1 6p1, 5p6 5d2, 5p6 4f1 7p1, 5p6 4f1 8s1, 5p6 4f1 7d1, 5p6 4f1 6f1, 5p6 4f1 5g1, 5p6 6p2, 5p6 5d1 6d1 \\
        ~ & ~ & IV & 36.906 & 10 & 10 & 42 & 1 & 5p6 4f1 & 5p6 5d1, 5p6 6s1, 5p6 6p1, 5p6 6d1, 5p6 7s1 \\
        59 & Pr & I & 5.4702 & 6516 & 3396 & 3237939 & 5 & 4f3 6s2 & 4f3 6s1 5d1, 4f3 6s1 6p1, 4f3 6s1 7s1, 4f3 6s1 8s1, 4f2 6s2 5d1, 4f2 6s2 6p1, 4f2 5d2 6s1, 4f2 5d2 6p1, 4f2 5d1 6s1 6p1 \\
        ~ & ~ & II & 10.631 & 2007 & 1983 & 1121572 & 1 & 4f3 6s1 & 4f3 5d1, 4f2 5d2, 4f2 5d1 6s1, 4f3 6p1, 4f2 5d1 6p1 \\
        ~ & ~ & III & 21.6237 & 653 & 653 & 131500 & 1 & 4f3 & 4f2 5d1, 4f2 6s1, 4f2 6p1, 4f1 5d2, 4f1 5d1 6s1, 4f2 7s1, 4f2 6d1, 4f2 5f1, 4f2 8s1 \\
        ~ & ~ & IV & 38.981 & 90 & 90 & 2941 & 1 & 4f2 & 4f1 5d1, 4f1 6s1, 4f1 6p1, 5d2, 4f1 6d1, 5d1 6p1 \\
        60 & Nd & I & 5.525 & 12215 & 3405 & 2932375 & 5 & 4f4 6s2 & 4f4 6s1 5d1, 4f4 6s1 6p1, 4f4 6s1 7s1, 4f4 6s1 8s1, 4f3 5d1 6s2, 4f3 5d2 6s1, 4f3 5d1 6s1 6p1 \\
        ~ & ~ & II & 10.783 & 6888 & 6052 & 9633647 & 1 & 4f4 6s1 & 4f4 5d1, 4f3 5d2, 4f3 5d1 6s1, 4f4 6p1, 4f3 5d1 6p1, 4f3 6s1 6p1 \\
        ~ & ~ & III & 22.09 & 2252 & 2185 & 1363594 & 1 & 4f4 & 4f3 5d1, 4f3 6s1, 4f3 6p1, 4f2 5d2, 4f2 5d1 6s1, 4f2 5d1 6p1, 4f2 6s1 6p1 \\
        ~ & ~ & IV & 40.6 & 474 & 474 & 72924 & 1 & 4f3 & 4f2 5d1, 4f2 6s1, 4f2 6p1, 4f1 5d2, 4f1 5d1 6s1, 4f1 5d1 6p1 \\
        61 & Pm & I & 5.577 & 16294 & 2870 & 1913953 & 1 & 4f5 6s2 & 4f5 6s1 5d1, 4f5 6s1 6p1, 4f5 6s1 7s1, 4f4 6s2 5d1, 4f4 6s1 5d2 \\
        ~ & ~ & II & 10.938 & 12372 & 7697 & 14408362 & 1 & 4f5 6s1 & 4f5 5d1, 4f5 6p1, 4f4 6s1 6p1, 4f4 6s1 5d1, 4f4 5d1 6p1 \\
        ~ & ~ & III & 22.44 & 1994 & 1992 & 1044907 & 1 & 4f5 & 4f4 5d1, 4f4 6s1, 4f4 6p1 \\
        ~ & ~ & IV & 41.17 & 817 & 817 & 185068 & 1 & 4f4 & 4f3 5d1, 4f3 6s1, 4f3 6p1 \\
        62 & Sm & I & 5.64371 & 28221 & 1821 & 757867 & 2 & 4f6 6s2 & 4f6 6s1 5d1, 4f6 6s1 6p1, 4f6 6s1 7s1, 4f5 5d1 6s2, 4f5 5d2 6s1 \\
        ~ & ~ & II & 11.078 & 9030 & 3793 & 3418712 & 2 & 4f6 6s1 & 4f7, 4f6 5d1, 4f6 6p1, 4f5 5d1 6s1 \\
        ~ & ~ & III & 23.55 & 3737 & 3717 & 3441421 & 1 & 4f6 & 4f5 5d1, 4f5 6s1, 4f5 6p1 \\
        ~ & ~ & IV & 41.64 & 1994 & 1994 & 1046409 & 1 & 4f5 & 4f4 5d1, 4f4 6s1, 4f4 6p1 \\
        63 & Eu & I & 5.670385 & 103229 & 519 & 79073 & 1 & 4f7 6s2 & 4f7 5d1 6s1, 4f7 6s1 6p1, 4f6 5d1 6s2, 4f7 5d1 6p1, 4f7 6s1 7s1, 4f6 5d2 6s1, 4f7 5d2, 4f7 6s1 7p1, 4f7 6s1 6d1, 4f7 6s1 8s1, 4f7 6s1 5f1, 4f7 6s1 8p1, 4f7 6s1 7d1, 4f7 6p2 \\
        ~ & ~ & II & 11.24 & 22973 & 4379 & 4350550 & 1 & 4f7 6s1 & 4f7 5d1, 4f7 6p1, 4f6 5d1 6s1, 4f6 5d2 \\
        ~ & ~ & III & 24.84 & 5323 & 5245 & 6757496 & 1 & 4f7 & 4f6 5d1, 4f6 6s1, 4f6 6p1 \\
        ~ & ~ & IV & 42.94 & 3737 & 3737 & 3481004 & 1 & 4f6 & 4f5 5d1, 4f5 6s1, 4f5 6p1 \\
        64 & Gd & I & 6.1498 & 103013 & 553 & 84153 & 1 & 4f7 5d1 6s2 & 4f7 5d2 6s1, 4f8 6s2, 4f7 6s2 6p1, 4f7 5d1 6s1 6p1, 4f7 5d3 \\
        ~ & ~ & II & 12.076 & 46733 & 9207 & 18705963 & 3 & 4f7 5d1 6s1 & 4f7 6s2, 4f7 5d2, 4f8 6s1, 4f8 5d1, 4f7 6s1 6p1, 4f7 5d1 6p1, 4f8 6p1 \\
        ~ & ~ & III & 20.54 & 6637 & 4976 & 5864629 & 1 & 4f7 5d1 & 4f8, 4f7 6s1, 4f7 6p1, 4f7 7s1 \\
        ~ & ~ & IV & 44.44 & 5323 & 5317 & 6959526 & 1 & 4f7 & 4f6 5d1, 4f6 6s1, 4f6 6p1 \\
        65 & Tb & I & 5.8638 & 65817 & 3984 & 3778522 & 1 & 4f9 6s2 & 4f8 5d1 6s2, 4f8 5d2 6s1, 4f8 6s2 6p1, 4f9 6s1 6p1, 4f8 5d1 6s1 6p1, 4f9 5d1 6s1 \\
        ~ & ~ & II & 11.513 & 19854 & 11561 & 29940347 & 1 & 4f9 6s1 & 4f8 5d1 6s1, 4f8 6s2, 4f8 5d2, 4f9 5d1 \\
        ~ & ~ & III & 21.82 & 5194 & 4995 & 6119561 & 1 & 4f9 & 4f8 5d1, 4f8 6s1, 4f8 6p1 \\
        ~ & ~ & IV & 39.33 & 5983 & 5951 & 8562557 & 1 & 4f8 & 4f7 5d1, 4f7 6s1, 4f7 6p1 \\
        66 & Dy & I & 5.93905 & 44669 & 2627 & 1690259 & 1 & 4f10 6s2 & 4f9 5d1 6s2, 4f10 6s1 6p1, 4f10 5d1 6s1, 4f9 5d2 6s1, 4f9 6s2 6p1, 4f9 5d1 6s1 6p1, 4f10 6s1 7s1 \\
        ~ & ~ & II & 11.647 & 16034 & 11034 & 28683287 & 2 & 4f10 6s1 & 4f10 5d1, 4f9 5d1 6s1, 4f9 6s2, 4f9 5d2, 4f10 6p1, 4f9 6s1 6p1 \\
        ~ & ~ & III & 22.89 & 3549 & 3510 & 3073722 & 1 & 4f10 & 4f9 5d1, 4f9 6s1, 4f9 6p1 \\
        ~ & ~ & IV & 41.23 & 5194 & 5188 & 6628942 & 1 & 4f9 & 4f8 5d1, 4f8 6s1, 4f8 6p1 \\
\bottomrule
\end{tabularx}
\end{table*}

\begin{table*}\centering
\scriptsize
\caption{Table \ref{tab:atomic_data} continued.}
\begin{tabularx}{\textwidth}{lrrrrrrrlXX}
\toprule
Z &El & Sp &$E^\mathrm{NIST}_\mathrm{i}$[eV] &$n_\mathrm{lev}$ &$n_\mathrm{lev}^\mathrm{red}$ &$n_\mathrm{trans}$ &$n^\mathrm{cfg}_\mathrm{opt}$ & Ground Conf. & Additional Configurations \\ \midrule
        67 & Ho & I & 6.0215 & 23182 & 1425 & 512211 & 1 & 4f11 6s2 & 4f10 5d1 6s2, 4f11 6s1 6p1, 4f10 6s2 6p1, 4f11 5d1 6s1, 4f10 5d2 6s1, 4f10 5d1 6s1 6p1, 4f11 6s1 7s1, 4f11 6s1 7p1 \\
        ~ & ~ & II & 11.781 & 9640 & 6379 & 10019305 & 2 & 4f11 6s1 & 4f11 5d1, 4f11 6p1, 4f10 6s1 6p1, 4f10 6s1 5d1, 4f10 5d1 6p1 \\
        ~ & ~ & III & 22.79 & 1837 & 1826 & 880639 & 1 & 4f11 & 4f10 5d1, 4f10 6s1, 4f10 6p1 \\
        ~ & ~ & IV & 42.52 & 3549 & 3549 & 3146149 & 1 & 4f10 & 4f9 5d1, 4f9 6s1, 4f9 6p1 \\
        68 & Er & I & 6.1077 & 1303 & 516 & 83524 & 5 & 4f12 6s2 & 4f12 6s1 6p1, 4f12 6s1 7s1, 4f12 6s1 6d1, 4f12 6s1 8s1, 4f11 5d1 6s2, 4f11 6s2 6p1, 4f12 5d1 6s1 \\
        ~ & ~ & II & 11.916 & 5333 & 4565 & 5519756 & 2 & 6s1 4f12 & 4f12 6p1, 4f12 5d1, 4f11 6s2, 4f11 5d1 6s1, 4f11 5d2, 4f11 6s1 6p1, 4f11 5d1 6p1 \\
        ~ & ~ & III & 22.7 & 723 & 723 & 145774 & 1 & 4f12 & 4f11 5d1, 4f11 6s1, 4f11 6p1 \\
        ~ & ~ & IV & 42.42 & 1837 & 1837 & 890171 & 1 & 4f11 & 4f10 6s1, 4f10 6p1, 4f10 5d1 \\
        69 & Tm & I & 6.18431 & 1716 & 302 & 34629 & 5 & 4f13 6s2 & 4f13 6s1 6p1, 4f13 5d1 6s1, 4f13 6s1 7s1, 4f13 6s1 8s1, 4f12 5d1 6s2, 4f12 6s2 6p1, 4f13 6s1 7p1, 4f13 5d1 6p1, 4f13 6s1 6d1, 4f12 5d1 6s1 6p1, 4f13 6p2, 4f13 6s1 8p1 \\
        ~ & ~ & II & 12.065 & 1484 & 1399 & 573865 & 2 & 4f13 6s1 & 4f12 6s2, 4f13 5d1, 4f13 6p1, 4f12 5d1 6s1, 4f12 5d2, 4f12 6s1 6p1, 4f12 5d1 6p1 \\
        ~ & ~ & III & 23.66 & 3666 & 2474 & 1639827 & 1 & 4f13 & 4f12 5d1, 4f12 6s1, 4f12 6p1, 4f11 5d1 6s1, 4f11 5d1 6p1, 4f11 6s1 6p1 \\
        ~ & ~ & IV & 42.41 & 723 & 723 & 145774 & 1 & 4f12 & 4f11 5d1, 4f11 6s1, 4f11 6p1 \\
        70 & Yb & I & 6.25416 & 446 & 26 & 290 & 5 & 4f14 6s2 & 4f14 6s1 6p1, 4f14 6s1 5d1, 4f14 6s1 7s1, 4f14 6s1 6d1, 4f14 6s1 7p1, 4f14 6s1 8s1, 4f13 6s2 5d1, 4f13 6s2 6p1, 4f13 6s1 5d2, 4f13 5d1 6s1 6p1, 4f14 6p2 \\
        ~ & ~ & II & 12.179185 & 265 & 262 & 23892 & 2 & 4f14 6s1 & 4f13 6s2, 4f14 5d1, 4f14 6p1, 4f14 7s1, 4f13 5d1 6s1, 4f13 5d2, 4f13 6s1 6p1, 4f13 5d1 6p1 \\
        ~ & ~ & III & 25.053 & 1039 & 788 & 187828 & 1 & 4f14 & 4f13 5d1, 4f13 6s1, 4f13 6p1, 4f13 7s1, 4f13 6d1, 4f12 5d1 6s1, 4f12 5d1 6p1, 4f12 6s1 6p1 \\
        ~ & ~ & IV & 43.61 & 202 & 202 & 12679 & 1 & 4f13 & 4f12 5d1, 4f12 6s1, 4f12 6p1 \\
        71 & Lu & I & 5.425871 & 61 & 58 & 1467 & 1 & 4f14 5d1 6s2 & 4f14 6s2 6p1, 4f14 5d1 6s1 6p1, 4f14 5d2 6s1, 4f14 6s2 7s1, 4f14 6s2 6d1, 4f14 6s2 8s1, 4f14 6s2 7p1, 4f14 6s1 6p2, 4f14 6s2 5f1, 4f14 6s2 7d1 \\
        ~ & ~ & II & 14.13 & 58 & 58 & 1365 & 1 & 4f14 6s2 & 4f14 5d1 6s1, 4f14 6s1 6p1, 4f14 5d2, 4f14 5d1 6p1, 4f14 6s1 7s1, 4f14 6s1 6d1, 4f14 5d1 7s1, 4f14 5d1 6d1 \\
        ~ & ~ & III & 20.9594 & 184 & 111 & 4146 & 1 & 4f14 6s1 & 4f14 5d1, 4f14 6p1, 4f14 7s1, 4f14 6d1, 4f13 5d1 6s1, 4f13 5d1 6p1, 4f13 6s1 6p1 \\
        ~ & ~ & IV & 45.249 & 61 & 61 & 1410 & 1 & 4f14 & 4f13 5d1, 4f13 6s1, 4f13 6p1, 4f13 6d1, 4f13 7s1 \\
        72 & Hf & I & 6.82507 & 313 & 222 & 17682 & 1 & 4f14 5d2 6s2 & 4f14 5d2 6s1 6p1, 4f14 5d3 6s1, 4f14 5d4, 4f14 5d3 6p1, 4f14 5d2 6s1 7s1 \\
        ~ & ~ & II & 14.61 & 129 & 129 & 6811 & 1 & 4f14 5d1 6s2 & 4f14 5d2 6s1, 4f14 5d3, 4f14 5d1 6s1 6p1, 4f14 5d2 6p1, 4f14 5d2 7s1, 4f14 5d1 6s1 7s1 \\
        ~ & ~ & III & 22.55 & 64 & 64 & 1680 & 1 & 4f14 5d2 & 4f14 5d1 6s1, 4f14 6s2, 4f14 5d1 6p1, 4f14 6s1 6p1, 4f14 5d1 6d1, 4f14 5d1 7s1, 4f14 5d1 7p1 \\
        ~ & ~ & IV & 33.37 & 14 & 14 & 87 & 1 & 4f14 5d1 & 4f14 6s1, 4f14 6p1, 4f14 6d1, 4f14 7s1, 4f14 5f1, 4f14 7p1, 4f14 7d1 \\
        73 & Ta & I & 7.549571 & 705 & 450 & 72410 & 2 & 5d3 6s2 & 5d5, 5d4 6s1, 5d3 6s1 6p1, 5d2 6s2 6p1, 5d4 6p1, 5d3 6s1 7s1, 5d3 6s1 8s1 \\
        ~ & ~ & II & 16.2 & 487 & 486 & 83641 & 3 & 5d3 6s1 & 5d2 6s2, 5d4, 5d3 6p1, 5d2 6s1 6p1, 5d3 7s1, 5d3 6d1 \\
        ~ & ~ & III & 23.1 & 188 & 188 & 13747 & 1 & 5d3 & 5d2 6s1, 5d2 6p1, 5d1 6s2, 5d1 6s1 6p1, 5d2 6d1, 5d2 7s1 \\
        ~ & ~ & IV & 35 & 52 & 52 & 1086 & 1 & 5d2 & 5d1 6s1, 5d1 6p1, 6s2, 6s1 6p1, 5d1 6d1, 5d1 7s1 \\
        74 & W & I & 7.86403 & 808 & 315 & 33704 & 1 & 5d4 6s2 & 5d5 6s1, 5d4 6s1 6p1, 5d5 6p1, 5d4 6s1 7s1 \\
        ~ & ~ & II & 16.37 & 851 & 814 & 231520 & 1 & 5d4 6s1 & 5d5, 5d3 6s2, 5d3 6s1 6p1, 5d4 6p1, 5d4 7s1, 5d4 6d1 \\
        ~ & ~ & III & 26 & 487 & 487 & 83833 & 1 & 5d4 & 5d3 6s1, 5d2 6s2, 5d3 6p1, 5d2 6s1 6p1, 5d3 7s1, 5d3 6d1 \\
        ~ & ~ & IV & 38.2 & 188 & 188 & 13747 & 1 & 5d3 & 5d2 6s1, 5d2 6p1, 5d1 6s2, 5d1 6s1 6p1, 5d2 7s1, 5d2 6d1 \\
        75 & Re & I & 7.83352 & 1875 & 501 & 87004 & 2 & 5d5 6s2 & 5d6 6s1, 5d5 6s1 6p1, 5d4 6s2 6p1, 5d5 6s1 7s1, 5d6 6p1, 5d5 6s1 6d1, 5d5 6s1 8s1, 5d4 6s2 7s1 \\
        ~ & ~ & II & 16.6 & 1122 & 860 & 246921 & 2 & 5d5 6s1 & 5d4 6s2, 5d5 6p1, 5d6, 5d4 6s1 6p1, 5d5 7s1, 5d5 6d1 \\
        ~ & ~ & III & 27 & 851 & 848 & 252095 & 1 & 5d5 & 5d4 6s1, 5d4 6p1, 5d3 6s1 6p1, 5d3 6s2, 5d4 7s1, 5d4 6d1 \\
        ~ & ~ & IV & 39.1 & 487 & 487 & 83833 & 1 & 5d4 & 5d3 6s1, 5d3 6p1, 5d2 6s2, 5d2 6s1 6p1, 5d3 7s1, 5d3 6d1 \\
        76 & Os & I & 8.43823 & 984 & 388 & 51476 & 1 & 5d6 6s2 & 5d7 6s1, 5d6 6s1 6p1, 5d6 6s1 7s1, 5d7 6p1, 5d7 7s1, 5d7 7p1, 5d7 6d1 \\
        ~ & ~ & II & 17 & 1435 & 960 & 316620 & 1 & 5d6 6s1 & 5d6 6p1, 5d7, 5d5 6s1 6p1, 5d6 7s1, 5d6 6d1, 5d5 6s1 7p1 \\
        ~ & ~ & III & 25 & 1088 & 1030 & 356508 & 1 & 5d6 & 5d5 6s1, 5d5 6p1, 5d4 6s1 6p1, 5d5 6d1, 5d5 7s1 \\
        ~ & ~ & IV & 41 & 851 & 851 & 253888 & 1 & 5d5 & 5d4 6s1, 5d4 6p1, 5d3 6s2, 5d3 6s1 6p1, 5d4 7s1, 5d4 6d1 \\
        77 & Ir & I & 8.96702 & 385 & 185 & 12521 & 2 & 5d7 6s2 & 5d9, 5d8 6s1, 5d7 6s1 6p1, 5d7 6s1 7s1, 5d8 6p1, 5d8 7s1 \\
        ~ & ~ & II & 17 & 699 & 580 & 115707 & 1 & 5d7 6s1 & 5d8, 5d6 6s2, 5d7 6p1, 5d6 6s1 6p1, 5d7 7s1, 5d7 7p1 \\
        ~ & ~ & III & 28 & 818 & 803 & 224115 & 1 & 5d7 & 5d6 6s1, 5d6 6p1, 5d5 6s2, 5d6 6d1, 5d6 7s1, 5d6 7p1 \\
        ~ & ~ & IV & 40 & 976 & 976 & 321490 & 1 & 5d6 & 5d5 6s1, 5d5 6p1, 5d4 6s2, 5d5 6d1, 5d5 7s1, 5d5 7p1 \\
        78 & Pt & I & 8.95883 & 152 & 110 & 4726 & 3 & 5d9 6s1 & 5d10, 5d9 6p1, 5d9 7s1, 5d8 6s2, 5d8 6s1 6p1, 5d8 6s1 7s1 \\
        ~ & ~ & II & 18.56 & 248 & 232 & 20285 & 1 & 5d9 & 5d8 6s1, 5d7 6s2, 5d8 6p1, 5d8 7s1, 5d8 6d1, 5d8 8s1, 5d8 7d1 \\
        ~ & ~ & III & 29 & 555 & 551 & 105890 & 1 & 5d8 & 5d7 6s1, 5d7 6p1, 5d7 7s1, 5d6 6s1 6p1 \\
        ~ & ~ & IV & 43 & 781 & 780 & 211123 & 1 & 5d7 & 5d6 6s1, 5d6 6p1, 5d6 6d1, 5d6 7s1, 5d6 7p1 \\
        79 & Au & I & 9.225554 & 36 & 33 & 472 & 1 & 5d10 6s1 & 5d9 6s2, 5d10 6p1, 5d9 6s1 6p1, 5d10 7s1, 5d10 7p1, 5d10 6d1, 5d10 8s1, 5d10 8p1 \\
        ~ & ~ & II & 20.203 & 60 & 60 & 1479 & 1 & 5d10 & 5d9 6s1, 5d8 6s2, 5d9 6p1, 5d9 7s1, 5d9 6d1, 5d9 7p1 \\
        ~ & ~ & III & 30 & 210 & 209 & 16830 & 1 & 5d9 & 5d8 6s1, 5d8 6p1, 5d7 6s2, 5d8 7s1, 5d8 6d1, 5d8 7p1 \\
        ~ & ~ & IV & 45 & 507 & 507 & 90263 & 1 & 5d8 & 5d7 6s1, 5d7 6p1, 5d6 6s2, 5d7 7s1, 5d7 6d1, 5d7 7p1 \\
        80 & Hg & I & 10.437504 & 41 & 31 & 404 & 1 & 5d10 6s2 & 5d10 6s1 6p1, 5d10 6s1 7s1, 5d9 6s2 6p1, 5d10 6s1 7p1, 5d10 6s1 6d1, 5d10 6s1 8s1, 5d10 6s1 8p1, 5d10 6s1 7d1, 5d10 6s1 5f1 \\
        ~ & ~ & II & 18.75687 & 38 & 38 & 637 & 1 & 5d10 6s1 & 5d9 6s2, 5d10 6p1, 5d9 6s1 6p1, 5d10 7s1, 5d10 6d1, 5d10 7p1, 5d10 8s1, 5d10 5f1, 5d10 7d1 \\
        ~ & ~ & III & 34.46 & 138 & 138 & 7316 & 1 & 5d10 & 5d9 6s1, 5d8 6s2, 5d9 6p1, 5d8 6s1 6p1, 5d9 7s1, 5d9 6d1 \\
        ~ & ~ & IV & 48.55 & 165 & 165 & 10423 & 1 & 5d9 & 5d8 6s1, 5d8 6p1, 5d7 6s2, 5d8 6d1, 5d8 7s1 \\
        81 & Tl & I & 6.1082873 & 12 & 12 & 66 & 3 & 5d10 6s2 6p1 & 5d10 6s2 7s1, 5d10 6s2 7p1, 5d10 6s2 6d1, 5d10 6s2 8s1, 5d10 6s2 8p1, 5d10 6s2 7d1 \\
        ~ & ~ & II & 20.4283 & 46 & 46 & 890 & 1 & 5d10 6s2 & 5d10 6s1 6p1, 5d10 6s1 7s1, 5d9 6s2 6p1, 5d10 6s1 6d1, 5d10 6p2, 5d10 6s1 7p1, 5d10 6s1 8s1, 5d10 6s1 5f1, 5d10 6s1 7d1, 5d10 6s1 8p1 \\
        ~ & ~ & III & 29.852 & 40 & 40 & 707 & 1 & 5d10 6s1 & 5d10 6p1, 5d9 6s2, 5d9 6s1 6p1, 5d10 7s1, 5d10 6d1, 5d10 7p1, 5d10 5f1, 5d10 8s1, 5d10 7d1, 5d10 8p1 \\
        ~ & ~ & IV & 51.14 & 43 & 43 & 773 & 1 & 5d10 & 5d9 6s1, 5d9 6p1, 5d9 6d1, 5d9 7s1, 5d9 8s1 \\
        82 & Pb & I & 7.4166799 & 95 & 41 & 730 & 6 & 5d10 6s2 6p2 & 5d10 6s2 6p1 7s1, 5d10 6s2 6p1 7p1, 5d10 6s2 6p1 6d1, 5d10 6s2 6p1 8s1, 5d10 6s2 6p1 8p1, 5d10 6s2 6p1 7d1, 5d10 6s2 6p1 5f1, 5d10 6s2 6p1 9s1, 5d10 6s2 6p1 9p1, 5d10 6s2 6p1 8d1 \\
        ~ & ~ & II & 15.032499 & 27 & 27 & 333 & 1 & 5d10 6s2 6p1 & 5d10 6s1 6p2, 5d10 6s2 7s1, 5d10 6s2 6d1, 5d10 6s2 7p1, 5d10 6s2 8s1, 5d10 6s2 5f1, 5d10 6s2 7d1, 5d10 6s2 8p1, 5d10 6s2 9s1, 5d10 6s2 6f1, 5d10 6s2 8d1 \\
        ~ & ~ & III & 31.9373 & 50 & 50 & 1039 & 1 & 5d10 6s2 & 5d10 6s1 6p1, 5d10 6p2, 5d10 6s1 7s1, 5d10 6s1 6d1, 5d9 6s2 6p1, 5d10 6s1 7p1, 5d10 6s1 5f1, 5d10 6s1 8s1, 5d10 6s1 7d1, 5d10 6s1 8p1, 5d10 6s1 6f1 \\
        ~ & ~ & IV & 42.33256 & 68 & 67 & 1956 & 1 & 5d10 6s1 & 5d10 6p1, 5d9 6s2, 5d9 6s1 6p1, 5d10 6d1, 5d10 7s1, 5d10 7p1, 5d10 5f1, 5d10 8s1, 5d10 7d1, 5d9 6p2, 5d10 8p1 \\
\bottomrule
\end{tabularx}
\end{table*}

\begin{table*}\centering
\scriptsize
\caption{Table \ref{tab:atomic_data} continued.}
\begin{tabularx}{\textwidth}{lrrrrrrrlXX}
\toprule
Z &El & Sp &$E^\mathrm{NIST}_\mathrm{i}$[eV] &$n_\mathrm{lev}$ &$n_\mathrm{lev}^\mathrm{red}$ &$n_\mathrm{trans}$ &$n^\mathrm{cfg}_\mathrm{opt}$ & Ground Conf. & Additional Configurations \\ \midrule
        83 & Bi & I & 7.285516 & 176 & 25 & 300 & 2 & 5d10 6s2 6p3 & 5d10 6s2 6p2 7s1, 5d10 6s2 6p2 7p1, 5d10 6s2 6p2 6d1, 5d10 6s2 6p2 8s1, 5d10 6s2 6p2 8p1, 5d10 6s2 6p2 7d1, 5d10 6s2 6p2 9s1, 5d10 6s2 6p2 9p1, 5d10 6s2 6p2 8d1 \\
        ~ & ~ & II & 16.703 & 107 & 85 & 3067 & 2 & 5d10 6s2 6p2 & 5d10 6s2 6p1 7s1, 5d10 6s1 6p3, 5d10 6s2 6p1 6d1, 5d10 6s2 6p1 7p1, 5d10 6s2 6p1 8s1, 5d10 6s2 6p1 5f1, 5d10 6s2 6p1 7d1, 5d10 6s2 6p1 8p1, 5d10 6s2 6p1 9s1, 5d10 6s2 6p1 6f1, 5d10 6s2 6p1 8d1 \\
        ~ & ~ & III & 25.563 & 24 & 24 & 260 & 1 & 5d10 6s2 6p1 & 5d10 6s1 6p2, 5d10 6s2 7s1, 5d10 6s2 6d1, 5d10 6s2 7p1, 5d10 6s2 5f1, 5d10 6s2 8s1, 5d10 6s2 7d1, 5d10 6s2 8p1, 5d10 6s2 6f1 \\
        ~ & ~ & IV & 45.37 & 42 & 42 & 737 & 1 & 5d10 6s2 & 5d10 6s1 6p1, 5d10 6p2, 5d10 6s1 6d1, 5d10 6s1 7s1, 5d9 6s2 6p1, 5d10 6s1 7p1, 5d10 6s1 5f1, 5d10 6s1 8s1, 5d10 6s1 7d1 \\
        84 & Po & I & 8.414 & 251 & 61 & 1614 & 5 & 6s2 6p4 & 6s2 6p3 7s1, 6s2 6p3 7p1, 6s2 6p3 6d1, 6s2 6p3 8s1, 6s2 6p3 8p1, 6s2 6p3 7d1, 6s2 6p3 9p1, 6s2 6p3 8d1, 6s2 6p3 10p1 \\
        ~ & ~ & II & 19.3 & 165 & 139 & 8209 & 1 & 6s2 6p3 & 6s1 6p4, 6s2 6p2 7s1, 6s2 6p2 6d1, 6s2 6p2 7p1, 6s2 6p2 8s1, 6s2 6p2 7d1, 6s2 6p2 5f1, 6s2 6p2 8p1, 6s2 6p2 9s1 \\
        ~ & ~ & III & 27.3 & 57 & 57 & 1432 & 1 & 6s2 6p2 & 6s1 6p3, 6s2 6p1 6d1, 6s2 6p1 7s1, 6s2 6p1 7p1, 6s2 6p1 7d1, 6s2 6p1 8s1 \\
        ~ & ~ & IV & 36 & 18 & 18 & 153 & 1 & 6s2 6p1 & 6s1 6p2, 6s2 6d1, 6s2 7s1, 6s2 7p1, 6s2 7d1, 6s2 8s1 \\
        85 & At & I & 9.31751 & 116 & 48 & 1000 & 1 & 6s2 6p5 & 6s2 6p4 7s1, 6s2 6p4 7p1, 6s2 6p4 6d1, 6s2 6p4 8s1, 6s2 6p4 8p1, 6s2 6p4 7d1 \\
        ~ & ~ & II & 17.88 & 289 & 138 & 7857 & 1 & 6s2 6p4 & 6s2 6p3 7s1, 6s1 6p5, 6s2 6p3 6d1, 6s2 6p3 7p1, 6s2 6p3 8s1, 6s2 6p3 7d1, 6s2 6p3 5f1, 6s2 6p3 8p1, 6s2 6p3 9s1, 6s2 6p3 8d1, 6s2 6p3 6f1 \\
        ~ & ~ & III & 26.58 & 121 & 93 & 3824 & 1 & 6s2 6p3 & 6s1 6p4, 6s2 6p2 6d1, 6s2 6p2 7s1, 6s2 6p2 8s1, 6s2 6p2 7d1, 6s2 6p2 9s1, 6s2 6p2 8d1 \\
        ~ & ~ & IV & 39.65 & 63 & 63 & 1740 & 1 & 6s2 6p2 & 6s1 6p3, 6s2 6p1 6d1, 6s2 6p1 7s1, 6s2 6p1 7d1, 6s2 6p1 8s1, 6s2 6p1 8d1, 6s2 6p1 9s1 \\
        86 & Rn & I & 10.7485 & 65 & 41 & 678 & 1 & 6s2 6p6 & 6s2 6p5 7s1, 6s2 6p5 7p1, 6s2 6p5 6d1, 6s2 6p5 8p1, 6s2 6p5 7d1, 6s2 6p5 9s1, 6s2 6p5 5f1 \\
        ~ & ~ & II & 21.4 & 155 & 144 & 8755 & 1 & 6s2 6p5 & 6s1 6p6, 6s2 6p4 7s1, 6s2 6p4 6d1, 6s2 6p4 7p1, 6s2 6p4 8s1, 6s2 6p4 7d1, 6s2 6p4 5f1, 6s2 6p4 8p1, 6s2 6p4 9s1 \\
        ~ & ~ & III & 29.4 & 214 & 193 & 14481 & 1 & 6s2 6p4 & 6s1 6p5, 6s2 6p3 6d1, 6s2 6p3 7s1, 6s2 6p3 7p1, 6s2 6p3 5f1, 6s2 6p3 7d1, 6s2 6p3 8s1, 6s2 6p3 6f1, 6p6 \\
        ~ & ~ & IV & 36.9 & 100 & 100 & 4148 & 1 & 6s2 6p3 & 6s1 6p4, 6s2 6p2 6d1, 6s2 6p2 7s1, 6s2 6p2 5f1, 6s2 6p2 7p1 \\
        87 & Fr & I & 4.072741 & 5 & 5 & 10 & 1 & 6s2 6p6 7s1 & 6s2 6p6 7p1, 6s2 6p6 6d1 \\
        ~ & ~ & II & 22.4 & 103 & 78 & 2484 & 1 & 6s2 6p6 & 6s2 6p5 7s1, 6s2 6p5 6d1, 6s2 6p5 7p1, 6s2 6p5 8s1, 6s2 6p5 7d1, 6s2 6p5 5f1, 6s2 6p5 8p1, 6s2 6p5 9s1, 6s2 6p5 6f1, 6s2 6p5 8d1, 6s2 6p5 9p1 \\
        ~ & ~ & III & 33.5 & 185 & 177 & 12894 & 1 & 6s2 6p5 & 6s1 6p6, 6s2 6p4 6d1, 6s2 6p4 7s1, 6s2 6p4 7p1, 6s2 6p4 5f1, 6s2 6p4 7d1, 6s2 6p4 8s1, 6s2 6p4 8p1, 6s2 6p4 6f1, 6s2 6p4 9s1 \\
        ~ & ~ & IV & 39.1 & 153 & 127 & 6626 & 1 & 6s2 6p4 & 6s1 6p5, 6s2 6p3 6d1, 6s2 6p3 7s1, 6s2 6p3 7d1, 6s2 6p3 8s1, 6s2 6p3 8d1, 6s2 6p3 9s1 \\
        88 & Ra & I & 5.2784239 & 26 & 23 & 221 & 1 & 6s2 6p6 7s2 & 6s2 6p6 7s1 7p1, 6s2 6p6 6d1 7s1, 6s2 6p6 6d1 7p1, 6s2 6p6 7p2 \\
        ~ & ~ & II & 10.14718 & 5 & 5 & 10 & 1 & 6s2 6p6 7s1 & 6s2 6p6 6d1, 6s2 6p6 7p1 \\
        ~ & ~ & III & 31 & 93 & 78 & 2487 & 1 & 6s2 6p6 & 6s2 6p5 6d1, 6s2 6p5 7s1, 6s2 6p5 5f1, 6s2 6p5 7p1, 6s2 6p5 7d1, 6s2 6p5 8s1, 6s2 6p5 6f1, 6s2 6p5 8p1, 6s2 6p5 8d1, 6s2 6p5 9s1 \\
        ~ & ~ & IV & 41 & 60 & 60 & 1586 & 1 & 6s2 6p5 & 6s1 6p6, 6s2 6p4 6d1, 6s2 6p4 7s1, 6s2 6p4 7p1 \\
        89 & Ac & I & 5.380226 & 170 & 112 & 4830 & 3 & 6p6 6d1 7s2 & 6p6 6d2 7s1, 6p6 6d1 7s1 7p1, 6p6 7s2 7p1, 6p6 6d2 7p1, 6p6 6d3, 6p6 5f1 6d1 7s1, 6p6 5f1 7s1 7p1 \\
        ~ & ~ & II & 11.75 & 68 & 68 & 1743 & 3 & 6p6 7s2 & 6p6 6d1 7s1, 6p6 6d2, 6p6 7s1 7p1, 6p6 6d1 7p1, 6p6 7s1 5f1, 6p6 6d1 5f1, 6p6 7s1 8s1, 6p6 5f1 7p1 \\
        ~ & ~ & III & 17.431 & 12 & 12 & 60 & 1 & 6p6 7s1 & 6p6 6d1, 6p6 5f1, 6p6 7p1, 6p6 8s1, 6p6 7d1, 6p6 6f1 \\
        ~ & ~ & IV & 44.8 & 55 & 55 & 1239 & 1 & 6p6 & 6p5 5f1, 6p5 6d1, 6p5 7s1, 6p5 7p1, 6p5 7d1, 6p5 8s1 \\
        90 & Th & I & 6.3067 & 822 & 590 & 114716 & 3 & 6p6 6d2 7s2 & 6p6 6d3 7s1, 6p6 5f1 6d2 7s1, 6p6 5f1 6d1 7s2, 6p6 6d1 7s2 7p1, 6p6 6d2 7s1 7p1, 6p6 5f1 7s2 7p1, 6p6 6d4, 6p6 5f1 6d1 7s1 7p1, 6p6 5f2 7s2, 6p6 5f1 6d3 \\
        ~ & ~ & II & 12.1 & 343 & 343 & 41678 & 7 & 6p6 6d1 7s2 & 6p6 6d2 7s1, 6p6 5f1 7s2, 6p6 5f1 6d1 7s1, 6p6 6d3, 6p6 5f1 6d2, 6p6 6d1 7s1 7p1, 6p6 5f2 7s1, 6p6 5f1 7s1 7p1, 6p6 5f1 6d1 7p1 \\
        ~ & ~ & III & 18.32 & 79 & 79 & 2269 & 1 & 6p6 5f1 6d1 & 6p6 6d2, 6p6 6d1 7s1, 6p6 7s2, 6p6 5f2, 6p6 5f1 7p1, 6p6 6d1 7p1, 6p6 7s1 7p1, 6p6 5f1 8s1 \\
        ~ & ~ & IV & 28.648 & 15 & 15 & 97 & 9 & 6p6 5f1 & 6p6 6d1, 6p6 7s1, 6p6 7p1, 6p6 7d1, 6p6 8s1, 6p6 6f1, 6p6 8d1, 6p6 9s1 \\
        91 & Pa & I & 5.89 & 6192 & 1990 & 1115150 & 3 & 5f2 6d1 7s2 & 5f3 7s2, 5f2 6d2 7s1, 5f3 6d1 7s1, 5f3 7s1 7p1, 5f2 6d1 7s1 7p1, 5f2 7s2 7p1, 5f2 6d2 7p1 \\
        ~ & ~ & II & 11.9 & 2020 & 2000 & 1141645 & 1 & 5f2 7s2 & 5f3 7s1, 5f3 6d1, 5f2 6d2, 5f2 6d1 7s1, 5f3 7p1, 5f2 6d1 7p1 \\
        ~ & ~ & III & 18.6 & 653 & 653 & 131500 & 1 & 5f2 6d1 & 5f3, 5f2 7s1, 5f2 7p1, 5f1 6d2, 5f1 6d1 7s1, 5f2 8s1, 5f2 7d1, 5f2 6f1, 5f2 9s1 \\
        ~ & ~ & IV & 30.9 & 90 & 90 & 2941 & 1 & 5f2 & 5f1 6d1, 5f1 7s1, 5f1 7p1, 6d2, 5f1 7d1, 6d1 7p1 \\
        92 & U & I & 6.19405 & 11383 & 2286 & 1315448 & 1 & 5f3 6d1 7s2 & 5f4 7s2, 5f4 7s1 6d1, 5f3 6d2 7s1, 5f4 7s1 7p1, 5f3 6d1 7s1 7p1 \\
        ~ & ~ & II & 11.6 & 6929 & 6101 & 9847373 & 1 & 5f3 7s2 & 5f4 7s1, 5f4 6d1, 5f3 6d2, 5f3 6d1 7s1, 5f4 7p1, 5f3 6d1 7p1, 5f3 7s1 7p1 \\
        ~ & ~ & III & 19.8 & 2252 & 2246 & 1441511 & 1 & 5f4 & 5f3 6d1, 5f3 7s1, 5f3 7p1, 5f2 6d2, 5f2 6d1 7s1, 5f2 6d1 7p1, 5f2 7s1 7p1 \\
        ~ & ~ & IV & 36.7 & 474 & 474 & 72924 & 1 & 5f3 & 5f2 6d1 \\
\bottomrule
\end{tabularx}
\end{table*}


\bsp	
\label{lastpage}
\end{document}